\documentclass[aps,prb,reprint,superscriptaddress]{revtex4-1}
\usepackage{graphicx,amssymb,amsfonts,amsmath,subfigure, upgreek}
\usepackage{natbib}
\usepackage{chemformula}
\usepackage{color,soul}
\usepackage{url}

\usepackage{breakurl}
\usepackage[colorlinks]{hyperref}
\usepackage{xspace}
\usepackage{xcolor}
\usepackage{comment}
\usepackage[normalem]{ulem}

\emergencystretch=\maxdimen
\newcommand*{\citen}[1]{%
  \begingroup
    \romannumeral-`\x 
    \setcitestyle{numbers}%
    \cite{#1}%
  \endgroup   
}

\newcommand{\thetaDL}{\ensuremath{\theta_\mathrm{DL}}\xspace}
\newcommand{\thetaFL}{\ensuremath{\theta_\mathrm{FL}}\xspace}
\newcommand{\sigmaDL}{\ensuremath{\sigma_\mathrm{DL}}\xspace}

\newcommand{\sigmaSH}{\ensuremath{\sigma_\mathrm{SH}}\xspace}
\newcommand{\tPt}{\ensuremath{t_\mathrm{Pt}}\xspace}
\newcommand{\tM}{\ensuremath{t_\mathrm{M}}\xspace}
\newcommand{\LW}{\ensuremath{\Delta H}\xspace}
\newcommand{\Hres}{\ensuremath{H_\mathrm{res}}\xspace}
\newcommand{\HOe}{\ensuremath{H_\mathrm{Oe}}\xspace}
\newcommand{\Jdc}{\ensuremath{J_\mathrm{dc}}\xspace}
\newcommand{\aeff}{\ensuremath{\alpha_\mathrm{eff}}\xspace}
\newcommand{\aSML}{\ensuremath{\alpha_\mathrm{SML}}\xspace}
\newcommand{\aSP}{\ensuremath{\alpha_\mathrm{SP}}}
\newcommand{\thetaInt}{\ensuremath{\theta_\mathrm{SH}}\xspace}
\newcommand{\thetaBulk}{\ensuremath{\theta^\mathrm{bulk}_\mathrm{SH}}\xspace}
\newcommand{\Ls}{\ensuremath{\lambda_\mathrm{s}}\xspace}
\newcommand{\Lsbulk}{\ensuremath{\lambda^\mathrm{bulk}_\mathrm{s}}\xspace}
\newcommand{\Gupdown}{\ensuremath{G_{\uparrow\downarrow}}\xspace}
\newcommand{\rbulk}{\ensuremath{\rho^\mathrm{bulk}_\mathrm{Pt}}\xspace}
\newcommand{\rPt}{\ensuremath{\rho_\mathrm{Pt}}\xspace}
\newcommand{\rint}{\ensuremath{\rho_\mathrm{int}}\xspace}
\newcommand{\Gprime}{\ensuremath{G'}\xspace}
\newcommand{\muB}{\ensuremath{\mu_\mathrm{B}}\xspace}
\newcommand{\Ms}{\ensuremath{M_\mathrm{s}}\xspace}

\begin{document}
\title{Charge-Spin Interconversion in Epitaxial Pt Probed by Spin-Orbit Torques in a Magnetic Insulator}

\author{Peng Li}
\affiliation{Geballe Laboratory for Advanced Materials, Stanford University, Stanford, CA 94305, USA}
\affiliation{Department of Applied Physics, Stanford University, Stanford, CA 94305, USA}
\author{Lauren J. Riddiford}
\affiliation{Geballe Laboratory for Advanced Materials, Stanford University, Stanford, CA 94305, USA}
\affiliation{Department of Applied Physics, Stanford University, Stanford, CA 94305, USA}
\author{Chong Bi}
\affiliation{Geballe Laboratory for Advanced Materials, Stanford University, Stanford, CA 94305, USA}
\affiliation{Department of Material Science, Stanford University, Stanford, CA 94305, USA}
\author{Jacob J. Wisser}
\affiliation{Geballe Laboratory for Advanced Materials, Stanford University, Stanford, CA 94305, USA}
\affiliation{Department of Applied Physics, Stanford University, Stanford, CA 94305, USA}
\author{Xiao-Qi Sun}
\affiliation{Geballe Laboratory for Advanced Materials, Stanford University, Stanford, CA 94305, USA}
\affiliation{Department of Physics, Stanford University, Stanford, CA 94305, USA}
\author{Arturas Vailionis}
\affiliation{Stanford Nano Shared Facilities, Stanford University, Stanford, CA 94305, USA}
\affiliation{Department of Physics, Kaunas University of Technology, Studentu Street 50, LT-51368 Kaunas, Lithuania}
\author{Michael J. Veit}
\author{Aaron Altman}
\affiliation{Geballe Laboratory for Advanced Materials, Stanford University, Stanford, CA 94305, USA}
\affiliation{Department of Applied Physics, Stanford University, Stanford, CA 94305, USA}
\author{Xiang Li}
\author{Mahendra DC}
\affiliation{Geballe Laboratory for Advanced Materials, Stanford University, Stanford, CA 94305, USA}
\affiliation{Department of Material Science, Stanford University, Stanford, CA 94305, USA}
\author{Shan X. Wang}
\affiliation{Geballe Laboratory for Advanced Materials, Stanford University, Stanford, CA 94305, USA}
\affiliation{Department of Material Science, Stanford University, Stanford, CA 94305, USA}
\author{Y. Suzuki}
\affiliation{Geballe Laboratory for Advanced Materials, Stanford University, Stanford, CA 94305, USA}
\affiliation{Department of Applied Physics, Stanford University, Stanford, CA 94305, USA}
\author{Satoru Emori}
\affiliation{Department of Physics, Virginia Tech, Blacksburg, VA 24061, USA}
\date{\today}

\begin{abstract}
We measure spin-orbit torques (SOTs) in a unique model system of all-epitaxial ferrite/Pt bilayers to gain insights into charge-spin interconversion in Pt. With negligible electronic conduction in the insulating ferrite, the crystalline Pt film acts as the sole source of charge-to-spin conversion. A small field-like SOT independent of Pt thickness suggests a weak Rashba-Edelstein effect at the ferrite/Pt interface. By contrast, we observe a sizable damping-like SOT that depends on the Pt thickness, from which we deduce the dominance of an extrinsic spin-Hall effect (skew scattering) and Dyakonov-Perel spin relaxation in the crystalline Pt film. Furthermore, our results point to a large internal spin-Hall ratio of $\approx$0.8 in epitaxial Pt. Our experimental work takes an essential step towards understanding the mechanisms of charge-spin interconversion and SOTs in Pt-based heterostructures, which are crucial for power-efficient spintronic devices.
\end{abstract}

\maketitle

\section{Introduction}
Spin-orbit torques (SOTs)~\cite{Sinova2015,Ramaswamy2018} have been recognized as a viable means to manipulate magnetization in thin-film heterostructures. A prototypical SOT-driven medium consists of a ferro(ferri)magnetic metal (FM) interfaced with a nonmagnetic heavy metal (HM) with strong spin-orbit coupling (e.g., Pt). In a conventional picture of SOTs in such a bilayer, an in-plane charge current through the HM (or its surface) generates non--equilibrium spin accumulation via the spin-Hall effect (or Rashba-Edelstein effect)~\cite{Sinova2015,Ramaswamy2018,Hoffmann2013,Manchon2015}. This charge-to-spin conversion then results in SOTs~\cite{Sinova2015,Ramaswamy2018,Brataas2014,Haney2013a}, typically classified into (1) a “damping-like” torque that either enhances or counteracts damping in the magnetic layer and (2) a “field-like” torque that acts similarly to a torque from an external magnetic field. 

Although SOTs are often attributed to charge-to-spin conversion effects in the HM, recent studies point to other effects that impact SOTs in metallic FM/HM bilayers~\cite{Taniguchi2015, Amin2016,  Amin2018, Humphries2017, Baek2018, Gibbons2018, Iihama2018, Safranski2019, Wang2019a, Wu2019, Zhou2015, Garello2013, Pai2015a, Rojas-Sanchez2014, Chen2015b, Dolui2017, Berger2018,Zhu2019, Zhang2015c, Caminale2016}. 
For example, current shunted through the FM can generate additional SOTs through spin-dependent scattering within the FM or across the FM/HM interface~\cite{Taniguchi2015, Amin2016,  Amin2018, Amin2019, Humphries2017, Baek2018, Gibbons2018, Iihama2018, Safranski2019, Wang2019a, Wu2019}. Roughness at the interfaces of FM/HM bilayers, which are typically disordered (i.e., polycrystalline or amorphous), may also contribute to SOTs~\cite{Zhou2015, Garello2013, Pai2015a}. Even with atomically sharp FM/HM interfaces, SOTs may be intrinsically impacted by spin-memory loss~\cite{Rojas-Sanchez2014, Chen2015b, Dolui2017, Berger2018,Zhu2019} and proximity-induced magnetism~\cite{Zhang2015c, Caminale2016} due to orbital hybridization. 

These possible complications in FM/HM bilayers make it difficult to elucidate the fundamental mechanisms of SOTs and, more generally, the underlying charge-to-spin conversion phenomena. These factors also impede reconciling the wide spread of reported spin transport parameters -- particularly for the often-used HM of Pt, with its spin diffusion length in the range $\sim$1-10 nm and its spin-Hall ratio $\sim$0.01-1~\cite{Liu2011a, Morota2011,  Kondou2012, Zhang2013, Niimi2013, Rojas-Sanchez2014, Wang2014, Isasa2015, Boone2015, Zhang2015e, Sagasta2016, Nguyen2016, Stamm2017, Berger2018, Tao2018, Swindells2019,Zhu2019c,Dai2019}. 

Here, we demonstrate a clean ferrimagnetic-insulator/heavy-metal (FI/HM) model system where SOTs originate solely in the HM layer, permitting a simpler analysis of charge-to-spin conversion mechanisms.
Specifically, we investigate SOTs at room temperature in FI/HM bilayers where the FI is an epitaxial spinel ferrite film of MgAl$_{0.5}$Fe$_{1.5}$O$_4$ (MAFO)~\cite{Emori2018a} and the HM is an epitaxial film of Pt, whose high crystallinity is enabled by its excellent lattice match to the spinel~\cite{Lee2019}. The insulating nature of MAFO removes all complications from electronic spin transport in the magnetic layer~\cite{Taniguchi2015, Amin2016,  Amin2018, Amin2019, Humphries2017, Baek2018, Gibbons2018, Iihama2018, Safranski2019, Wang2019a, Wu2019}, and the Pt layer with a sharp crystalline interface minimizes roughness-induced mechanisms~\cite{Zhou2015, Garello2013, Pai2015a}. Spin-memory loss and proximity-induced magnetism are also expected to be significantly weaker in FI/HM~\cite{Emori2018, Valvidares2016, Gray2018a, Riddiford2019} compared to FM/HM~\cite{Rojas-Sanchez2014, Chen2015b, Dolui2017, Berger2018, Zhu2019, Caminale2016, Zhang2015c} due to weaker interfacial hybridization~\cite{Dolui2017}. 

We leverage the low damping of MAFO~\cite{Emori2018a} to quantify both the damping-like and field-like SOTs in a straightforward manner through dc-biased spin-torque ferromagnetic resonance (ST-FMR)~\cite{Liu2011,Kasai2014, Nan2015a, Tiwari2017, Kim2018b}. We observe a large damping-like SOT due to the spin-Hall effect in the bulk of Pt~\cite{Hoffmann2013, Sinova2015}, along with an order-of-magnitude smaller field-like SOT attributed to the interfacial Rashba-Edelstein effect~\cite{Manchon2015,Gambardella2011a}. Modeling the Pt thickness dependence of the damping-like SOT and spin-pumping damping indicates that the skew scattering~\cite{Hoffmann2013, Sinova2015, Sagasta2016, Karnad2018a} and Dyakonov-Perel~\cite{Ryu2016, Freeman2018} mechanisms primarily govern charge-to-spin conversion and spin relaxation, respectively, in epitaxial Pt. 
This combination of mechanisms is distinct from the intrinsic spin-Hall effect and Elliott-Yafet spin relaxation often found in Pt-based systems~\cite{Nguyen2016, Stamm2017, Swindells2019,Zhu2019c,Du2020}. Our modeling results point to a large internal spin-Hall ratio of $\approx$0.8 in Pt, while a small spin-mixing conductance of $\approx$1$\times$$10^{14}$ $\Omega^{-1}$m$^{-2}$ primarily limits the efficiency of the damping-like SOT in the MAFO/Pt bilayer. Our work demonstrates a unique material system and experimental approach to uncover the mechanisms of charge-spin interconversion in Pt, with minimal spurious influence from the adjacent magnetic layer.

\section{Film growth and structural properties}
\label{sec:structure}
MAFO is a low-damping FI with a Curie temperature of $\approx$400 K, which can be grown epitaxially on spinel MgAl$_2$O$_4$ (MAO) substrates~\cite{Emori2018a}. 
We first deposit epitaxial MAFO films on (001)-oriented single-crystal MAO by pulsed laser ablation. A sintered ceramic target of stoichiometric MgAl$_{0.5}$Fe$_{1.5}$O$_4$ is ablated in 10 mTorr of O$_2$ at a fluence of $\approx$2 J/cm$^2$, repetition rate of 1 Hz, target-to-substrate separation of $\approx$75 mm, and substrate temperature of 450 $^{\circ}$C. No post-annealing at a higher temperature is performed. All MAFO films are grown to be 13 nm thick, which is within the optimal thickness range that ensures coherently strained growth and low Gilbert damping~\cite{Emori2018a, Wisser2019}. Broadband ferromagnetic resonance (FMR) measurements confirm a Gilbert damping parameter of $\alpha \approx 0.0017$ for these MAFO films, similar to prior reports~\cite{Emori2018a, Wisser2019, Riddiford2019}. Then, 3-19 nm thick Pt layers are sputtered onto the MAFO films in 3 mTorr of Ar at room temperature. To avoid surface damage, we used a low dc power of 15 W. 

\begin{figure}[tb]
 \centering
 \includegraphics[width=1.00\columnwidth]{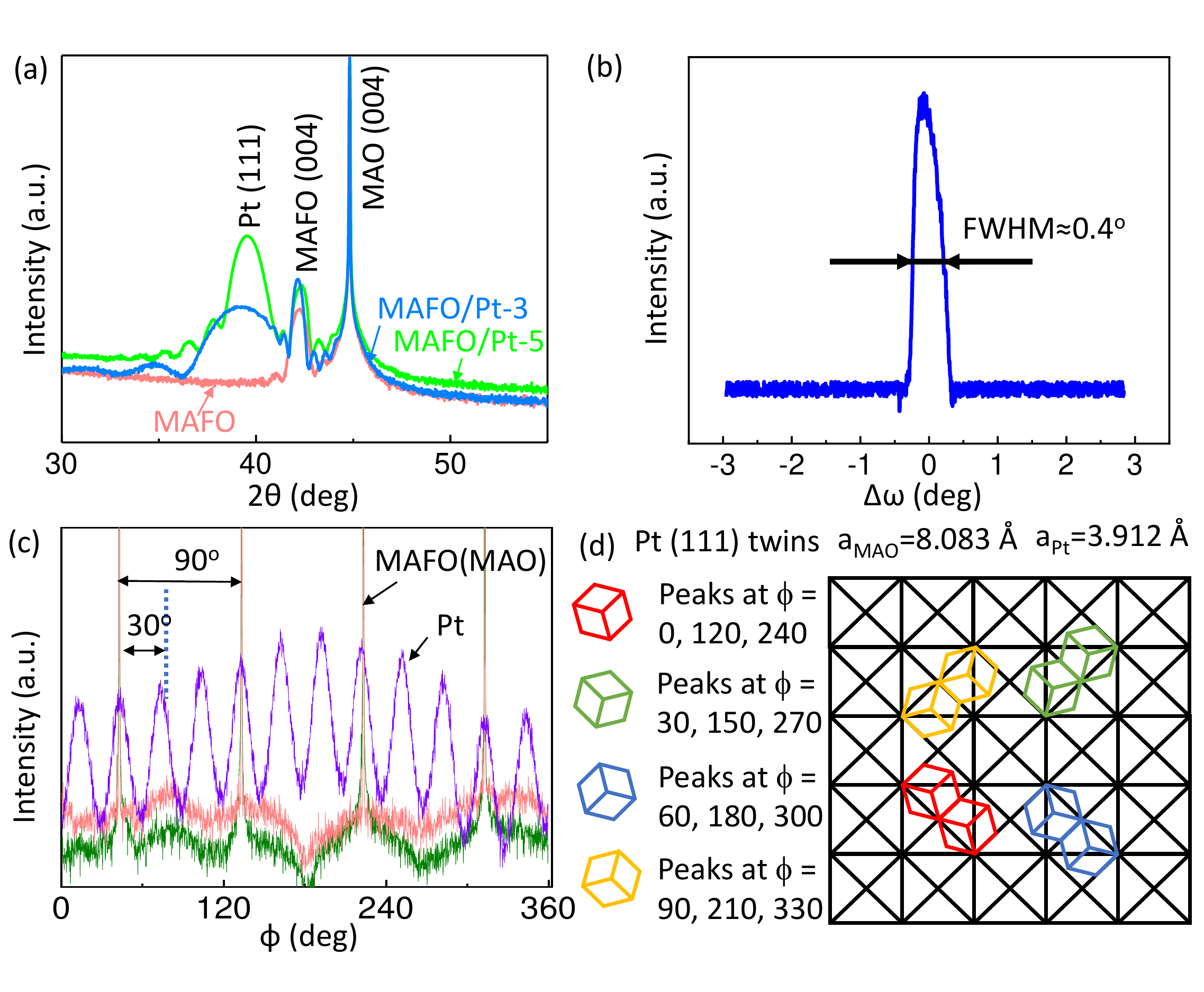}
 \caption{XRD analysis of samples. (a) XRD 2$\theta/\omega$ scans of MAFO (13 nm)/Pt (5 nm), MAFO (13 nm)/Pt (3 nm), and MAFO (13 nm). (b) Rocking curve scan about the Pt (111) peak for the MAFO (13 nm)/Pt (5 nm) sample shown in (a), with FWHM $\approx 0.4^\circ$. (c) XRD $\phi$ scans on the (113) plane of the MAFO (13 nm)/Pt (5 nm) sample. Pink: MAFO. Green: MAO. (d) Lattice matching relationship between the Pt and MAFO (MAO) unit cells.}
 \label{fig:XRD}
\end{figure}

X--ray diffraction (XRD) measurements indicate epitaxy and high crystallinity of our MAFO/Pt samples. Figure~\ref{fig:XRD}(a) shows symmetrical scans for MAFO/Pt and MAFO samples. Strong Pt(111) and MAFO(004) Bragg peaks indicate a high degree of out-of-plane epitaxy. The visible Laue oscillations around the Pt(111) peak for the MAFO/Pt bilayers further indicate high structural quality of the Pt film. 
The degree of crystallinity of the Pt layer is determined by performing a rocking curve measurement around the Pt(111) peak. The narrow rocking curve width of $\approx$0.4$^\circ$ (Fig.~\ref{fig:XRD}(b)) indicates a uniform out-of-plane orientation of Pt crystals with an only small mosaic spread. 

The in-plane orientation of MAFO/Pt is investigated by measuring asymmetrical (113) Bragg peaks for Pt, MAFO, and MAO layers. The MAFO layer is fully coherently strained to the MAO substrate as indicated in the previous study~\cite{Emori2018a}. As can be seen from Fig.~\ref{fig:XRD}(c), the MAFO layer and MAO substrate exhibit four-fold symmetry that is expected from their cubic structures. The Pt(113) peak exhibits twelve maxima indicating a rather complex epitaxial relationship. Careful analysis of the Pt in-plane orientation on MAFO reveals a twinning pattern of the Pt domains, which is presented in Fig.~\ref{fig:XRD}(d). One can distinguish four Pt domains that match MAFO epitaxially and produce in total twelve Pt(113) peaks as shown in Fig.~\ref{fig:XRD}(c). 

It should be noted that the epitaxial growth of Pt on MAFO is in contrast to polycrystalline or amorphous Pt on iron garnets~\cite{Wang2014, Lustikova2014,Chang2017a}. 
Further, X-ray reflectivity indicates a small roughness of $<$0.2 nm at the MAFO/Pt interface. Our structural characterization thus confirms that MAFO/Pt is a high-quality model system with a highly crystalline structure and sharp interface.

\begin{figure}[tb]
 \centering
 \includegraphics[width=1\columnwidth]{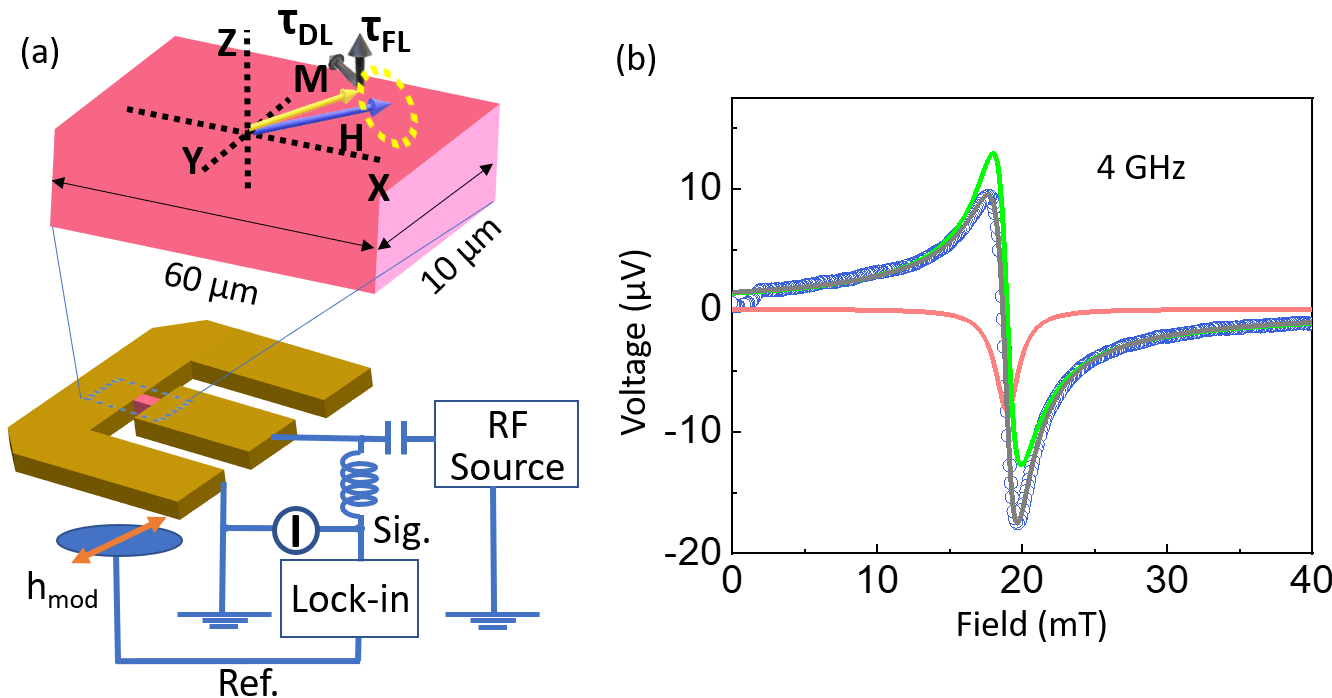}
 \caption{ST-FMR measurement setup. (a) MAFO/Pt stack etched to a 60 $\mu$m $\times$ 10 $\mu$m strip. Magnetization, external field, rf field, and SOTs are shown as the arrows. The ground-signal-ground Au electrode connects MAFO/Pt to the external circuit. (b) FMR spectrum at 4 GHz. Red curve: symmetric Lorentzian contribution. Green curve: antisymmetric Lorentzian contribution. Blue curve: total fit.}
 \label{fig:ST-FMR}
\end{figure}

\section{Results and Discussion}
\subsection{DC-Biased Spin-Torque Ferromagnetic Resonance}\label{subsec:STFMR}
The MAFO/Pt bilayers are lithographically patterned and ion-milled to 60 $\mu$m $\times$ 10 $\mu$m strips with the edges parallel to the in-plane $\left<110\right>$ axes of MAFO. They are then contacted by Ti (5 nm)/Au (120 nm) ground-signal-ground electrodes to allow input of a microwave current for our ST-FMR measurements at room temperature, as illustrated in Fig.~\ref{fig:ST-FMR}(a). We have verified that the magnetic properties of MAFO/Pt are unchanged by the lithographic patterning process (see Appendix~\ref{app:FMRcomparison}).

The microwave current in Pt induces SOTs and a classical Oersted field torque on the magnetization in the MAFO layer. ST-FMR spectra are obtained from the rectified voltage due to magnetoresistance and spin-pumping signals~\cite{Schreier2015, Sklenar2015a} with field modulation~\cite{Goncalves2013b}.  Each integrated ST-FMR spectrum (e.g., Fig.~\ref{fig:ST-FMR}(c)) can be fit with a superposition of symmetric and antisymmetric Lorentzians to extract the  half-width-at-half-maximum linewidth \LW\ and  resonance field \Hres.

\begin{figure}[tb]
 \centering
 \includegraphics[width=1.00\columnwidth]{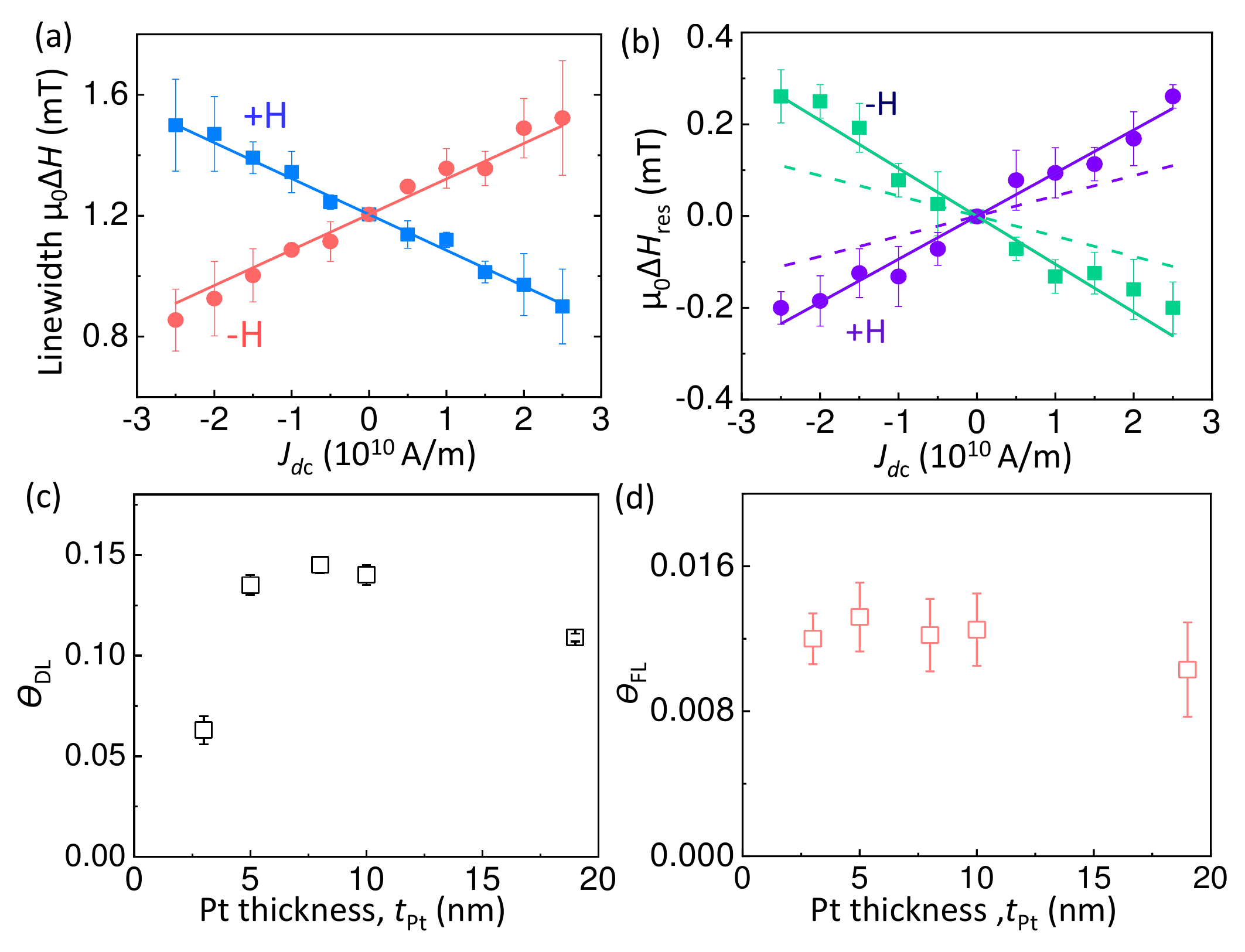}
 \caption{Measurement of SOT efficiencies. (a) Dependence of linewidth $\Delta H$ on dc bias current density \Jdc\ for MAFO (13 nm)/Pt (5 nm). Linewidths and linear fits under positive (blue boxes and line) and negative (red dots and line) magnetic fields are shown. (b) Resonance field change $\Delta \Hres$ as a function of \Jdc\ for the MAFO (13 nm)/Pt (5 nm). Resonance fields and linear fits under positive (purple dots and line) and negative (green dots and line) magnetic fields are shown. The Oersted field contributions are shown as purple (positive) and green (negative) dashed lines. (c,d) Pt thickness dependence of (c) \thetaDL\ and (d) \thetaFL\ for MAFO/Pt. Note the different vertical scales for \thetaDL\  and \thetaFL. The error bars in (c) and (d) are derived from the linear fits of linewidth and resonance field change vs. \Jdc.  } 
 \label{fig:DC-analysis}
\end{figure}

We use an additional dc bias current to directly extract the damping-like and field-like SOTs~\cite{Liu2011,Kasai2014, Nan2015a, Tiwari2017, Kim2018b} in MAFO/Pt. 
This dc bias approach circumvents ambiguities of the oft-used symmetric/antisymmetric Lorentzian ST-FMR lineshape analysis (e.g., where the symmetric Lorentzian can contain voltage signals from spin pumping and thermoelectric effects~\cite{Schreier2015, Sklenar2015a, Okada2019, Schultheiss2012}) and instead probes both SOTs in a direct manner.
In particular, the dc damping-like SOT modifies the effective damping ($\propto$ linewidth \LW) linearly with the dc bias current density \Jdc; the dc field-like torque shifts the resonance field \Hres\ linearly with \Jdc.  Since all of the current  flows in the Pt layer, the classical Oersted field \HOe\ is easily determined from $\HOe/\Jdc=\tPt/2$, where $t_{\text{Pt}}$ is the Pt thickness, and subtracted from $d\Hres/d\Jdc$ to extract the field-like SOT.

Figure~\ref{fig:DC-analysis}(a,b) shows the effect of \Jdc\ on \LW\ and \Hres. 
The linear dependence on current indicates that Joule heating contributions~\cite{Safranski2017} are minimal in these measurements. 
By reversing the magnetization direction (external magnetic field direction), we observe a reversal in the slope for \LW\ (or \Hres) versus \Jdc\, consistent with the symmetry of the SOTs~\cite{Sinova2015,Ramaswamy2018}. 

From the linear slope of linewidth \LW\ versus \Jdc\ (Fig.~\ref{fig:DC-analysis}(a)), the damping-like SOT efficiency \thetaDL\ is readily quantified with~\cite{Liu2011,Nan2015a}

\begin{equation}\label{eq:DLefficiency}
	|\thetaDL|=\frac{2|e|}{\hbar} \frac{(\Hres+M_{\text{eff}}/2)\mu_0 M_s t_{\text{M}}}{|\sin\phi|}\left|\frac{d\alpha_{\text{eff}}}{d J_\text{dc}}\right|,
	\end{equation}
\noindent where $\aeff =|\gamma|\Delta H/(2\pi f)$, $|\gamma|/(2\pi) = 29$ GHz/T  is the gyromagnetic ratio of MAFO~\cite{Emori2018a}, $f$ is the microwave frequency (e.g., $f = 4$ GHz in Figs.~\ref{fig:ST-FMR} and \ref{fig:DC-analysis}), $t_{\text{M}} = 13$ nm is the MAFO thickness, and $\phi = 45^\circ$ or $225^\circ$ is the in-plane magnetization orientation with respect to the current axis ($x$-axis in Fig.~\ref{fig:ST-FMR}(a)). In applying Eq.~\ref{eq:DLefficiency}, we account for the sample-to-sample variation in the saturation magnetization $M_\text{s} = 90-95$ kA/m (determined by SQUID magnetometry) and the effective magnetization $\mu_0 M_\text{eff} = 1.2-1.5$ T (determined by fitting the frequency dependence of resonance field~\cite{Emori2018a}). The large effective magnetization of epitaxial MAFO arises due to significant magnetoelastic easy-plane anisotropy~\cite{Emori2018a}. 

The \tPt\ dependence of \thetaDL\ is summarized in Fig.~\ref{fig:DC-analysis}(c). The increase in $\theta_{\text{DL}}$ with \tPt\ up to $\approx$5~nm (Fig.~\ref{fig:DC-analysis}(c)) suggests that the spin-Hall effect in the Pt bulk is the dominant source of the damping-like SOT~\cite{Haney2013a, Nguyen2016}. The decrease in \thetaDL\ at higher \tPt\ might seem surprising, but a similar trend has been observed in prior experiments~\cite{Nguyen2016}. 

We also quantify the field-like SOT efficiency \thetaFL\ from the linear shift of \Hres\ with \Jdc\ (Fig.~\ref{fig:DC-analysis}(b)) and subtracting the Oersted field contribution~\cite{Nan2015a, Pai2015a}

\begin{equation}\label{eq:FLefficiency}
    |\theta_{\text{FL}}|=\frac{2|e|}{\hbar} \frac{\mu_0 M_s t_{\text{M}}}{|\sin\phi|}\left(\left|\frac{d H_\text{res}}{dJ_{\text{dc}}}\right| - \frac{t_\text{Pt}}{2}|\sin\phi| \right),
	\end{equation}
\noindent where the term proportional to \tPt\ accounts for the Oersted field. 
As shown in Fig.~\ref{fig:DC-analysis}(d), the constant value of \thetaFL\ with Pt thickness implies that the field-like SOT arises from the MAFO/Pt interface, e.g., via the Rashba-Edelestein effect~\cite{Kalitsov2017,Manchon2015,Gambardella2011a}. However, this field-like SOT is weak, i.e., similar in magnitude to the Oersted field (Fig.~\ref{fig:DC-analysis}(b)). Indeed, we find that $\thetaFL\sim0.01$ is about an order of magnitude smaller than \thetaDL. 
 
Based on the dominance of the strongly \tPt-dependent damping-like SOT over the \tPt-independent field-like SOT, we conclude that charge-spin interconversion processes in the bulk of Pt dominate over those at the MAFO/Pt interface. It has been proposed that a field-like SOT could arise from the bulk of Pt in the presence of an imaginary part of the spin-mixing conductance, Im$[\Gupdown]$~\cite{Roy2020}. A substantial Im$[\Gupdown]$ would manifest in a shift in the gyromagnetic ratio (or $g$-factor) in MAFO with the addition of a Pt overlayer~\cite{Tserkovnyak2002}. Since such a shift is not observed, we rule out this scenario of a field-like SOT of ``bulk" origin. In other words, the damping-like torque is the predominant type of SOT that arises from the bulk of Pt. Therefore, in the following sections, we use the damping-like SOT as a measure of charge-to-spin conversion in Pt. 

\subsection{Modeling the Pt-Thickness Dependence of the Spin-Pumping Damping and Damping-Like Spin-Orbit Torque }\label{subsec:model}

We employ a model similar to the one used by Berger \textit{et al.}~\cite{Berger2018} to assess charge-spin interconversion mechanisms in Pt. This model estimates key parameters that govern charge-spin interconversion by fitting the \tPt\ dependence of two experimentally measured quantities: the Gilbert damping parameter $\alpha$ and the damping-like SOT conductivity \sigmaDL.

We have measured the damping parameter $\alpha$ by coplanar-waveguide-based FMR and ST-FMR, which yield consistent results for unpatterned and patterned MAFO/Pt  (see Appendix~\ref{app:FMRcomparison}). As can be seen in Fig.~\ref{fig:model}(b,c), MAFO/Pt bilayers exhibit higher $\alpha$ than the bare MAFO films with \tPt = 0. In Sec.~\ref{subsec:zeroSML}, we attribute this damping enhancement to spin pumping~\cite{Tserkovnyak2002}, i.e., due to the loss of spin angular momentum pumped from the resonantly excited MAFO layer to the adjacent spin sink layer of Pt. In Sec.~\ref{subsec:finiteSML}, we further consider an additional contribution to the enhancement of $\alpha$ due to spin-memory loss or two-magnon scattering.

To parameterize the strength of the damping-like SOT, we employ the ``SOT conductivity," $\sigmaDL=\thetaDL/\rPt$.  
Normalizing \thetaDL\ by the Pt resistivity \rPt makes explicit the relationship between the SOT and electronic transport. We also remark that \sigmaDL is equivalent to the SOT efficiency per unit electric field $\xi^E_{DL}$ in Refs.~\citen{Nguyen2016, Zhu2019c}. The \tPt\ dependence of \rPt\ (fit curve in Fig.~\ref{fig:model}(a)) is interpolated by using the empirical model outlined in Appendix~\ref{app:resistivity}. 

In contrast to Ref.~\citen{Berger2018} that studies FM/Pt bilayers where electronic spin transport in the FM can generally yield additional effects that impact SOTs, our MAFO/Pt system restricts the source of SOTs to Pt. We are therefore able to examine the spin-Hall effect of Pt without any complications from an electrically conductive FM.

To model our experimental results, we consider two types of spin-Hall effect~\cite{Hoffmann2013, Sinova2015}:
\begin{itemize}
\item the \emph{intrinsic} mechanism, where the internal spin-Hall ratio \thetaInt\ -- i.e., the charge-to-spin conversion efficiency within the Pt layer itself -- is proportional to \rPt, with a constant internal spin-Hall conductivity \sigmaSH = \thetaInt/\rPt;
\item the \emph{skew scattering} mechanism, where \thetaInt\ is independent of \rPt. 
\end{itemize}
We also consider two mechanisms of spin relaxation that govern the spin diffusion length \Ls\ in Pt~\cite{Boone2015, Ryu2016, Freeman2018}: 
\begin{itemize}
\item Elliott-Yafet (EY) spin relaxation, where spins depolarize \emph{during} scattering such that \Ls\ scales inversely with \rPt, i.e., $\Ls = \Lsbulk \rbulk/\rPt$; 
\item Dyakonov-Perel (DP) spin relaxation, where spins depolarize \emph{between} scattering events such that \Ls\ is independent of \rPt\ (as outlined by Boone \textit{et al.}~\cite{Boone2015}).
\end{itemize}
Thus, we model four combinations of the above-listed charge-to-spin conversion and spin relaxation mechanisms, as shown in Fig.~\ref{fig:model}(b-e).

Similar to Ref.~\citen{Berger2018}, we self-consistently fit $\alpha$ vs.\tPt\ (Fig.~\ref{fig:model}(b,c)) and $\sigmaDL$ vs. \tPt\ (Fig.~\ref{fig:model}(d,e)) by using standard spin diffusion models~\cite{Tserkovnyak2002, Boone2015,Haney2013a}, as elaborated in Appendix~\ref{app:diffusionEqs}, with four free parameters: 
\begin{itemize}
\item spin diffusion length \Ls\ in the case of DP spin relaxation, or its bulk-limit value \Lsbulk in the case of EY spin relaxation;
\item internal spin-Hall ratio \thetaInt of Pt in the case of skew scattering, or its bulk-limit value $\thetaBulk = \sigmaSH \rbulk$ in the case of intrinsic spin-Hall effect;
\item real part of the spin-mixing conductance \Gupdown at the MAFO/Pt interface, neglecting the imaginary part as justified in Sec.~\ref{subsec:STFMR};
\item effective damping enhancement \aSML\ due to interfacial spin-memory loss or two-magnon scattering, as discussed in detail in Sec.~\ref{subsec:finiteSML}.
\end{itemize}
A key assumption here is that the spin-pumping damping and damping-like SOT share the same values of \Ls, \Gupdown, and \aSML. This is justified by the enforcement of Onsager reciprocity on the charge-spin interconversion processes of spin pumping and SOT~\cite{Brataas2011, Berger2018}. We also assume a negligible interfacial contribution to the spin-Hall effect in Pt~\cite{Wang2016}, which would yield a finite value of \sigmaDL\ when \tPt\ is extrapolated to zero. Indeed, as shown in Fig.~\ref{fig:model}, the \tPt\ dependence of \sigmaDL\ is adequately modeled without incorporating the interfacial spin-Hall effect.

For simplicity, we first proceed by setting $\aSML = 0$ in Sec.~\ref{subsec:zeroSML}. This is a reasonable assumption considering that interfacial spin-memory loss is likely much weaker in MAFO/Pt than in all-metallic FM/Pt systems~\cite{Rojas-Sanchez2014,Chen2015b, Dolui2017, Berger2018, Zhu2019, Caminale2016, Zhang2015c}. Nevertheless, we also discuss the consequence of $\aSML > 0$ in Sec.~\ref{subsec:finiteSML}.

\subsection{Mechanisms and Parameters for Charge-Spin Interconversion in Pt: Zero Spin-Memory Loss}
\label{subsec:zeroSML}
\begin{figure}[t]
 \centering
 \includegraphics[width=1.00\columnwidth]{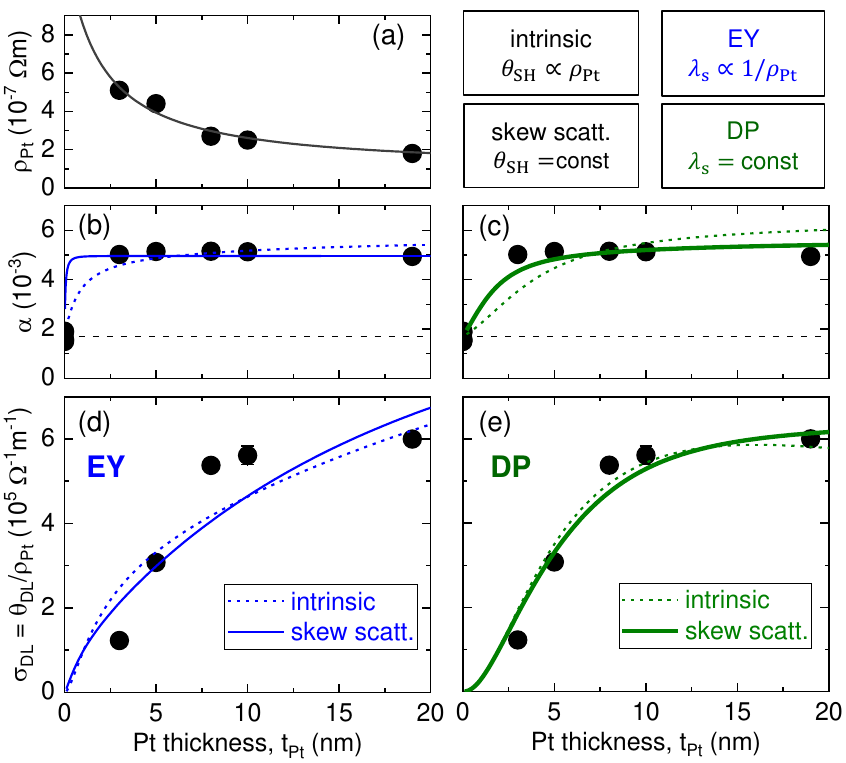}
 \caption{Pt thickness dependence of: (a) resistivity \rPt, with the solid curve showing the fit obtained with the model described in Appendix~\ref{app:resistivity}; (b,c) Gilbert damping parameter $\alpha$, with the black horizontal dashed line indicating the average damping parameter of uncapped MAFO; (d,e) damping-like SOT conductivity \sigmaDL. 
 Modeling results based on Elliott-Yafet (EY) spin relaxation are shown in (b,d), whereas those based on Dyakonov-Perel (DP) spin relaxation are shown in (c,e). The dotted curves are based on the intrinsic spin-Hall effect, and the solid curves are based on skew scattering. The modeling results in (b-e) are obtained by assuming zero spin-memory loss and two-magnon scattering (i.e., \aSML = 0). In (b-e), the error bars are comparable to or smaller than the symbol size and are derived from the linear fits of FMR linewidth vs. frequency (b,c) and dc bias current density (d,e).} 
 \label{fig:model}
\end{figure}

Our modeling results under the assumption of zero spin-memory loss are summarized in Fig.~\ref{fig:model} and Table~\ref{table:model}. We find that the combination of skew scattering and DP spin relaxation (solid green curves in Fig.~\ref{fig:model}(c,e)) best reproduces the \tPt\ dependence of both $\alpha$ and \sigmaDL. Although this observation does not necessarily rule out the coexistence of other mechanisms~\cite{Ryu2016, Freeman2018, Dai2019, Berger2018}, it suggests the dominance of the skew scattering + DP combination in the epitaxial Pt film. Skew scattering in highly crystalline Pt is consistent with what is expected for ``superclean" Pt, in contrast to the intrinsic spin-Hall effect that is dominant in ``moderately dirty" Pt~\cite{Sagasta2016}. 

The dominance of DP spin relaxation -- i.e., spin depolarization (dephasing) from precession about effective spin-orbit fields -- is perhaps surprising, since it is usually thought to be inactive in centrosymmetric metals (e.g., Pt). Indeed, in the context of spin transport in Pt, it is typical to assume EY spin relaxation where spins depolarize when their carriers (e.g., electrons) are scattered~\cite{Nguyen2016, Stamm2017, Swindells2019,Zhu2019c,Du2020}. However, a recent quantum transport study indicates the dominance of DP spin relaxation in crystalline Pt~\cite{Ryu2016}, which is in line with our conclusion here. Possible origins of the DP mechanism include symmetry breaking between the substrate and the surface of the crystalline Pt film~\cite{Long2016} and strong spin mixing caused by the distinct band structure (large spin Berry curvature) of Pt~\cite{Freeman2018}. DP spin relaxation may also be more pronounced when proximity-induced magnetism in Pt is negligible~\cite{Freeman2018}, as is likely the case for Pt interfaced with the insulating MAFO~\cite{Gray2018b}. We also note that DP spin relaxation has been previously used to model the angular dependence of spin-Hall magnetoresistance~\cite{Althammer2013,Chen2013} in MAFO/Pt~\cite{Riddiford2019}. The combination of skew scattering \emph{and} DP spin relaxation, though not reported in prior SOT experiments, is reasonable for MAFO/Pt. 

\begin{table}[b]
\begin{tabular}{lcccc}
\hline
model            &\aSML & \Gupdown ($\Omega^{-1}$m$^{-2}$) & \Ls (nm) & \thetaInt \\ \hline
intrinsic + EY     & 0    & $2.5\times 10^{14}$                     & 21                      & 0.21                     \\
skew scatt. + EY   & 0    & $1.1\times 10^{14}$                     & 4.7                     & 1.2                     \\
intrinsic + DP     & 0    & $1.8\times 10^{14}$                     & 5.7                     & 0.25                     \\
\textbf{skew scatt. + DP}     & \textbf{0}    & \boldmath{$1.3\times 10^{14}$}                     & \textbf{3.3}                     & \textbf{0.83}   \\ \hline                  
\end{tabular}
\caption{Parameters for the modeled curves in Fig.~\ref{fig:model}. For charge-to-spin conversion = intrinsic (for spin relaxation = EY), \thetaInt\ (\Ls) is the value at $\rPt = \rbulk = 1.1\times10^{-7}$ $\Omega$m.}
\label{table:model}
\end{table}

We now discuss the parameters quantified with our model, as summarized in the ``skew scatt.+DP" row in Table~\ref{table:model}. The value of $\Gupdown\approx 1 \times 10^{14}$ $\Omega^{-1}$m$^{-2}$ is comparable to those previously reported for FI/Pt interfaces~\cite{Sun2013, Hahn2013a, Wang2014, Riddiford2019}, and $\Ls \approx 3$ nm is in the intermediate regime of $\Ls \sim1-10$ nm in prior reports on Pt~\cite{Liu2011a, Morota2011,  Kondou2012, Zhang2013, Niimi2013, Rojas-Sanchez2014, Wang2014, Isasa2015, Boone2015, Zhang2015e, Sagasta2016, Nguyen2016, Stamm2017, Berger2018, Tao2018, Swindells2019,Zhu2019c,Dai2019}. 

We find a large internal spin-Hall ratio of $\thetaInt\approx 0.8$. While a few studies have alluded to \thetaInt\ on the order of unity in transition metals~\cite{Zhu2019c, Berger2018, Wang2014, Weiler2014b, Keller2019}, our experimental study is the first to derive such a large value in Pt without uncertainties from a conductive FM~\cite{Weiler2014b,Zhu2019c, Berger2018, Keller2019} or microwave calibration~\cite{Wang2014,Weiler2014b, Berger2018, Keller2019}. Our finding of $\thetaInt$ approaching unity is also distinct from previously reported spin-Hall ratios $<~0.1$ in all-epitaxial FM/Pt~\cite{Huo2017, Keller2018, Guillemard2018, Ryu2019, Du2020}.
This discrepancy may be partially explained by the conductive FM reducing the apparent charge-to-spin conversion efficiency, or by the indirect nature of the measurements in these reports.
With direct SOT measurements on the model-system MAFO/Pt bilayers, our study points to the possibility of a strong spin-Hall effect in highly crystalline Pt in the skew-scattering regime, where the charge-to-spin conversion efficiency could be greater than the limit set by the intrinsic spin-Hall effect~\cite{Hoffmann2013, Sinova2015, Sagasta2016, Zhu2019c}.

\subsection{Mechanisms and Parameters for Charge-Spin Interconversion in Pt: Finite Spin-Memory Loss}
\label{subsec:finiteSML}
A natural question at this point is how finite spin-memory loss at the MAFO/Pt interface impacts the parameters quantified in our modeling. Moreover, while bare MAFO exhibits negligible two-magnon scattering~\cite{Emori2018a}, an overlayer (Pt in this case) on top of MAFO may give rise to two-magnon scattering at the interface~\cite{Wisser2019b}. Both spin-memory loss and two-magnon scattering would have the same consequence in that they enhance the apparent damping parameter, $\alpha$, independent of \tPt~\cite{Berger2018, Zhu2019b}. We therefore model spin-memory loss and two-magnon scattering with a phenomenological parameter, \aSML. 

\begin{figure*}[t!]
 \centering
 \includegraphics[width=2.00\columnwidth]{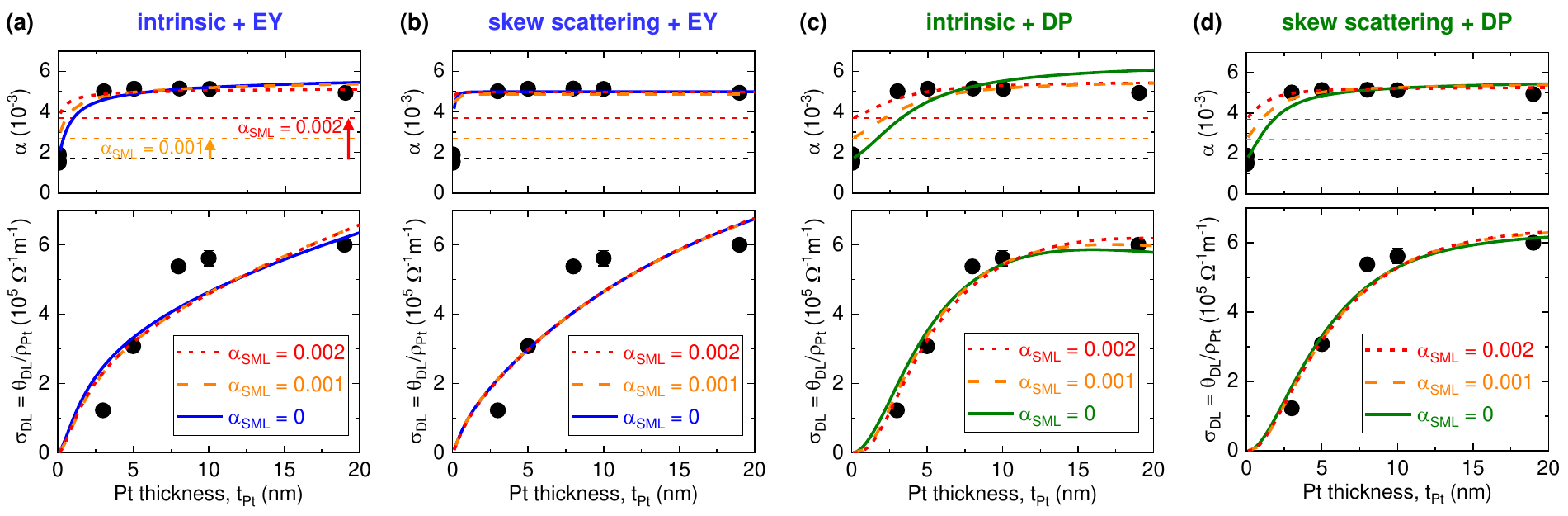}
 \caption{Pt thickness dependence of the Gilbert damping parameter $\alpha$ and the damping-like SOT conductivity \sigmaDL, taking into account different strengths of spin-memory loss or two-magnon scattering (parameterized by \aSML), for the four combinations of charge-to-spin conversion and spin relaxation mechanisms: (a) intrinsic spin-Hall effect + Elliott-Yafet (EY), (b) skew scattering + EY, (c) intrinsic spin-Hall effect + Dyakonov-Perel (DP), and (d) skew scattering + DP. The error bars are comparable to or smaller than the symbol size; they are derived from the linear fits of FMR linewidth vs. frequency (for $\alpha$) and ST-FMR linewidth vs. dc bias current density (for \sigmaDL).}
 \label{fig:finiteSML}
\end{figure*}

\begin{table*}[]
\begin{tabular}{|c|ccc|ccc|ccc|ccc|}
\hline
      & \multicolumn{3}{c|}{intrinsic + EY} & \multicolumn{3}{c|}{skew scatt. + EY} & \multicolumn{3}{c|}{intrinsic + DP} & \multicolumn{3}{c|}{skew scatt. + DP} \\ \hline
\aSML  & \Gupdown ($\Omega^{-1}$m$^{-2}$)     & \Ls (nm)     & \thetaInt     & \Gupdown ($\Omega^{-1}$m$^{-2}$)     & \Ls (nm)     & \thetaInt     & \Gupdown ($\Omega^{-1}$m$^{-2}$)     & \Ls (nm)     & \thetaInt    & \Gupdown ($\Omega^{-1}$m$^{-2}$)     & \Ls (nm)     & \thetaInt     \\ \hline
0     & $2.5\times10^{14}$         & 21     & 0.21         & $1.1\times10^{14}$          & 4.7     & 1.2          & $1.8\times10^{14}$         & 5.7     & 0.25        & $\mathbf{1.3\times10^{14}}$          & \textbf{3.3}     & \textbf{0.83}         \\
0.001 & $1.5\times10^{14}$        & 23     & 0.40         & $0.7\times10^{14}$          & 4.7     & 2.7          & $1.0\times10^{14}$         & 6.2     & 0.53        & $0.9\times10^{14}$          & 3.6     & 1.5          \\
0.002 & $0.6\times10^{14}$         & 26     & 1.3          & $0.4\times10^{14}$          & 5.0     & 7.5          & $0.6\times10^{14}$         & 7.1     & 1.2         & $0.5\times10^{14}$          & 3.8     & 4.1          \\ \hline
\end{tabular}
\caption{Parameters for the modeled curves in Fig.~\ref{fig:finiteSML}. For charge-to-spin conversion = intrinsic (for spin relaxation = EY), \thetaInt\ (\Ls) is the value at $\rPt = \rbulk = 1.1\times10^{-7}$ $\Omega$m.}
\label{table:finiteSML}
\end{table*}

Figure~\ref{fig:finiteSML} and Table~\ref{table:finiteSML} summarize our modeling results incorporating finite spin-memory loss or two-magnon scattering (i.e., $\aSML > 0$). 
Finite \aSML\ does not improve the fit quality in \sigmaDL vs. \tPt\ of the EY models (Fig.~\ref{fig:finiteSML}(a,b)). By contrast, the fit quality is improved for the DP models with increasing \aSML, particularly in $\alpha$ vs. \tPt (Fig.~\ref{fig:finiteSML}(c,d)). We therefore focus on the results for the DP models. 

As shown in Table~\ref{table:finiteSML}, increasing \aSML\ significantly decreases \Gupdown, consistent with the reduced share of spin pumping in the damping enhancement. To compensate for the smaller \Gupdown, the internal spin-Hall ratio \thetaInt\ must increase to reproduce the \tPt\ dependence of \sigmaDL (Ref.~\citen{Zhu2019b}). 
In the ``skew scattering + DP" model, shown to be most plausible in Sec.~\ref{subsec:zeroSML}, \thetaInt\ increases to values \emph{exceeding} unity with finite \aSML.
At a sufficiently large \aSML\ of $\gtrsim 0.002$, the ``intrinsic + DP" model appears to becomes plausible (see Fig.~\ref{fig:finiteSML}(c)), but this scenario also yields $\thetaInt > 1$. 

In both of the above DP scenarios, substantial spin-memory loss or two-magnon scattering apparently leads to an unphysically large value of internal spin-Hall ratio in Pt exceeding unity. It is then reasonable to conclude that spin-memory loss and two-magnon scattering is negligibly small in epitaxial MAFO/Pt. This is in stark contrast to the large spin-memory loss deduced for all-metallic FM/Pt bilayers~\cite{Berger2018}. The small spin-memory loss in MAFO/Pt also suggests fundamentally different spin-transport mechanisms between FM/Pt and FI/Pt systems, which could be exploited for more efficient SOT devices in the future. Our finding motivates further studies to test whether the negligible spin-memory loss is due to the crystalline growth or due to the absence of proximity-induced magnetism.

\subsection{Implications of the Large Internal Spin-Hall Ratio in Pt}

From our analysis in Sec.~\ref{subsec:zeroSML}, we have arrived at a large internal spin-Hall ratio of $\thetaInt \approx 0.8$ in epitaxial Pt. Yet, the observed spin-torque efficiency of $\thetaDL \lesssim 0.15$ implies an interfacial spin transparency ratio $\thetaDL/\thetaInt$ of $\lesssim 0.2$. In other words, at most only 20\% of the spin accumulation generated by the spin-Hall effect in Pt is transferred to the magnetic MAFO layer as the damping-like SOT. The origin of this inefficient spin transfer, according to the spin diffusion model employed here, is the small spin-mixing conductance of $\Gupdown \approx 1\times 10^{14}$ $\Omega^{-1}$m$^{-2}$, which is several times lower than $\Gupdown$ computationally predicted for FM/Pt interfaces~\cite{Liu2014c, Mahfouzi2017, Zhao2018}.  The small \Gupdown\ results in a substantial spin backflow~\cite{Zhu2019b, Zhu2021} that prevents efficient transmission of spin angular momentum across the MAFO/Pt interface. We emphasize that spin-memory loss is likely negligible at the MAFO/Pt interface (see Sec.~\ref{subsec:finiteSML}) and hence not responsible for the inefficient spin transfer. 

There may be an opportunity to enhance the spin transparency -- and hence the SOT efficiency -- by engineering the interface. One possible approach is to use an ultrathin insertion layer of NiO, which has been reported to increase the spin transparency ratio to essentially unity in FM/Pt systems~\cite{Zhu2021}. However, it remains to be explored whether the ultrathin NiO insertion layer can increase the spin transparency without causing substantial interfacial spin scattering~\cite{Wisser2019b} in FI/Pt bilayers. An increased spin transparency (via enhanced \Gupdown) also leads to higher spin-pumping damping~\cite{Tserkovnyak2002, Brataas2012a}, which may not be desirable for applications driven by precessional switching or auto-oscillations.  

Another striking implication of the large internal spin-Hall ratio is a large maximum spin-Hall conductivity $\sigmaSH = \thetaInt/\rbulk$ of $\approx 8\times10^{6}$ $\Omega^{-1}$m$^{-1}$, which is at least an order of magnitude greater than $\sigmaSH \sim 10^{4}-10^{5}$ $\Omega^{-1}$m$^{-1}$ typically predicted from band structure calculations~\cite{Tanaka2008, Guo2008, Obstbaum2016, Go2018, Derunova2019}. It should be noted, however, that these calculations are for the \emph{intrinsic} spin-Hall effect, whereas our experimental data are best captured by the \emph{extrinsic} spin-Hall effect of skew scattering. We thus speculate that this difference in mechanism could account for the discrepancy in \sigmaSH\ derived from our experimental work and from prior calculations. 

Finally, we comment on remaining open fundamental questions.
Comparing MAFO/epitaxial-Pt and MAFO/\emph{polycrystalline}-Pt could reveal the critical role of crystallinity in charge-spin interconversion, spin relaxation, and the internal spin-Hall ratio in Pt. This comparison study is precluded here due to the difficulty in growing polycrystalline Pt on MAFO; Pt has a strong tendency to be epitaxial on MAFO due to the excellent lattice match, even when Pt is sputter-deposited with the substrate at room temperature.
Moreover, while the epitaxial Pt film on MAFO is single-crystalline in the sense that its out-of-plane crytallographic orientation is exclusively (111), it is yet unclear how the twin domains (discussed in Sec.~\ref{sec:structure}) influence charge-spin interconversion in Pt. Determining the impact of microstructure on spin-Hall and related effects in Pt remains a subject of future work.

Furthermore, we acknowledge the possibility that the model employed in our present study (outlined in Sec.~\ref{subsec:model} and Appendix~\ref{app:diffusionEqs}) is incomplete.
For instance, we have assumed that the damping-like SOT and spin-pumping damping are reciprocal phenomena with shared \Gupdown and \Ls. This commonly made assumption~\cite{Berger2018} -- with prior studies suggesting that such reciprocity holds~\cite{Nan2015a, Emori2018} -- is necessary for constraining the fitting of the limited number of experimental data points. Further studies are required for confirming whether the damping-like SOT and spin-pumping damping can be captured by the same values of \Gupdown and \Ls.

\section{Summary}
We have measured SOTs in a low-damping, epitaxial insulating spinel ferrite (MgAl$_{0.5}$Fe$_{1.5}$O$_4$, MAFO) interfaced with epitaxial Pt. This model-system bilayer enables a unique opportunity to examine charge-spin interconversion mechanisms in highly crystalline Pt, while eliminating complications from electronic transport in (or hybridization with) a magnetic metal. Our key findings are as follow.
\begin{enumerate}
    \item Charge-to-spin conversion in Pt appears to be primarily a bulk effect, rather than an interfacial effect. A sizable damping-like SOT, which depends strongly on the Pt thickness, arises from the spin-Hall effect within Pt. An order-of-magnitude smaller field-like SOT, independent of the Pt thickness, is attributed to the Rashba-Edelstein effect at the MAFO/Pt interface.
    \item In crystalline Pt, the extrinsic spin-Hall effect of skew scattering and the Dyakonov-Perel spin relaxation mechanism likely dominate. This is in contrast to the combination of the intrinsic spin-Hall effect and Elliott-Yafet spin relaxation typically reported for Pt-based systems.
    \item The internal spin-Hall ratio deduced for crystalline Pt is large, i.e., $\thetaInt\approx0.8$. While a similar magnitude has been suggested before from experiments on all-metallic FM/Pt bilayers, greater confidence may be placed in our result owing to the cleanliness of the MAFO/Pt system, the direct nature of the SOT measurement method, and the self-consistent modeling of the SOT and spin-pumping damping.
    \item Spin-memory loss appears to be minimal in the epitaxial MAFO/Pt system. Modeled scenarios with substantial spin-memory loss yield unphysically large internal spin-Hall ratios that exceed unity.
    \item The factor limiting the damping-like SOT efficiency in the MAFO/Pt bilayer, despite the apparently large \thetaInt, is the small spin-mixing conductance \Gupdown. Enhancing \Gupdown\ while keeping spin-memory loss minimal could increase the SOT efficiency.
\end{enumerate}

\noindent Overall, our work demonstrates the utility of epitaxial insulating-ferrite-based heterostructures for understanding spin-transport phenomena in the widely-used spin-Hall metal of Pt, as well as for engineering  materials for efficient spintronic devices.

\begin{acknowledgments}
This work was funded by the Vannevar Bush Faculty Fellowship of the Department of Defense under Contract No. N00014-15-1-0045. L.J.R. acknowledges support from the Air Force Office of Scientific Research under Grant No. FA 9550-20-1-0293. J.J.W. acknowledges support from the U.S. Department of Energy, Director, Office of Science, Office of Basic Energy Sciences, Division of Materials Sciences and Engineering under Contract No. DESC0008505. S.X.W. acknowledges funding from NSF Center for Energy Efficient Electronics Science (E3S) and ASCENT, one of six centers in JUMP, a Semiconductor Research Corporation (SRC) program sponsored by DARPA. L.J.R. and M.J.V. acknowledge the National Science Foundation Graduate Fellowships. X.-Q.S. acknowledges support from DOE Office of Science, Office of High Energy Physics under Grant NO. DE-SC0019380. S.E. acknowledges support from the National Science Foundation, Award No. DMR-2003914. Part of this work was performed at the Stanford Nano Shared Facilities (SNSF)/Stanford Nanofabrication Facility (SNF), supported by the National Science Foundation under award ECCS-1542152. S.E. acknowledges Makoto Kohda, Xin Fan, and Vivek Amin for fruitful discussions. P.L. acknowledges Wei Zhang for valuable suggestions in building the ST-FMR system.
\end{acknowledgments}

\newpage
\appendix
\section{Effect of Sample Processing on the Magnetic Properties of MAFO}
\label{app:FMRcomparison}
We have used both broadband FMR (i.e., with unpatterned films placed on a coplanar waveguide, see Ref.~\citen{Emori2018a} for details) and ST-FMR (i.e., with microwave current injected through patterned 10-$\mu$m-wide strips) to measure the frequency dependence of FMR linewidth and resonance field. Thus, it is important to confirm the consistency of measurements between the two techniques. 

Figure~\ref{fig:BBFMR}(a) plots the linewidth vs. frequency data for a bare MAFO film (13 nm) that we started with, the MAFO (13 nm)/Pt (5 nm) film after Pt deposition, and the ST-FMR pattern with MAFO (13 nm)/Pt (5 nm) after the microfabrication processes. The damping parameters of the MAFO/Pt unpatterned film and patterned strip are essentially identical, confirming the consistency of the broadband FMR and ST-FMR measurements.

We also show in Fig.~\ref{fig:BBFMR}(b) that the frequency dependence of resonance field is unaltered before and after microfabrication. 
The fit using the Kittel equation~\cite{Emori2018a} indicates negligible ($\ll5\%$) difference in the effective magnetization (dominated by mangnetoelastic easy-plane anisotropy) and gyromagnetic ratio for the unpatterned film and patterned strip. The results in Fig.~\ref{fig:BBFMR} therefore confirm that the microfabrication processes have little to no effect on the essential magnetic properties of MAFO/Pt.

\begin{figure}[h]
\includegraphics[width=1\columnwidth]{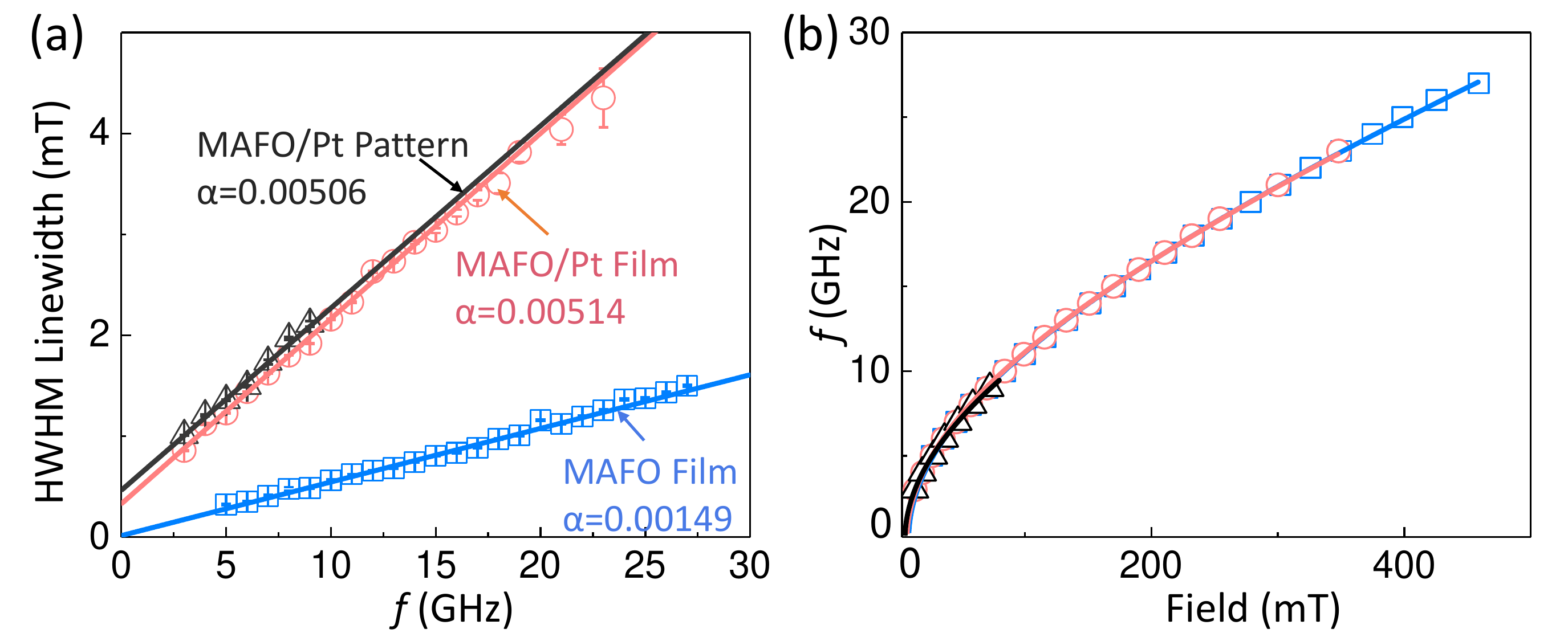}
\caption{\label{fig:BBFMR} Frequency dependence of (a) linewidth and (b) resonance field a bare MAFO film (13 nm), unpatterned MAFO (13 nm)/Pt (5 nm) film, and patterned MAFO (13 nm)/Pt (5 nm) ST-FMR strip.}
\label{fig:BBFMR}
\end{figure}

\section{Microwave Power Dependence of the Spin-Torque Ferromagnetic Resonance Signal}
\begin{figure}[h]
\includegraphics[width=1\columnwidth]{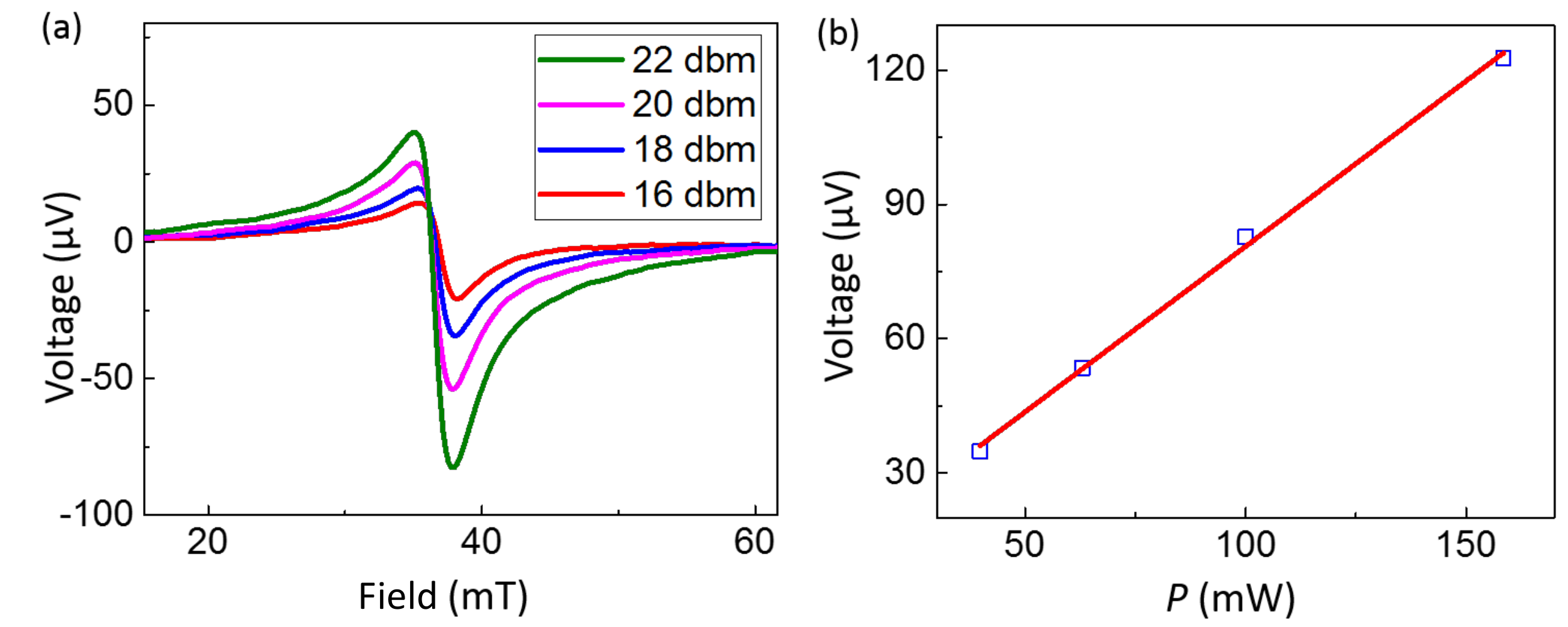}
\caption{\label{fig:py} (a) Exemplary ST-FMR spectra at different microwave powers. (b) ST-FMR amplitude vs. microwave power.} 
\label{fig:RFpower}
\end{figure}

Figure~\ref{fig:RFpower} shows the dependence of the ST-FMR signal amplitude on the microwave power. The ST-FMR voltage increases linearly with the microwave power, indicating that all the measurements are done in the linear regime in this present study.

\section{Frequency Dependence of the Spin-Orbit Torque Efficiencies}

\begin{figure}[h]
\includegraphics[width=1.00\columnwidth]{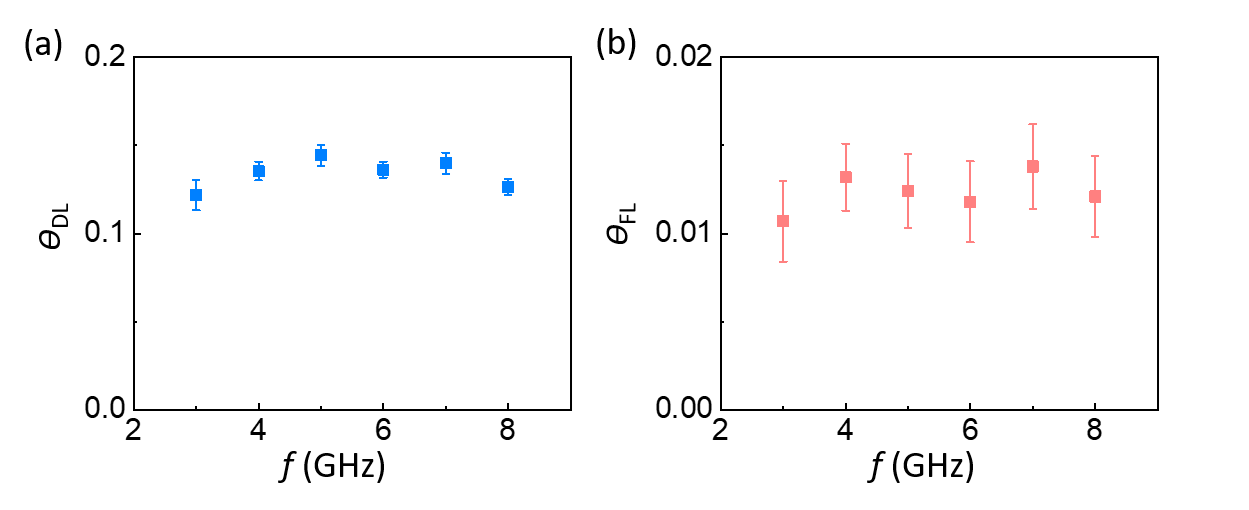}
\caption{\label{fig:py} Dependence of SOTs in MAFO (13 nm)/ Pt (5 nm). (a) Damping-like torque efficiency $\theta_\textrm{DL}$. (b) Field-like torque efficiency $\theta_\textrm{FL}$.} 
\label{fig:freqDep}
\end{figure}

We have carried out a frequency dependence study of damping-like and field-like SOT efficiencies. The dc-biased ST-FMR method is used to extract each data point. Figure~\ref{fig:freqDep} shows that both the damping-like and field-like SOT efficiencies are nearly constant across the frequency range of 3-8 GHz. This verifies that the SOT efficiencies are independent of the microwave frequency.

\section{Model for the Pt Thickness Dependence of Resistivity}
\label{app:resistivity}
We model the Pt thickness dependence of resistivity \rPt\ by using an approach similar to that reported by Berger \textit{et al.}~\cite{Berger2018}. This model takes into account the conductivity $\sigma$ as a function of position along the film thickness axis $z$, expressed as the sum of bulk and interfacial contributions, 
\begin{equation}
    \sigma(z) = \frac{1-\exp{\left(-\frac{z}{L}\right)}}{\rbulk} + \frac{\exp{\left(-\frac{z}{L}\right)}}{\rint}, 
\end{equation}
where $\rbulk = 1.1\times10^{-7}$ $\Omega$m is the bulk resistivity of Pt, \rint\ is the interfacial resistivity, and $L$ is an empirical characteristic length scale capturing the decay of the interfacial contribution to resistivity. The resistivity of the Pt film with thickness \tPt\ is then given by, 
\begin{equation}
\begin{split}
    \rPt(\tPt) & = \left(\frac{1}{\tPt}\int_{0}^{\tPt} \sigma(z) dz \right)^{-1} \\
    & = \frac{\rbulk}{1+\frac{L}{\tPt}\left(\frac{\rbulk}{\rint}-1\right)\left(1-\exp{\left(-\frac{\tPt}{L}\right)}\right)}.
    \end{split}
\label{eq:resistivity}
\end{equation}
The fit curve for the experimentally measured \tPt-dependence of \rPt (Fig.~\ref{fig:model}(a)) is obtained with Eq.~\ref{eq:resistivity} with $\rint = 1.3\times10^{-6}$ $\Omega$m and $L~=~10.$~nm.

\begin{figure*}[t!]
 \centering
 \includegraphics[width=2.00\columnwidth]{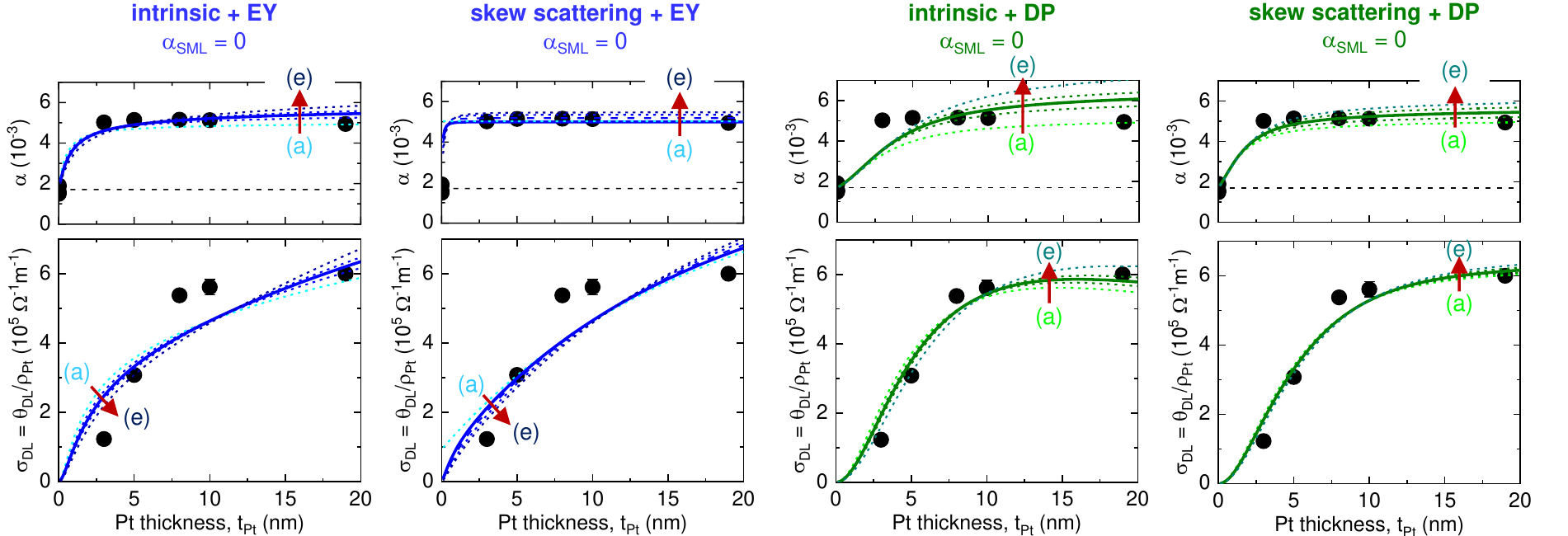}
 \caption{Exemplary fit results for the Pt thickness dependence of the Gilbert damping parameter $\alpha$ and the damping-like SOT conductivity \sigmaDL with several values of \Gupdown. The solid curves (parameterized by the values in bold font in Table~\ref{table:examples}) are the curves shown in Fig.~\ref{fig:model}. The error bars are comparable to or smaller than the symbol size; they are derived from the linear fits of FMR linewidth vs. frequency (for $\alpha$) and ST-FMR linewidth vs. dc bias current density (for \sigmaDL).}
 \label{fig:fitExamples}
\end{figure*}

\begin{table*}[]
\begin{tabular}{ccccccccccccc}
                         & \multicolumn{3}{c}{intrinsic + EY}           & \multicolumn{3}{c}{skew scatt. + EY}          & \multicolumn{3}{c}{intrinsic + DP}            & \multicolumn{3}{c}{skew scatt. + DP} \\ \cline{2-13} 
                         & \Gupdown ($\Omega^{-1}$m$^{-2}$) & \Ls (nm) & \multicolumn{1}{c|}{\thetaInt} & \Gupdown ($\Omega^{-1}$m$^{-2}$) & \Ls (nm)  & \multicolumn{1}{c|}{\thetaInt} & \Gupdown ($\Omega^{-1}$m$^{-2}$) & \Ls (nm)  & \multicolumn{1}{c|}{\thetaInt} & \Gupdown ($\Omega^{-1}$m$^{-2}$)     & \Ls (nm)      & \thetaInt     \\ \hline
\multicolumn{1}{c|}{(a)} & 1.5$\times10^{14}$    & 16 & \multicolumn{1}{c|}{0.24}     & 1.0$\times10^{14}$    & 0.1 & \multicolumn{1}{c|}{48}       & 1.2$\times10^{14}$    & 5.7 & \multicolumn{1}{c|}{0.32}     & 1.1$\times10^{14}$        & 3.2     & 0.96         \\
\multicolumn{1}{c|}{(b)} & 2.0$\times10^{14}$    & 19 & \multicolumn{1}{c|}{0.22}     & \textbf{1.1}$\mathbf{\times10^{14}}$    & \textbf{4.7} & \multicolumn{1}{c|}{\textbf{1.2}}      & 1.6$\times10^{14}$    & 5.6 & \multicolumn{1}{c|}{0.27}     & 1.2$\times10^{14}$        & 3.3     & 0.89         \\
\multicolumn{1}{c|}{(c)} & \textbf{2.5}$\mathbf{\times10^{14}}$    & \textbf{21} & \multicolumn{1}{c|}{\textbf{0.21}}     & 1.2$\times10^{14}$    & 5.7 & \multicolumn{1}{c|}{0.96}     & \textbf{1.8}$\mathbf{\times10^{14}}$    & \textbf{5.7} & \multicolumn{1}{c|}{\textbf{0.25}}     & \textbf{1.3}$\mathbf{\times10^{14}}$        & \textbf{3.3}     & \textbf{0.83}         \\
\multicolumn{1}{c|}{(d)} & 3.0$\times10^{14}$    & 22 & \multicolumn{1}{c|}{0.20}     & 1.3$\times10^{14}$    & 6.7 & \multicolumn{1}{c|}{0.80}     & 2.0$\times10^{14}$    & 5.8 & \multicolumn{1}{c|}{0.23}     & 1.4$\times10^{14}$        & 3.4     & 0.77         \\
\multicolumn{1}{c|}{(e)} & 5.0$\times10^{14}$    & 24 & \multicolumn{1}{c|}{0.19}     & 1.4$\times10^{14}$    & 6.1 & \multicolumn{1}{c|}{0.68}     & 2.5$\times10^{14}$    & 6.1 & \multicolumn{1}{c|}{0.21}     & 1.5$\times10^{14}$        & 3.5     & 0.72         \\ \hline
\end{tabular}
\caption{Parameters for the modeled curves in Fig.~\ref{fig:fitExamples}. For charge-to-spin conversion = intrinsic (for spin relaxation = EY), \thetaInt\ (\Ls) is the value at $\rPt = \rbulk = 1.1\times10^{-7}$ $\Omega$m.}
\label{table:examples}
\end{table*}

\vspace{20pt}
\section{Equations for the Pt Thickness Dependence of $\alpha$ and \sigmaDL}
\label{app:diffusionEqs}

We fit the \tPt dependence of $\alpha$ with~\cite{Boone2015}

\begin{equation}\label{eq:deltaA}
\begin{split}
\alpha(\tPt) & = \alpha_0+\aSML + \aSP\\
& = \alpha_0+\aSML + \frac{g\muB\hbar}{2e^2\Ms \tM}{\left[\frac{1}{\Gupdown}+{2\rPt\Ls}\coth\left(\frac{\tPt}{\Ls}\right)\right]}^{-1},
\end{split}
\end{equation}
where $\alpha_0 = 0.0017$ is the mean value for the five bare MAFO films ($\tPt = 0$) prior to Pt deposition for the MAFO/Pt bilayers, \aSP\ is the spin-pumping contribution to Gilbert damping, \aSML\ is the phenomenological parameter capturing the \tPt-independent enhancement of damping (from interfacial spin-memory loss or two-magnon scattering), $g = 2.05$ is the spectroscopic $g$-factor~\cite{Emori2018a}, $\Ms = 93$ kA/m is the mean value of the saturation magnetization of MAFO used in this study, and $\tM = 13$~nm is the thickness of MAFO. 

The \tPt\ dependence of \sigmaDL is fit with~\cite{Haney2013a}
\begin{widetext}
\begin{equation}
\sigmaDL(\tPt) = \frac{\thetaDL}{\rPt} = \frac{\thetaInt}{\rPt} \left\{\frac{(1 - e^{-\tPt/\Ls})^2}{(1 + e^{-2 \tPt/\Ls})} 
\frac{\Gprime}
{{\Gprime} + \tanh^2\left(\dfrac{\tPt}{\Ls}\right)}\right\}
\frac{\aSP}{\aSML+\aSP}, 
\label{eq:DLSOT}
\end{equation}
\end{widetext}
where $\Gprime =\Gupdown 2\rPt\Ls \tanh(\tPt/\Ls)$.  We also remark that \rPt\ is dependent on \tPt\ as given by Eq.~\ref{eq:resistivity}.

By using Equations~\ref{eq:deltaA} and \ref{eq:DLSOT}, along with the interpolated \rPt (Eq.~\ref{eq:resistivity}), we find the values of \Gupdown,  \Ls, and \thetaInt\ that adequately capture the experimentally measured $\alpha(\tPt)$ and $\sigmaDL(\tPt)$. In particular, $\alpha(\tPt)$ and $\sigmaDL(\tPt)$ are fit simultaneously~\footnote{We used the MATLAB function of nlinmultifit, which is a wrapper for nlinfit that allows for simultaneous fitting of multiple data sets with shared parameters, available at: \url{https://www.mathworks.com/matlabcentral/fileexchange/40613-multiple-curve-fitting-with-common-parameters-using-nlinfit}} with a series of fixed values for \Gupdown (e.g., Figs.~\ref{fig:fitExamples} and Table~\ref{table:examples}). 


\begin{thebibliography}{98}%
\makeatletter
\providecommand \@ifxundefined [1]{%
 \@ifx{#1\undefined}
}%
\providecommand \@ifnum [1]{%
 \ifnum #1\expandafter \@firstoftwo
 \else \expandafter \@secondoftwo
 \fi
}%
\providecommand \@ifx [1]{%
 \ifx #1\expandafter \@firstoftwo
 \else \expandafter \@secondoftwo
 \fi
}%
\providecommand \natexlab [1]{#1}%
\providecommand \enquote  [1]{``#1''}%
\providecommand \bibnamefont  [1]{#1}%
\providecommand \bibfnamefont [1]{#1}%
\providecommand \citenamefont [1]{#1}%
\providecommand \href@noop [0]{\@secondoftwo}%
\providecommand \href [0]{\begingroup \@sanitize@url \@href}%
\providecommand \@href[1]{\@@startlink{#1}\@@href}%
\providecommand \@@href[1]{\endgroup#1\@@endlink}%
\providecommand \@sanitize@url [0]{\catcode `\\12\catcode `\$12\catcode
  `\&12\catcode `\#12\catcode `\^12\catcode `\_12\catcode `\%12\relax}%
\providecommand \@@startlink[1]{}%
\providecommand \@@endlink[0]{}%
\providecommand \url  [0]{\begingroup\@sanitize@url \@url }%
\providecommand \@url [1]{\endgroup\@href {#1}{\urlprefix }}%
\providecommand \urlprefix  [0]{URL }%
\providecommand \Eprint [0]{\href }%
\providecommand \doibase [0]{http://dx.doi.org/}%
\providecommand \selectlanguage [0]{\@gobble}%
\providecommand \bibinfo  [0]{\@secondoftwo}%
\providecommand \bibfield  [0]{\@secondoftwo}%
\providecommand \translation [1]{[#1]}%
\providecommand \BibitemOpen [0]{}%
\providecommand \bibitemStop [0]{}%
\providecommand \bibitemNoStop [0]{.\EOS\space}%
\providecommand \EOS [0]{\spacefactor3000\relax}%
\providecommand \BibitemShut  [1]{\csname bibitem#1\endcsname}%
\let\auto@bib@innerbib\@empty
\bibitem [{\citenamefont {Sinova}\ \emph {et~al.}(2015)\citenamefont {Sinova},
  \citenamefont {Valenzuela}, \citenamefont {Wunderlich}, \citenamefont
  {Back},\ and\ \citenamefont {Jungwirth}}]{Sinova2015}%
  \BibitemOpen
  \bibfield  {author} {\bibinfo {author} {\bibfnamefont {J.}~\bibnamefont
  {Sinova}}, \bibinfo {author} {\bibfnamefont {S.~O.}\ \bibnamefont
  {Valenzuela}}, \bibinfo {author} {\bibfnamefont {J.}~\bibnamefont
  {Wunderlich}}, \bibinfo {author} {\bibfnamefont {C.~H.}\ \bibnamefont
  {Back}}, \ and\ \bibinfo {author} {\bibfnamefont {T.}~\bibnamefont
  {Jungwirth}},\ }\href {\doibase 10.1103/RevModPhys.87.1213} {\bibfield
  {journal} {\bibinfo  {journal} {Rev. Mod. Phys.}\ }\textbf {\bibinfo {volume}
  {87}},\ \bibinfo {pages} {1213} (\bibinfo {year} {2015})}\BibitemShut
  {NoStop}%
\bibitem [{\citenamefont {Ramaswamy}\ \emph {et~al.}(2018)\citenamefont
  {Ramaswamy}, \citenamefont {Lee}, \citenamefont {Cai},\ and\ \citenamefont
  {Yang}}]{Ramaswamy2018}%
  \BibitemOpen
  \bibfield  {author} {\bibinfo {author} {\bibfnamefont {R.}~\bibnamefont
  {Ramaswamy}}, \bibinfo {author} {\bibfnamefont {J.~M.}\ \bibnamefont {Lee}},
  \bibinfo {author} {\bibfnamefont {K.}~\bibnamefont {Cai}}, \ and\ \bibinfo
  {author} {\bibfnamefont {H.}~\bibnamefont {Yang}},\ }\href {\doibase
  10.1063/1.5041793} {\bibfield  {journal} {\bibinfo  {journal} {Appl. Phys.
  Rev.}\ }\textbf {\bibinfo {volume} {5}},\ \bibinfo {pages} {031107} (\bibinfo
  {year} {2018})}\BibitemShut {NoStop}%
\bibitem [{\citenamefont {Hoffmann}(2013)}]{Hoffmann2013}%
  \BibitemOpen
  \bibfield  {author} {\bibinfo {author} {\bibfnamefont {A.}~\bibnamefont
  {Hoffmann}},\ }\href {\doibase 10.1109/TMAG.2013.2262947} {\bibfield
  {journal} {\bibinfo  {journal} {IEEE Trans. Magn.}\ }\textbf {\bibinfo
  {volume} {49}},\ \bibinfo {pages} {5172} (\bibinfo {year}
  {2013})}\BibitemShut {NoStop}%
\bibitem [{\citenamefont {Manchon}\ \emph {et~al.}(2015)\citenamefont
  {Manchon}, \citenamefont {Koo}, \citenamefont {Nitta}, \citenamefont
  {Frolov},\ and\ \citenamefont {Duine}}]{Manchon2015}%
  \BibitemOpen
  \bibfield  {author} {\bibinfo {author} {\bibfnamefont {A.}~\bibnamefont
  {Manchon}}, \bibinfo {author} {\bibfnamefont {H.~C.}\ \bibnamefont {Koo}},
  \bibinfo {author} {\bibfnamefont {J.}~\bibnamefont {Nitta}}, \bibinfo
  {author} {\bibfnamefont {S.~M.}\ \bibnamefont {Frolov}}, \ and\ \bibinfo
  {author} {\bibfnamefont {R.~A.}\ \bibnamefont {Duine}},\ }\href {\doibase
  10.1038/nmat4360} {\bibfield  {journal} {\bibinfo  {journal} {Nat. Mater.}\
  }\textbf {\bibinfo {volume} {14}},\ \bibinfo {pages} {871} (\bibinfo {year}
  {2015})}\BibitemShut {NoStop}%
\bibitem [{\citenamefont {Brataas}\ and\ \citenamefont
  {Hals}(2014)}]{Brataas2014}%
  \BibitemOpen
  \bibfield  {author} {\bibinfo {author} {\bibfnamefont {A.}~\bibnamefont
  {Brataas}}\ and\ \bibinfo {author} {\bibfnamefont {K.~M.~D.}\ \bibnamefont
  {Hals}},\ }\href {\doibase 10.1038/nnano.2014.8} {\bibfield  {journal}
  {\bibinfo  {journal} {Nat. Nanotechnol.}\ }\textbf {\bibinfo {volume} {9}},\
  \bibinfo {pages} {86} (\bibinfo {year} {2014})}\BibitemShut {NoStop}%
\bibitem [{\citenamefont {Haney}\ \emph {et~al.}(2013)\citenamefont {Haney},
  \citenamefont {Lee}, \citenamefont {Lee}, \citenamefont {Manchon},\ and\
  \citenamefont {Stiles}}]{Haney2013a}%
  \BibitemOpen
  \bibfield  {author} {\bibinfo {author} {\bibfnamefont {P.~M.}\ \bibnamefont
  {Haney}}, \bibinfo {author} {\bibfnamefont {H.-W.}\ \bibnamefont {Lee}},
  \bibinfo {author} {\bibfnamefont {K.-J.}\ \bibnamefont {Lee}}, \bibinfo
  {author} {\bibfnamefont {A.}~\bibnamefont {Manchon}}, \ and\ \bibinfo
  {author} {\bibfnamefont {M.~D.}\ \bibnamefont {Stiles}},\ }\href {\doibase
  10.1103/PhysRevB.87.174411} {\bibfield  {journal} {\bibinfo  {journal} {Phys.
  Rev. B}\ }\textbf {\bibinfo {volume} {87}},\ \bibinfo {pages} {174411}
  (\bibinfo {year} {2013})}\BibitemShut {NoStop}%
\bibitem [{\citenamefont {Taniguchi}\ \emph {et~al.}(2015)\citenamefont
  {Taniguchi}, \citenamefont {Grollier},\ and\ \citenamefont
  {Stiles}}]{Taniguchi2015}%
  \BibitemOpen
  \bibfield  {author} {\bibinfo {author} {\bibfnamefont {T.}~\bibnamefont
  {Taniguchi}}, \bibinfo {author} {\bibfnamefont {J.}~\bibnamefont {Grollier}},
  \ and\ \bibinfo {author} {\bibfnamefont {M.~D.}\ \bibnamefont {Stiles}},\
  }\href {\doibase 10.1103/PhysRevApplied.3.044001} {\bibfield  {journal}
  {\bibinfo  {journal} {Phys. Rev. Appl.}\ }\textbf {\bibinfo {volume} {3}},\
  \bibinfo {pages} {044001} (\bibinfo {year} {2015})}\BibitemShut {NoStop}%
\bibitem [{\citenamefont {Amin}\ and\ \citenamefont {Stiles}(2016)}]{Amin2016}%
  \BibitemOpen
  \bibfield  {author} {\bibinfo {author} {\bibfnamefont {V.~P.}\ \bibnamefont
  {Amin}}\ and\ \bibinfo {author} {\bibfnamefont {M.~D.}\ \bibnamefont
  {Stiles}},\ }\href {\doibase 10.1103/PhysRevB.94.104420} {\bibfield
  {journal} {\bibinfo  {journal} {Phys. Rev. B}\ }\textbf {\bibinfo {volume}
  {94}},\ \bibinfo {pages} {104420} (\bibinfo {year} {2016})}\BibitemShut
  {NoStop}%
\bibitem [{\citenamefont {Amin}\ \emph {et~al.}(2018)\citenamefont {Amin},
  \citenamefont {Zemen},\ and\ \citenamefont {Stiles}}]{Amin2018}%
  \BibitemOpen
  \bibfield  {author} {\bibinfo {author} {\bibfnamefont {V.~P.}\ \bibnamefont
  {Amin}}, \bibinfo {author} {\bibfnamefont {J.}~\bibnamefont {Zemen}}, \ and\
  \bibinfo {author} {\bibfnamefont {M.~D.}\ \bibnamefont {Stiles}},\ }\href
  {\doibase 10.1103/PhysRevLett.121.136805} {\bibfield  {journal} {\bibinfo
  {journal} {Phys. Rev. Lett.}\ }\textbf {\bibinfo {volume} {121}},\ \bibinfo
  {pages} {136805} (\bibinfo {year} {2018})}\BibitemShut {NoStop}%
\bibitem [{\citenamefont {Humphries}\ \emph {et~al.}(2017)\citenamefont
  {Humphries}, \citenamefont {Wang}, \citenamefont {Edwards}, \citenamefont
  {Allen}, \citenamefont {Shaw}, \citenamefont {Nembach}, \citenamefont {Xiao},
  \citenamefont {Silva},\ and\ \citenamefont {Fan}}]{Humphries2017}%
  \BibitemOpen
  \bibfield  {author} {\bibinfo {author} {\bibfnamefont {A.~M.}\ \bibnamefont
  {Humphries}}, \bibinfo {author} {\bibfnamefont {T.}~\bibnamefont {Wang}},
  \bibinfo {author} {\bibfnamefont {E.~R.~J.}\ \bibnamefont {Edwards}},
  \bibinfo {author} {\bibfnamefont {S.~R.}\ \bibnamefont {Allen}}, \bibinfo
  {author} {\bibfnamefont {J.~M.}\ \bibnamefont {Shaw}}, \bibinfo {author}
  {\bibfnamefont {H.~T.}\ \bibnamefont {Nembach}}, \bibinfo {author}
  {\bibfnamefont {J.~Q.}\ \bibnamefont {Xiao}}, \bibinfo {author}
  {\bibfnamefont {T.~J.}\ \bibnamefont {Silva}}, \ and\ \bibinfo {author}
  {\bibfnamefont {X.}~\bibnamefont {Fan}},\ }\href {\doibase
  10.1038/s41467-017-00967-w} {\bibfield  {journal} {\bibinfo  {journal} {Nat.
  Commun.}\ }\textbf {\bibinfo {volume} {8}},\ \bibinfo {pages} {911} (\bibinfo
  {year} {2017})}\BibitemShut {NoStop}%
\bibitem [{\citenamefont {Baek}\ \emph {et~al.}(2018)\citenamefont {Baek},
  \citenamefont {Amin}, \citenamefont {Oh}, \citenamefont {Go}, \citenamefont
  {Lee}, \citenamefont {Lee}, \citenamefont {Kim}, \citenamefont {Stiles},
  \citenamefont {Park},\ and\ \citenamefont {Lee}}]{Baek2018}%
  \BibitemOpen
  \bibfield  {author} {\bibinfo {author} {\bibfnamefont {S.-h.~C.}\
  \bibnamefont {Baek}}, \bibinfo {author} {\bibfnamefont {V.~P.}\ \bibnamefont
  {Amin}}, \bibinfo {author} {\bibfnamefont {Y.-W.}\ \bibnamefont {Oh}},
  \bibinfo {author} {\bibfnamefont {G.}~\bibnamefont {Go}}, \bibinfo {author}
  {\bibfnamefont {S.-J.}\ \bibnamefont {Lee}}, \bibinfo {author} {\bibfnamefont
  {G.-H.}\ \bibnamefont {Lee}}, \bibinfo {author} {\bibfnamefont {K.-J.}\
  \bibnamefont {Kim}}, \bibinfo {author} {\bibfnamefont {M.~D.}\ \bibnamefont
  {Stiles}}, \bibinfo {author} {\bibfnamefont {B.-G.}\ \bibnamefont {Park}}, \
  and\ \bibinfo {author} {\bibfnamefont {K.-J.}\ \bibnamefont {Lee}},\ }\href
  {\doibase 10.1038/s41563-018-0041-5} {\bibfield  {journal} {\bibinfo
  {journal} {Nat. Mater. 2018}\ ,\ \bibinfo {pages} {1}} (\bibinfo {year}
  {2018})}\BibitemShut {NoStop}%
\bibitem [{\citenamefont {Gibbons}\ \emph {et~al.}(2018)\citenamefont
  {Gibbons}, \citenamefont {MacNeill}, \citenamefont {Buhrman},\ and\
  \citenamefont {Ralph}}]{Gibbons2018}%
  \BibitemOpen
  \bibfield  {author} {\bibinfo {author} {\bibfnamefont {J.~D.}\ \bibnamefont
  {Gibbons}}, \bibinfo {author} {\bibfnamefont {D.}~\bibnamefont {MacNeill}},
  \bibinfo {author} {\bibfnamefont {R.~A.}\ \bibnamefont {Buhrman}}, \ and\
  \bibinfo {author} {\bibfnamefont {D.~C.}\ \bibnamefont {Ralph}},\ }\href
  {\doibase 10.1103/PhysRevApplied.9.064033} {\bibfield  {journal} {\bibinfo
  {journal} {Phys. Rev. Appl.}\ }\textbf {\bibinfo {volume} {9}},\ \bibinfo
  {pages} {064033} (\bibinfo {year} {2018})}\BibitemShut {NoStop}%
\bibitem [{\citenamefont {Iihama}\ \emph {et~al.}(2018)\citenamefont {Iihama},
  \citenamefont {Taniguchi}, \citenamefont {Yakushiji}, \citenamefont
  {Fukushima}, \citenamefont {Shiota}, \citenamefont {Tsunegi}, \citenamefont
  {Hiramatsu}, \citenamefont {Yuasa}, \citenamefont {Suzuki},\ and\
  \citenamefont {Kubota}}]{Iihama2018}%
  \BibitemOpen
  \bibfield  {author} {\bibinfo {author} {\bibfnamefont {S.}~\bibnamefont
  {Iihama}}, \bibinfo {author} {\bibfnamefont {T.}~\bibnamefont {Taniguchi}},
  \bibinfo {author} {\bibfnamefont {K.}~\bibnamefont {Yakushiji}}, \bibinfo
  {author} {\bibfnamefont {A.}~\bibnamefont {Fukushima}}, \bibinfo {author}
  {\bibfnamefont {Y.}~\bibnamefont {Shiota}}, \bibinfo {author} {\bibfnamefont
  {S.}~\bibnamefont {Tsunegi}}, \bibinfo {author} {\bibfnamefont
  {R.}~\bibnamefont {Hiramatsu}}, \bibinfo {author} {\bibfnamefont
  {S.}~\bibnamefont {Yuasa}}, \bibinfo {author} {\bibfnamefont
  {Y.}~\bibnamefont {Suzuki}}, \ and\ \bibinfo {author} {\bibfnamefont
  {H.}~\bibnamefont {Kubota}},\ }\href {\doibase 10.1038/s41928-018-0026-z}
  {\bibfield  {journal} {\bibinfo  {journal} {Nat. Electron.}\ }\textbf
  {\bibinfo {volume} {1}},\ \bibinfo {pages} {120} (\bibinfo {year}
  {2018})}\BibitemShut {NoStop}%
\bibitem [{\citenamefont {Safranski}\ \emph {et~al.}(2019)\citenamefont
  {Safranski}, \citenamefont {Montoya},\ and\ \citenamefont
  {Krivorotov}}]{Safranski2019}%
  \BibitemOpen
  \bibfield  {author} {\bibinfo {author} {\bibfnamefont {C.}~\bibnamefont
  {Safranski}}, \bibinfo {author} {\bibfnamefont {E.~A.}\ \bibnamefont
  {Montoya}}, \ and\ \bibinfo {author} {\bibfnamefont {I.~N.}\ \bibnamefont
  {Krivorotov}},\ }\href {\doibase 10.1038/s41565-018-0282-0} {\bibfield
  {journal} {\bibinfo  {journal} {Nat. Nanotechnol.}\ }\textbf {\bibinfo
  {volume} {14}},\ \bibinfo {pages} {27} (\bibinfo {year} {2019})}\BibitemShut
  {NoStop}%
\bibitem [{\citenamefont {Wang}\ \emph {et~al.}(2019)\citenamefont {Wang},
  \citenamefont {Wang}, \citenamefont {Amin}, \citenamefont {Wang},
  \citenamefont {Radhakrishnan}, \citenamefont {Davidson}, \citenamefont
  {Allen}, \citenamefont {Silva}, \citenamefont {Ohldag}, \citenamefont
  {Balzar}, \citenamefont {Zink}, \citenamefont {Haney}, \citenamefont {Xiao},
  \citenamefont {Cahill}, \citenamefont {Lorenz},\ and\ \citenamefont
  {Fan}}]{Wang2019a}%
  \BibitemOpen
  \bibfield  {author} {\bibinfo {author} {\bibfnamefont {W.}~\bibnamefont
  {Wang}}, \bibinfo {author} {\bibfnamefont {T.}~\bibnamefont {Wang}}, \bibinfo
  {author} {\bibfnamefont {V.~P.}\ \bibnamefont {Amin}}, \bibinfo {author}
  {\bibfnamefont {Y.}~\bibnamefont {Wang}}, \bibinfo {author} {\bibfnamefont
  {A.}~\bibnamefont {Radhakrishnan}}, \bibinfo {author} {\bibfnamefont
  {A.}~\bibnamefont {Davidson}}, \bibinfo {author} {\bibfnamefont {S.~R.}\
  \bibnamefont {Allen}}, \bibinfo {author} {\bibfnamefont {T.~J.}\ \bibnamefont
  {Silva}}, \bibinfo {author} {\bibfnamefont {H.}~\bibnamefont {Ohldag}},
  \bibinfo {author} {\bibfnamefont {D.}~\bibnamefont {Balzar}}, \bibinfo
  {author} {\bibfnamefont {B.~L.}\ \bibnamefont {Zink}}, \bibinfo {author}
  {\bibfnamefont {P.~M.}\ \bibnamefont {Haney}}, \bibinfo {author}
  {\bibfnamefont {J.~Q.}\ \bibnamefont {Xiao}}, \bibinfo {author}
  {\bibfnamefont {D.~G.}\ \bibnamefont {Cahill}}, \bibinfo {author}
  {\bibfnamefont {V.~O.}\ \bibnamefont {Lorenz}}, \ and\ \bibinfo {author}
  {\bibfnamefont {X.}~\bibnamefont {Fan}},\ }\href {\doibase
  10.1038/s41565-019-0504-0} {\bibfield  {journal} {\bibinfo  {journal} {Nat.
  Nanotechnol.}\ }\textbf {\bibinfo {volume} {14}},\ \bibinfo {pages} {819}
  (\bibinfo {year} {2019})}\BibitemShut {NoStop}%
\bibitem [{\citenamefont {Wu}\ \emph {et~al.}(2019)\citenamefont {Wu},
  \citenamefont {Razavi}, \citenamefont {Shao}, \citenamefont {Li},
  \citenamefont {Wong}, \citenamefont {Liu}, \citenamefont {Yin},\ and\
  \citenamefont {Wang}}]{Wu2019}%
  \BibitemOpen
  \bibfield  {author} {\bibinfo {author} {\bibfnamefont {H.}~\bibnamefont
  {Wu}}, \bibinfo {author} {\bibfnamefont {S.~A.}\ \bibnamefont {Razavi}},
  \bibinfo {author} {\bibfnamefont {Q.}~\bibnamefont {Shao}}, \bibinfo {author}
  {\bibfnamefont {X.}~\bibnamefont {Li}}, \bibinfo {author} {\bibfnamefont
  {K.~L.}\ \bibnamefont {Wong}}, \bibinfo {author} {\bibfnamefont
  {Y.}~\bibnamefont {Liu}}, \bibinfo {author} {\bibfnamefont {G.}~\bibnamefont
  {Yin}}, \ and\ \bibinfo {author} {\bibfnamefont {K.~L.}\ \bibnamefont
  {Wang}},\ }\href {\doibase 10.1103/PhysRevB.99.184403} {\bibfield  {journal}
  {\bibinfo  {journal} {Phys. Rev. B}\ }\textbf {\bibinfo {volume} {99}},\
  \bibinfo {pages} {184403} (\bibinfo {year} {2019})}\BibitemShut {NoStop}%
\bibitem [{\citenamefont {Zhou}\ \emph {et~al.}(2015)\citenamefont {Zhou},
  \citenamefont {Grigoryan}, \citenamefont {Maekawa}, \citenamefont {Wang},\
  and\ \citenamefont {Xiao}}]{Zhou2015}%
  \BibitemOpen
  \bibfield  {author} {\bibinfo {author} {\bibfnamefont {L.}~\bibnamefont
  {Zhou}}, \bibinfo {author} {\bibfnamefont {V.~L.}\ \bibnamefont {Grigoryan}},
  \bibinfo {author} {\bibfnamefont {S.}~\bibnamefont {Maekawa}}, \bibinfo
  {author} {\bibfnamefont {X.}~\bibnamefont {Wang}}, \ and\ \bibinfo {author}
  {\bibfnamefont {J.}~\bibnamefont {Xiao}},\ }\href {\doibase
  10.1103/PhysRevB.91.045407} {\bibfield  {journal} {\bibinfo  {journal} {Phys.
  Rev. B}\ }\textbf {\bibinfo {volume} {91}},\ \bibinfo {pages} {045407}
  (\bibinfo {year} {2015})}\BibitemShut {NoStop}%
\bibitem [{\citenamefont {Garello}\ \emph {et~al.}(2013)\citenamefont
  {Garello}, \citenamefont {Miron}, \citenamefont {Avci}, \citenamefont
  {Freimuth}, \citenamefont {Mokrousov}, \citenamefont {Bl{\"{u}}gel},
  \citenamefont {Auffret}, \citenamefont {Boulle}, \citenamefont {Gaudin},\
  and\ \citenamefont {Gambardella}}]{Garello2013}%
  \BibitemOpen
  \bibfield  {author} {\bibinfo {author} {\bibfnamefont {K.}~\bibnamefont
  {Garello}}, \bibinfo {author} {\bibfnamefont {I.~M.}\ \bibnamefont {Miron}},
  \bibinfo {author} {\bibfnamefont {C.~O.}\ \bibnamefont {Avci}}, \bibinfo
  {author} {\bibfnamefont {F.}~\bibnamefont {Freimuth}}, \bibinfo {author}
  {\bibfnamefont {Y.}~\bibnamefont {Mokrousov}}, \bibinfo {author}
  {\bibfnamefont {S.}~\bibnamefont {Bl{\"{u}}gel}}, \bibinfo {author}
  {\bibfnamefont {S.}~\bibnamefont {Auffret}}, \bibinfo {author} {\bibfnamefont
  {O.}~\bibnamefont {Boulle}}, \bibinfo {author} {\bibfnamefont
  {G.}~\bibnamefont {Gaudin}}, \ and\ \bibinfo {author} {\bibfnamefont
  {P.}~\bibnamefont {Gambardella}},\ }\href {\doibase 10.1038/nnano.2013.145}
  {\bibfield  {journal} {\bibinfo  {journal} {Nat. Nanotechnol.}\ }\textbf
  {\bibinfo {volume} {8}},\ \bibinfo {pages} {587} (\bibinfo {year}
  {2013})}\BibitemShut {NoStop}%
\bibitem [{\citenamefont {Pai}\ \emph {et~al.}(2015)\citenamefont {Pai},
  \citenamefont {Ou}, \citenamefont {Vilela-Le{\~{a}}o}, \citenamefont
  {Ralph},\ and\ \citenamefont {Buhrman}}]{Pai2015a}%
  \BibitemOpen
  \bibfield  {author} {\bibinfo {author} {\bibfnamefont {C.-F.}\ \bibnamefont
  {Pai}}, \bibinfo {author} {\bibfnamefont {Y.}~\bibnamefont {Ou}}, \bibinfo
  {author} {\bibfnamefont {L.~H.}\ \bibnamefont {Vilela-Le{\~{a}}o}}, \bibinfo
  {author} {\bibfnamefont {D.~C.}\ \bibnamefont {Ralph}}, \ and\ \bibinfo
  {author} {\bibfnamefont {R.~A.}\ \bibnamefont {Buhrman}},\ }\href {\doibase
  10.1103/PhysRevB.92.064426} {\bibfield  {journal} {\bibinfo  {journal} {Phys.
  Rev. B}\ }\textbf {\bibinfo {volume} {92}},\ \bibinfo {pages} {064426}
  (\bibinfo {year} {2015})}\BibitemShut {NoStop}%
\bibitem [{\citenamefont {Rojas-S{\'{a}}nchez}\ \emph
  {et~al.}(2014)\citenamefont {Rojas-S{\'{a}}nchez}, \citenamefont {Reyren},
  \citenamefont {Laczkowski}, \citenamefont {Savero}, \citenamefont
  {Attan{\'{e}}}, \citenamefont {Deranlot}, \citenamefont {Jamet},
  \citenamefont {George}, \citenamefont {Vila},\ and\ \citenamefont
  {Jaffr{\`{e}}s}}]{Rojas-Sanchez2014}%
  \BibitemOpen
  \bibfield  {author} {\bibinfo {author} {\bibfnamefont {J.-C.}\ \bibnamefont
  {Rojas-S{\'{a}}nchez}}, \bibinfo {author} {\bibfnamefont {N.}~\bibnamefont
  {Reyren}}, \bibinfo {author} {\bibfnamefont {P.}~\bibnamefont {Laczkowski}},
  \bibinfo {author} {\bibfnamefont {W.}~\bibnamefont {Savero}}, \bibinfo
  {author} {\bibfnamefont {J.-P.}\ \bibnamefont {Attan{\'{e}}}}, \bibinfo
  {author} {\bibfnamefont {C.}~\bibnamefont {Deranlot}}, \bibinfo {author}
  {\bibfnamefont {M.}~\bibnamefont {Jamet}}, \bibinfo {author} {\bibfnamefont
  {J.-M.}\ \bibnamefont {George}}, \bibinfo {author} {\bibfnamefont
  {L.}~\bibnamefont {Vila}}, \ and\ \bibinfo {author} {\bibfnamefont
  {H.}~\bibnamefont {Jaffr{\`{e}}s}},\ }\href {\doibase
  10.1103/PhysRevLett.112.106602} {\bibfield  {journal} {\bibinfo  {journal}
  {Phys. Rev. Lett.}\ }\textbf {\bibinfo {volume} {112}},\ \bibinfo {pages}
  {106602} (\bibinfo {year} {2014})}\BibitemShut {NoStop}%
\bibitem [{\citenamefont {Chen}\ and\ \citenamefont {Zhang}(2015)}]{Chen2015b}%
  \BibitemOpen
  \bibfield  {author} {\bibinfo {author} {\bibfnamefont {K.}~\bibnamefont
  {Chen}}\ and\ \bibinfo {author} {\bibfnamefont {S.}~\bibnamefont {Zhang}},\
  }\href {\doibase 10.1103/PhysRevLett.114.126602} {\bibfield  {journal}
  {\bibinfo  {journal} {Phys. Rev. Lett.}\ }\textbf {\bibinfo {volume} {114}},\
  \bibinfo {pages} {126602} (\bibinfo {year} {2015})}\BibitemShut {NoStop}%
\bibitem [{\citenamefont {Dolui}\ and\ \citenamefont
  {Nikoli{\'{c}}}(2017)}]{Dolui2017}%
  \BibitemOpen
  \bibfield  {author} {\bibinfo {author} {\bibfnamefont {K.}~\bibnamefont
  {Dolui}}\ and\ \bibinfo {author} {\bibfnamefont {B.~K.}\ \bibnamefont
  {Nikoli{\'{c}}}},\ }\href {\doibase 10.1103/PhysRevB.96.220403} {\bibfield
  {journal} {\bibinfo  {journal} {Phys. Rev. B}\ }\textbf {\bibinfo {volume}
  {96}},\ \bibinfo {pages} {220403} (\bibinfo {year} {2017})}\BibitemShut
  {NoStop}%
\bibitem [{\citenamefont {Berger}\ \emph {et~al.}(2018)\citenamefont {Berger},
  \citenamefont {Edwards}, \citenamefont {Nembach}, \citenamefont {Karis},
  \citenamefont {Weiler},\ and\ \citenamefont {Silva}}]{Berger2018}%
  \BibitemOpen
  \bibfield  {author} {\bibinfo {author} {\bibfnamefont {A.~J.}\ \bibnamefont
  {Berger}}, \bibinfo {author} {\bibfnamefont {E.~R.~J.}\ \bibnamefont
  {Edwards}}, \bibinfo {author} {\bibfnamefont {H.~T.}\ \bibnamefont
  {Nembach}}, \bibinfo {author} {\bibfnamefont {O.}~\bibnamefont {Karis}},
  \bibinfo {author} {\bibfnamefont {M.}~\bibnamefont {Weiler}}, \ and\ \bibinfo
  {author} {\bibfnamefont {T.~J.}\ \bibnamefont {Silva}},\ }\href {\doibase
  10.1103/PhysRevB.98.024402} {\bibfield  {journal} {\bibinfo  {journal} {Phys.
  Rev. B}\ }\textbf {\bibinfo {volume} {98}},\ \bibinfo {pages} {024402}
  (\bibinfo {year} {2018})}\BibitemShut {NoStop}%
\bibitem [{\citenamefont {Zhu}\ \emph {et~al.}(2019{\natexlab{a}})\citenamefont
  {Zhu}, \citenamefont {Ralph},\ and\ \citenamefont {Buhrman}}]{Zhu2019}%
  \BibitemOpen
  \bibfield  {author} {\bibinfo {author} {\bibfnamefont {L.}~\bibnamefont
  {Zhu}}, \bibinfo {author} {\bibfnamefont {D.~C.}\ \bibnamefont {Ralph}}, \
  and\ \bibinfo {author} {\bibfnamefont {R.~A.}\ \bibnamefont {Buhrman}},\
  }\href {\doibase 10.1103/PhysRevLett.122.077201} {\bibfield  {journal}
  {\bibinfo  {journal} {Phys. Rev. Lett.}\ }\textbf {\bibinfo {volume} {122}},\
  \bibinfo {pages} {077201} (\bibinfo {year} {2019}{\natexlab{a}})}\BibitemShut
  {NoStop}%
\bibitem [{\citenamefont {Zhang}\ \emph
  {et~al.}(2015{\natexlab{a}})\citenamefont {Zhang}, \citenamefont
  {Jungfleisch}, \citenamefont {Jiang}, \citenamefont {Liu}, \citenamefont
  {Pearson}, \citenamefont {te~Velthuis}, \citenamefont {Hoffmann},
  \citenamefont {Freimuth},\ and\ \citenamefont {Mokrousov}}]{Zhang2015c}%
  \BibitemOpen
  \bibfield  {author} {\bibinfo {author} {\bibfnamefont {W.}~\bibnamefont
  {Zhang}}, \bibinfo {author} {\bibfnamefont {M.~B.}\ \bibnamefont
  {Jungfleisch}}, \bibinfo {author} {\bibfnamefont {W.}~\bibnamefont {Jiang}},
  \bibinfo {author} {\bibfnamefont {Y.}~\bibnamefont {Liu}}, \bibinfo {author}
  {\bibfnamefont {J.~E.}\ \bibnamefont {Pearson}}, \bibinfo {author}
  {\bibfnamefont {S.~G.~E.}\ \bibnamefont {te~Velthuis}}, \bibinfo {author}
  {\bibfnamefont {A.}~\bibnamefont {Hoffmann}}, \bibinfo {author}
  {\bibfnamefont {F.}~\bibnamefont {Freimuth}}, \ and\ \bibinfo {author}
  {\bibfnamefont {Y.}~\bibnamefont {Mokrousov}},\ }\href {\doibase
  10.1103/PhysRevB.91.115316} {\bibfield  {journal} {\bibinfo  {journal} {Phys.
  Rev. B}\ }\textbf {\bibinfo {volume} {91}},\ \bibinfo {pages} {115316}
  (\bibinfo {year} {2015}{\natexlab{a}})}\BibitemShut {NoStop}%
\bibitem [{\citenamefont {Caminale}\ \emph {et~al.}(2016)\citenamefont
  {Caminale}, \citenamefont {Ghosh}, \citenamefont {Auffret}, \citenamefont
  {Ebels}, \citenamefont {Ollefs}, \citenamefont {Wilhelm}, \citenamefont
  {Rogalev},\ and\ \citenamefont {Bailey}}]{Caminale2016}%
  \BibitemOpen
  \bibfield  {author} {\bibinfo {author} {\bibfnamefont {M.}~\bibnamefont
  {Caminale}}, \bibinfo {author} {\bibfnamefont {A.}~\bibnamefont {Ghosh}},
  \bibinfo {author} {\bibfnamefont {S.}~\bibnamefont {Auffret}}, \bibinfo
  {author} {\bibfnamefont {U.}~\bibnamefont {Ebels}}, \bibinfo {author}
  {\bibfnamefont {K.}~\bibnamefont {Ollefs}}, \bibinfo {author} {\bibfnamefont
  {F.}~\bibnamefont {Wilhelm}}, \bibinfo {author} {\bibfnamefont
  {A.}~\bibnamefont {Rogalev}}, \ and\ \bibinfo {author} {\bibfnamefont
  {W.~E.}\ \bibnamefont {Bailey}},\ }\href {\doibase
  10.1103/PhysRevB.94.014414} {\bibfield  {journal} {\bibinfo  {journal} {Phys.
  Rev. B}\ }\textbf {\bibinfo {volume} {94}},\ \bibinfo {pages} {014414}
  (\bibinfo {year} {2016})}\BibitemShut {NoStop}%
\bibitem [{\citenamefont {Amin}\ \emph {et~al.}(2019)\citenamefont {Amin},
  \citenamefont {Li}, \citenamefont {Stiles},\ and\ \citenamefont
  {Haney}}]{Amin2019}%
  \BibitemOpen
  \bibfield  {author} {\bibinfo {author} {\bibfnamefont {V.~P.}\ \bibnamefont
  {Amin}}, \bibinfo {author} {\bibfnamefont {J.}~\bibnamefont {Li}}, \bibinfo
  {author} {\bibfnamefont {M.~D.}\ \bibnamefont {Stiles}}, \ and\ \bibinfo
  {author} {\bibfnamefont {P.~M.}\ \bibnamefont {Haney}},\ }\href {\doibase
  10.1103/PhysRevB.99.220405} {\bibfield  {journal} {\bibinfo  {journal} {Phys.
  Rev. B}\ }\textbf {\bibinfo {volume} {99}},\ \bibinfo {pages} {220405}
  (\bibinfo {year} {2019})}\BibitemShut {NoStop}%
\bibitem [{\citenamefont {Liu}\ \emph {et~al.}(2011{\natexlab{a}})\citenamefont
  {Liu}, \citenamefont {Buhrman},\ and\ \citenamefont {Ralph}}]{Liu2011a}%
  \BibitemOpen
  \bibfield  {author} {\bibinfo {author} {\bibfnamefont {L.}~\bibnamefont
  {Liu}}, \bibinfo {author} {\bibfnamefont {R.~A.}\ \bibnamefont {Buhrman}}, \
  and\ \bibinfo {author} {\bibfnamefont {D.~C.}\ \bibnamefont {Ralph}},\ }\href
  {http://arxiv.org/abs/1111.3702} {\ ,\ \bibinfo {pages} {32} (\bibinfo {year}
  {2011}{\natexlab{a}})},\ \Eprint {http://arxiv.org/abs/1111.3702}
  {arXiv:1111.3702} \BibitemShut {NoStop}%
\bibitem [{\citenamefont {Morota}\ \emph {et~al.}(2011)\citenamefont {Morota},
  \citenamefont {Niimi}, \citenamefont {Ohnishi}, \citenamefont {Wei},
  \citenamefont {Tanaka}, \citenamefont {Kontani}, \citenamefont {Kimura},\
  and\ \citenamefont {Otani}}]{Morota2011}%
  \BibitemOpen
  \bibfield  {author} {\bibinfo {author} {\bibfnamefont {M.}~\bibnamefont
  {Morota}}, \bibinfo {author} {\bibfnamefont {Y.}~\bibnamefont {Niimi}},
  \bibinfo {author} {\bibfnamefont {K.}~\bibnamefont {Ohnishi}}, \bibinfo
  {author} {\bibfnamefont {D.~H.}\ \bibnamefont {Wei}}, \bibinfo {author}
  {\bibfnamefont {T.}~\bibnamefont {Tanaka}}, \bibinfo {author} {\bibfnamefont
  {H.}~\bibnamefont {Kontani}}, \bibinfo {author} {\bibfnamefont
  {T.}~\bibnamefont {Kimura}}, \ and\ \bibinfo {author} {\bibfnamefont
  {Y.}~\bibnamefont {Otani}},\ }\href {\doibase 10.1103/PhysRevB.83.174405}
  {\bibfield  {journal} {\bibinfo  {journal} {Phys. Rev. B}\ }\textbf {\bibinfo
  {volume} {83}},\ \bibinfo {pages} {174405} (\bibinfo {year}
  {2011})}\BibitemShut {NoStop}%
\bibitem [{\citenamefont {Kondou}\ \emph {et~al.}(2012)\citenamefont {Kondou},
  \citenamefont {Sukegawa}, \citenamefont {Mitani}, \citenamefont
  {Tsukagoshi},\ and\ \citenamefont {Kasai}}]{Kondou2012}%
  \BibitemOpen
  \bibfield  {author} {\bibinfo {author} {\bibfnamefont {K.}~\bibnamefont
  {Kondou}}, \bibinfo {author} {\bibfnamefont {H.}~\bibnamefont {Sukegawa}},
  \bibinfo {author} {\bibfnamefont {S.}~\bibnamefont {Mitani}}, \bibinfo
  {author} {\bibfnamefont {K.}~\bibnamefont {Tsukagoshi}}, \ and\ \bibinfo
  {author} {\bibfnamefont {S.}~\bibnamefont {Kasai}},\ }\href {\doibase
  10.1143/APEX.5.073002} {\bibfield  {journal} {\bibinfo  {journal} {Appl.
  Phys. Express}\ }\textbf {\bibinfo {volume} {5}},\ \bibinfo {pages} {073002}
  (\bibinfo {year} {2012})}\BibitemShut {NoStop}%
\bibitem [{\citenamefont {Zhang}\ \emph {et~al.}(2013)\citenamefont {Zhang},
  \citenamefont {Vlaminck}, \citenamefont {Pearson}, \citenamefont {Divan},
  \citenamefont {Bader},\ and\ \citenamefont {Hoffmann}}]{Zhang2013}%
  \BibitemOpen
  \bibfield  {author} {\bibinfo {author} {\bibfnamefont {W.}~\bibnamefont
  {Zhang}}, \bibinfo {author} {\bibfnamefont {V.}~\bibnamefont {Vlaminck}},
  \bibinfo {author} {\bibfnamefont {J.~E.}\ \bibnamefont {Pearson}}, \bibinfo
  {author} {\bibfnamefont {R.}~\bibnamefont {Divan}}, \bibinfo {author}
  {\bibfnamefont {S.~D.}\ \bibnamefont {Bader}}, \ and\ \bibinfo {author}
  {\bibfnamefont {A.}~\bibnamefont {Hoffmann}},\ }\href {\doibase
  10.1063/1.4848102} {\bibfield  {journal} {\bibinfo  {journal} {Appl. Phys.
  Lett.}\ }\textbf {\bibinfo {volume} {103}},\ \bibinfo {pages} {242414}
  (\bibinfo {year} {2013})}\BibitemShut {NoStop}%
\bibitem [{\citenamefont {Niimi}\ \emph {et~al.}(2013)\citenamefont {Niimi},
  \citenamefont {Wei}, \citenamefont {Idzuchi}, \citenamefont {Wakamura},
  \citenamefont {Kato},\ and\ \citenamefont {Otani}}]{Niimi2013}%
  \BibitemOpen
  \bibfield  {author} {\bibinfo {author} {\bibfnamefont {Y.}~\bibnamefont
  {Niimi}}, \bibinfo {author} {\bibfnamefont {D.}~\bibnamefont {Wei}}, \bibinfo
  {author} {\bibfnamefont {H.}~\bibnamefont {Idzuchi}}, \bibinfo {author}
  {\bibfnamefont {T.}~\bibnamefont {Wakamura}}, \bibinfo {author}
  {\bibfnamefont {T.}~\bibnamefont {Kato}}, \ and\ \bibinfo {author}
  {\bibfnamefont {Y.}~\bibnamefont {Otani}},\ }\href {\doibase
  10.1103/PhysRevLett.110.016805} {\bibfield  {journal} {\bibinfo  {journal}
  {Phys. Rev. Lett.}\ }\textbf {\bibinfo {volume} {110}},\ \bibinfo {pages}
  {016805} (\bibinfo {year} {2013})}\BibitemShut {NoStop}%
\bibitem [{\citenamefont {Wang}\ \emph {et~al.}(2014)\citenamefont {Wang},
  \citenamefont {Du}, \citenamefont {Pu}, \citenamefont {Adur}, \citenamefont
  {Hammel},\ and\ \citenamefont {Yang}}]{Wang2014}%
  \BibitemOpen
  \bibfield  {author} {\bibinfo {author} {\bibfnamefont {H.~L.}\ \bibnamefont
  {Wang}}, \bibinfo {author} {\bibfnamefont {C.~H.}\ \bibnamefont {Du}},
  \bibinfo {author} {\bibfnamefont {Y.}~\bibnamefont {Pu}}, \bibinfo {author}
  {\bibfnamefont {R.}~\bibnamefont {Adur}}, \bibinfo {author} {\bibfnamefont
  {P.~C.}\ \bibnamefont {Hammel}}, \ and\ \bibinfo {author} {\bibfnamefont
  {F.~Y.}\ \bibnamefont {Yang}},\ }\href {\doibase
  10.1103/PhysRevLett.112.197201} {\bibfield  {journal} {\bibinfo  {journal}
  {Phys. Rev. Lett.}\ }\textbf {\bibinfo {volume} {112}},\ \bibinfo {pages}
  {197201} (\bibinfo {year} {2014})}\BibitemShut {NoStop}%
\bibitem [{\citenamefont {Isasa}\ \emph {et~al.}(2015)\citenamefont {Isasa},
  \citenamefont {Villamor}, \citenamefont {Hueso}, \citenamefont {Gradhand},\
  and\ \citenamefont {Casanova}}]{Isasa2015}%
  \BibitemOpen
  \bibfield  {author} {\bibinfo {author} {\bibfnamefont {M.}~\bibnamefont
  {Isasa}}, \bibinfo {author} {\bibfnamefont {E.}~\bibnamefont {Villamor}},
  \bibinfo {author} {\bibfnamefont {L.~E.}\ \bibnamefont {Hueso}}, \bibinfo
  {author} {\bibfnamefont {M.}~\bibnamefont {Gradhand}}, \ and\ \bibinfo
  {author} {\bibfnamefont {F.}~\bibnamefont {Casanova}},\ }\href {\doibase
  10.1103/PhysRevB.91.024402} {\bibfield  {journal} {\bibinfo  {journal} {Phys.
  Rev. B}\ }\textbf {\bibinfo {volume} {91}},\ \bibinfo {pages} {024402}
  (\bibinfo {year} {2015})}\BibitemShut {NoStop}%
\bibitem [{\citenamefont {Boone}\ \emph {et~al.}(2015)\citenamefont {Boone},
  \citenamefont {Shaw}, \citenamefont {Nembach},\ and\ \citenamefont
  {Silva}}]{Boone2015}%
  \BibitemOpen
  \bibfield  {author} {\bibinfo {author} {\bibfnamefont {C.~T.}\ \bibnamefont
  {Boone}}, \bibinfo {author} {\bibfnamefont {J.~M.}\ \bibnamefont {Shaw}},
  \bibinfo {author} {\bibfnamefont {H.~T.}\ \bibnamefont {Nembach}}, \ and\
  \bibinfo {author} {\bibfnamefont {T.~J.}\ \bibnamefont {Silva}},\ }\href
  {\doibase 10.1063/1.4922581} {\bibfield  {journal} {\bibinfo  {journal} {J.
  Appl. Phys.}\ }\textbf {\bibinfo {volume} {117}},\ \bibinfo {pages} {223910}
  (\bibinfo {year} {2015})}\BibitemShut {NoStop}%
\bibitem [{\citenamefont {Zhang}\ \emph
  {et~al.}(2015{\natexlab{b}})\citenamefont {Zhang}, \citenamefont {Han},
  \citenamefont {Jiang}, \citenamefont {Yang},\ and\ \citenamefont {{S. P.
  Parkin}}}]{Zhang2015e}%
  \BibitemOpen
  \bibfield  {author} {\bibinfo {author} {\bibfnamefont {W.}~\bibnamefont
  {Zhang}}, \bibinfo {author} {\bibfnamefont {W.}~\bibnamefont {Han}}, \bibinfo
  {author} {\bibfnamefont {X.}~\bibnamefont {Jiang}}, \bibinfo {author}
  {\bibfnamefont {S.-H.}\ \bibnamefont {Yang}}, \ and\ \bibinfo {author}
  {\bibfnamefont {S.}~\bibnamefont {{S. P. Parkin}}},\ }\href {\doibase
  10.1038/nphys3304} {\bibfield  {journal} {\bibinfo  {journal} {Nat. Phys.}\
  }\textbf {\bibinfo {volume} {11}},\ \bibinfo {pages} {496} (\bibinfo {year}
  {2015}{\natexlab{b}})}\BibitemShut {NoStop}%
\bibitem [{\citenamefont {Sagasta}\ \emph {et~al.}(2016)\citenamefont
  {Sagasta}, \citenamefont {Omori}, \citenamefont {Isasa}, \citenamefont
  {Gradhand}, \citenamefont {Hueso}, \citenamefont {Niimi}, \citenamefont
  {Otani},\ and\ \citenamefont {Casanova}}]{Sagasta2016}%
  \BibitemOpen
  \bibfield  {author} {\bibinfo {author} {\bibfnamefont {E.}~\bibnamefont
  {Sagasta}}, \bibinfo {author} {\bibfnamefont {Y.}~\bibnamefont {Omori}},
  \bibinfo {author} {\bibfnamefont {M.}~\bibnamefont {Isasa}}, \bibinfo
  {author} {\bibfnamefont {M.}~\bibnamefont {Gradhand}}, \bibinfo {author}
  {\bibfnamefont {L.~E.}\ \bibnamefont {Hueso}}, \bibinfo {author}
  {\bibfnamefont {Y.}~\bibnamefont {Niimi}}, \bibinfo {author} {\bibfnamefont
  {Y.}~\bibnamefont {Otani}}, \ and\ \bibinfo {author} {\bibfnamefont
  {F.}~\bibnamefont {Casanova}},\ }\href {\doibase 10.1103/PhysRevB.94.060412}
  {\bibfield  {journal} {\bibinfo  {journal} {Phys. Rev. B}\ }\textbf {\bibinfo
  {volume} {94}},\ \bibinfo {pages} {060412} (\bibinfo {year}
  {2016})}\BibitemShut {NoStop}%
\bibitem [{\citenamefont {Nguyen}\ \emph {et~al.}(2016)\citenamefont {Nguyen},
  \citenamefont {Ralph},\ and\ \citenamefont {Buhrman}}]{Nguyen2016}%
  \BibitemOpen
  \bibfield  {author} {\bibinfo {author} {\bibfnamefont {M.-H.}\ \bibnamefont
  {Nguyen}}, \bibinfo {author} {\bibfnamefont {D.~C.}\ \bibnamefont {Ralph}}, \
  and\ \bibinfo {author} {\bibfnamefont {R.~A.}\ \bibnamefont {Buhrman}},\
  }\href {\doibase 10.1103/PhysRevLett.116.126601} {\bibfield  {journal}
  {\bibinfo  {journal} {Phys. Rev. Lett.}\ }\textbf {\bibinfo {volume} {116}},\
  \bibinfo {pages} {126601} (\bibinfo {year} {2016})}\BibitemShut {NoStop}%
\bibitem [{\citenamefont {Stamm}\ \emph {et~al.}(2017)\citenamefont {Stamm},
  \citenamefont {Murer}, \citenamefont {Berritta}, \citenamefont {Feng},
  \citenamefont {Gabureac}, \citenamefont {Oppeneer},\ and\ \citenamefont
  {Gambardella}}]{Stamm2017}%
  \BibitemOpen
  \bibfield  {author} {\bibinfo {author} {\bibfnamefont {C.}~\bibnamefont
  {Stamm}}, \bibinfo {author} {\bibfnamefont {C.}~\bibnamefont {Murer}},
  \bibinfo {author} {\bibfnamefont {M.}~\bibnamefont {Berritta}}, \bibinfo
  {author} {\bibfnamefont {J.}~\bibnamefont {Feng}}, \bibinfo {author}
  {\bibfnamefont {M.}~\bibnamefont {Gabureac}}, \bibinfo {author}
  {\bibfnamefont {P.~M.}\ \bibnamefont {Oppeneer}}, \ and\ \bibinfo {author}
  {\bibfnamefont {P.}~\bibnamefont {Gambardella}},\ }\href {\doibase
  10.1103/PhysRevLett.119.087203} {\bibfield  {journal} {\bibinfo  {journal}
  {Phys. Rev. Lett.}\ }\textbf {\bibinfo {volume} {119}},\ \bibinfo {pages}
  {087203} (\bibinfo {year} {2017})}\BibitemShut {NoStop}%
\bibitem [{\citenamefont {Tao}\ \emph {et~al.}(2018)\citenamefont {Tao},
  \citenamefont {Liu}, \citenamefont {Miao}, \citenamefont {Yu}, \citenamefont
  {Feng}, \citenamefont {Sun}, \citenamefont {You}, \citenamefont {Du},
  \citenamefont {Chen}, \citenamefont {Zhang}, \citenamefont {Zhang},
  \citenamefont {Yuan}, \citenamefont {Wu},\ and\ \citenamefont
  {Ding}}]{Tao2018}%
  \BibitemOpen
  \bibfield  {author} {\bibinfo {author} {\bibfnamefont {X.}~\bibnamefont
  {Tao}}, \bibinfo {author} {\bibfnamefont {Q.}~\bibnamefont {Liu}}, \bibinfo
  {author} {\bibfnamefont {B.}~\bibnamefont {Miao}}, \bibinfo {author}
  {\bibfnamefont {R.}~\bibnamefont {Yu}}, \bibinfo {author} {\bibfnamefont
  {Z.}~\bibnamefont {Feng}}, \bibinfo {author} {\bibfnamefont {L.}~\bibnamefont
  {Sun}}, \bibinfo {author} {\bibfnamefont {B.}~\bibnamefont {You}}, \bibinfo
  {author} {\bibfnamefont {J.}~\bibnamefont {Du}}, \bibinfo {author}
  {\bibfnamefont {K.}~\bibnamefont {Chen}}, \bibinfo {author} {\bibfnamefont
  {S.}~\bibnamefont {Zhang}}, \bibinfo {author} {\bibfnamefont
  {L.}~\bibnamefont {Zhang}}, \bibinfo {author} {\bibfnamefont
  {Z.}~\bibnamefont {Yuan}}, \bibinfo {author} {\bibfnamefont {D.}~\bibnamefont
  {Wu}}, \ and\ \bibinfo {author} {\bibfnamefont {H.}~\bibnamefont {Ding}},\
  }\href {\doibase 10.1126/sciadv.aat1670} {\bibfield  {journal} {\bibinfo
  {journal} {Sci. Adv.}\ }\textbf {\bibinfo {volume} {4}},\ \bibinfo {pages}
  {eaat1670} (\bibinfo {year} {2018})}\BibitemShut {NoStop}%
\bibitem [{\citenamefont {Swindells}\ \emph {et~al.}(2019)\citenamefont
  {Swindells}, \citenamefont {Hindmarch}, \citenamefont {Gallant},\ and\
  \citenamefont {Atkinson}}]{Swindells2019}%
  \BibitemOpen
  \bibfield  {author} {\bibinfo {author} {\bibfnamefont {C.}~\bibnamefont
  {Swindells}}, \bibinfo {author} {\bibfnamefont {A.~T.}\ \bibnamefont
  {Hindmarch}}, \bibinfo {author} {\bibfnamefont {A.~J.}\ \bibnamefont
  {Gallant}}, \ and\ \bibinfo {author} {\bibfnamefont {D.}~\bibnamefont
  {Atkinson}},\ }\href {\doibase 10.1103/PhysRevB.99.064406} {\bibfield
  {journal} {\bibinfo  {journal} {Phys. Rev. B}\ }\textbf {\bibinfo {volume}
  {99}},\ \bibinfo {pages} {064406} (\bibinfo {year} {2019})}\BibitemShut
  {NoStop}%
\bibitem [{\citenamefont {Zhu}\ \emph {et~al.}(2019{\natexlab{b}})\citenamefont
  {Zhu}, \citenamefont {Zhu}, \citenamefont {Sui}, \citenamefont {Ralph},\ and\
  \citenamefont {Buhrman}}]{Zhu2019c}%
  \BibitemOpen
  \bibfield  {author} {\bibinfo {author} {\bibfnamefont {L.}~\bibnamefont
  {Zhu}}, \bibinfo {author} {\bibfnamefont {L.}~\bibnamefont {Zhu}}, \bibinfo
  {author} {\bibfnamefont {M.}~\bibnamefont {Sui}}, \bibinfo {author}
  {\bibfnamefont {D.~C.}\ \bibnamefont {Ralph}}, \ and\ \bibinfo {author}
  {\bibfnamefont {R.~A.}\ \bibnamefont {Buhrman}},\ }\href {\doibase
  10.1126/sciadv.aav8025} {\bibfield  {journal} {\bibinfo  {journal} {Sci.
  Adv.}\ }\textbf {\bibinfo {volume} {5}},\ \bibinfo {pages} {eaav8025}
  (\bibinfo {year} {2019}{\natexlab{b}})}\BibitemShut {NoStop}%
\bibitem [{\citenamefont {Dai}\ \emph {et~al.}(2019)\citenamefont {Dai},
  \citenamefont {Xu}, \citenamefont {Chen}, \citenamefont {Fan}, \citenamefont
  {Yang}, \citenamefont {Xue}, \citenamefont {Song}, \citenamefont {Zhu},
  \citenamefont {Zhou},\ and\ \citenamefont {Qiu}}]{Dai2019}%
  \BibitemOpen
  \bibfield  {author} {\bibinfo {author} {\bibfnamefont {Y.}~\bibnamefont
  {Dai}}, \bibinfo {author} {\bibfnamefont {S.~J.}\ \bibnamefont {Xu}},
  \bibinfo {author} {\bibfnamefont {S.~W.}\ \bibnamefont {Chen}}, \bibinfo
  {author} {\bibfnamefont {X.~L.}\ \bibnamefont {Fan}}, \bibinfo {author}
  {\bibfnamefont {D.~Z.}\ \bibnamefont {Yang}}, \bibinfo {author}
  {\bibfnamefont {D.~S.}\ \bibnamefont {Xue}}, \bibinfo {author} {\bibfnamefont
  {D.~S.}\ \bibnamefont {Song}}, \bibinfo {author} {\bibfnamefont
  {J.}~\bibnamefont {Zhu}}, \bibinfo {author} {\bibfnamefont {S.~M.}\
  \bibnamefont {Zhou}}, \ and\ \bibinfo {author} {\bibfnamefont
  {X.}~\bibnamefont {Qiu}},\ }\href {\doibase 10.1103/PhysRevB.100.064404}
  {\bibfield  {journal} {\bibinfo  {journal} {Phys. Rev. B}\ }\textbf {\bibinfo
  {volume} {100}},\ \bibinfo {pages} {064404} (\bibinfo {year}
  {2019})}\BibitemShut {NoStop}%
\bibitem [{\citenamefont {Emori}\ \emph
  {et~al.}(2018{\natexlab{a}})\citenamefont {Emori}, \citenamefont {Yi},
  \citenamefont {Crossley}, \citenamefont {Wisser}, \citenamefont
  {Balakrishnan}, \citenamefont {Shafer}, \citenamefont {Klewe}, \citenamefont
  {N'Diaye}, \citenamefont {Urwin}, \citenamefont {Mahalingam}, \citenamefont
  {Howe}, \citenamefont {Hwang}, \citenamefont {Arenholz},\ and\ \citenamefont
  {Suzuki}}]{Emori2018a}%
  \BibitemOpen
  \bibfield  {author} {\bibinfo {author} {\bibfnamefont {S.}~\bibnamefont
  {Emori}}, \bibinfo {author} {\bibfnamefont {D.}~\bibnamefont {Yi}}, \bibinfo
  {author} {\bibfnamefont {S.}~\bibnamefont {Crossley}}, \bibinfo {author}
  {\bibfnamefont {J.~J.}\ \bibnamefont {Wisser}}, \bibinfo {author}
  {\bibfnamefont {P.~P.}\ \bibnamefont {Balakrishnan}}, \bibinfo {author}
  {\bibfnamefont {P.}~\bibnamefont {Shafer}}, \bibinfo {author} {\bibfnamefont
  {C.}~\bibnamefont {Klewe}}, \bibinfo {author} {\bibfnamefont {A.~T.}\
  \bibnamefont {N'Diaye}}, \bibinfo {author} {\bibfnamefont {B.~T.}\
  \bibnamefont {Urwin}}, \bibinfo {author} {\bibfnamefont {K.}~\bibnamefont
  {Mahalingam}}, \bibinfo {author} {\bibfnamefont {B.~M.}\ \bibnamefont
  {Howe}}, \bibinfo {author} {\bibfnamefont {H.~Y.}\ \bibnamefont {Hwang}},
  \bibinfo {author} {\bibfnamefont {E.}~\bibnamefont {Arenholz}}, \ and\
  \bibinfo {author} {\bibfnamefont {Y.}~\bibnamefont {Suzuki}},\ }\href
  {https://pubs.acs.org/doi/10.1021/acs.nanolett.8b01261} {\bibfield  {journal}
  {\bibinfo  {journal} {Nano Lett.}\ }\textbf {\bibinfo {volume} {18}},\
  \bibinfo {pages} {4273} (\bibinfo {year} {2018}{\natexlab{a}})}\BibitemShut
  {NoStop}%
\bibitem [{\citenamefont {Lee}\ \emph {et~al.}(2019)\citenamefont {Lee},
  \citenamefont {Ahmed}, \citenamefont {Guo}, \citenamefont {Esser},
  \citenamefont {McComb},\ and\ \citenamefont {Yang}}]{Lee2019}%
  \BibitemOpen
  \bibfield  {author} {\bibinfo {author} {\bibfnamefont {A.~J.}\ \bibnamefont
  {Lee}}, \bibinfo {author} {\bibfnamefont {A.~S.}\ \bibnamefont {Ahmed}},
  \bibinfo {author} {\bibfnamefont {S.}~\bibnamefont {Guo}}, \bibinfo {author}
  {\bibfnamefont {B.~D.}\ \bibnamefont {Esser}}, \bibinfo {author}
  {\bibfnamefont {D.~W.}\ \bibnamefont {McComb}}, \ and\ \bibinfo {author}
  {\bibfnamefont {F.}~\bibnamefont {Yang}},\ }\href {\doibase
  10.1063/1.5093503} {\bibfield  {journal} {\bibinfo  {journal} {J. Appl.
  Phys.}\ }\textbf {\bibinfo {volume} {125}},\ \bibinfo {pages} {183903}
  (\bibinfo {year} {2019})}\BibitemShut {NoStop}%
\bibitem [{\citenamefont {Emori}\ \emph
  {et~al.}(2018{\natexlab{b}})\citenamefont {Emori}, \citenamefont {Matyushov},
  \citenamefont {Jeon}, \citenamefont {Babroski}, \citenamefont {Nan},
  \citenamefont {Belkessam}, \citenamefont {Jones}, \citenamefont {McConney},
  \citenamefont {Brown}, \citenamefont {Howe},\ and\ \citenamefont
  {Sun}}]{Emori2018}%
  \BibitemOpen
  \bibfield  {author} {\bibinfo {author} {\bibfnamefont {S.}~\bibnamefont
  {Emori}}, \bibinfo {author} {\bibfnamefont {A.}~\bibnamefont {Matyushov}},
  \bibinfo {author} {\bibfnamefont {H.-M.}\ \bibnamefont {Jeon}}, \bibinfo
  {author} {\bibfnamefont {C.~J.}\ \bibnamefont {Babroski}}, \bibinfo {author}
  {\bibfnamefont {T.}~\bibnamefont {Nan}}, \bibinfo {author} {\bibfnamefont
  {A.~M.}\ \bibnamefont {Belkessam}}, \bibinfo {author} {\bibfnamefont {J.~G.}\
  \bibnamefont {Jones}}, \bibinfo {author} {\bibfnamefont {M.~E.}\ \bibnamefont
  {McConney}}, \bibinfo {author} {\bibfnamefont {G.~J.}\ \bibnamefont {Brown}},
  \bibinfo {author} {\bibfnamefont {B.~M.}\ \bibnamefont {Howe}}, \ and\
  \bibinfo {author} {\bibfnamefont {N.~X.}\ \bibnamefont {Sun}},\ }\href
  {\doibase 10.1063/1.5025623} {\bibfield  {journal} {\bibinfo  {journal}
  {Appl. Phys. Lett.}\ }\textbf {\bibinfo {volume} {112}},\ \bibinfo {pages}
  {182406} (\bibinfo {year} {2018}{\natexlab{b}})}\BibitemShut {NoStop}%
\bibitem [{\citenamefont {Valvidares}\ \emph {et~al.}(2016)\citenamefont
  {Valvidares}, \citenamefont {Dix}, \citenamefont {Isasa}, \citenamefont
  {Ollefs}, \citenamefont {Wilhelm}, \citenamefont {Rogalev}, \citenamefont
  {S{\'{a}}nchez}, \citenamefont {Pellegrin}, \citenamefont {Bedoya-Pinto},
  \citenamefont {Gargiani}, \citenamefont {Hueso}, \citenamefont {Casanova},\
  and\ \citenamefont {Fontcuberta}}]{Valvidares2016}%
  \BibitemOpen
  \bibfield  {author} {\bibinfo {author} {\bibfnamefont {M.}~\bibnamefont
  {Valvidares}}, \bibinfo {author} {\bibfnamefont {N.}~\bibnamefont {Dix}},
  \bibinfo {author} {\bibfnamefont {M.}~\bibnamefont {Isasa}}, \bibinfo
  {author} {\bibfnamefont {K.}~\bibnamefont {Ollefs}}, \bibinfo {author}
  {\bibfnamefont {F.}~\bibnamefont {Wilhelm}}, \bibinfo {author} {\bibfnamefont
  {A.}~\bibnamefont {Rogalev}}, \bibinfo {author} {\bibfnamefont
  {F.}~\bibnamefont {S{\'{a}}nchez}}, \bibinfo {author} {\bibfnamefont
  {E.}~\bibnamefont {Pellegrin}}, \bibinfo {author} {\bibfnamefont
  {A.}~\bibnamefont {Bedoya-Pinto}}, \bibinfo {author} {\bibfnamefont
  {P.}~\bibnamefont {Gargiani}}, \bibinfo {author} {\bibfnamefont {L.~E.}\
  \bibnamefont {Hueso}}, \bibinfo {author} {\bibfnamefont {F.}~\bibnamefont
  {Casanova}}, \ and\ \bibinfo {author} {\bibfnamefont {J.}~\bibnamefont
  {Fontcuberta}},\ }\href {\doibase 10.1103/PhysRevB.93.214415} {\bibfield
  {journal} {\bibinfo  {journal} {Phys. Rev. B}\ }\textbf {\bibinfo {volume}
  {93}},\ \bibinfo {pages} {214415} (\bibinfo {year} {2016})}\BibitemShut
  {NoStop}%
\bibitem [{\citenamefont {Gray}\ \emph
  {et~al.}(2018{\natexlab{a}})\citenamefont {Gray}, \citenamefont {Emori},
  \citenamefont {Gray}, \citenamefont {Jeon}, \citenamefont {{van 't Erve}},
  \citenamefont {Jonker}, \citenamefont {Kim}, \citenamefont {Suzuki},
  \citenamefont {Ono}, \citenamefont {Howe},\ and\ \citenamefont
  {Suzuki}}]{Gray2018a}%
  \BibitemOpen
  \bibfield  {author} {\bibinfo {author} {\bibfnamefont {M.~T.}\ \bibnamefont
  {Gray}}, \bibinfo {author} {\bibfnamefont {S.}~\bibnamefont {Emori}},
  \bibinfo {author} {\bibfnamefont {B.~A.}\ \bibnamefont {Gray}}, \bibinfo
  {author} {\bibfnamefont {H.}~\bibnamefont {Jeon}}, \bibinfo {author}
  {\bibfnamefont {O.~M.~J.}\ \bibnamefont {{van 't Erve}}}, \bibinfo {author}
  {\bibfnamefont {B.~T.}\ \bibnamefont {Jonker}}, \bibinfo {author}
  {\bibfnamefont {S.}~\bibnamefont {Kim}}, \bibinfo {author} {\bibfnamefont
  {M.}~\bibnamefont {Suzuki}}, \bibinfo {author} {\bibfnamefont
  {T.}~\bibnamefont {Ono}}, \bibinfo {author} {\bibfnamefont {B.~M.}\
  \bibnamefont {Howe}}, \ and\ \bibinfo {author} {\bibfnamefont
  {Y.}~\bibnamefont {Suzuki}},\ }\href {\doibase
  10.1103/PhysRevApplied.9.064039} {\bibfield  {journal} {\bibinfo  {journal}
  {Phys. Rev. Appl.}\ }\textbf {\bibinfo {volume} {9}},\ \bibinfo {pages}
  {064039} (\bibinfo {year} {2018}{\natexlab{a}})}\BibitemShut {NoStop}%
\bibitem [{\citenamefont {Riddiford}\ \emph {et~al.}(2019)\citenamefont
  {Riddiford}, \citenamefont {Wisser}, \citenamefont {Emori}, \citenamefont
  {Li}, \citenamefont {Roy}, \citenamefont {Cogulu}, \citenamefont {{van 't
  Erve}}, \citenamefont {Deng}, \citenamefont {Wang}, \citenamefont {Jonker},
  \citenamefont {Kent},\ and\ \citenamefont {Suzuki}}]{Riddiford2019}%
  \BibitemOpen
  \bibfield  {author} {\bibinfo {author} {\bibfnamefont {L.~J.}\ \bibnamefont
  {Riddiford}}, \bibinfo {author} {\bibfnamefont {J.~J.}\ \bibnamefont
  {Wisser}}, \bibinfo {author} {\bibfnamefont {S.}~\bibnamefont {Emori}},
  \bibinfo {author} {\bibfnamefont {P.}~\bibnamefont {Li}}, \bibinfo {author}
  {\bibfnamefont {D.}~\bibnamefont {Roy}}, \bibinfo {author} {\bibfnamefont
  {E.}~\bibnamefont {Cogulu}}, \bibinfo {author} {\bibfnamefont
  {O.}~\bibnamefont {{van 't Erve}}}, \bibinfo {author} {\bibfnamefont
  {Y.}~\bibnamefont {Deng}}, \bibinfo {author} {\bibfnamefont {S.~X.}\
  \bibnamefont {Wang}}, \bibinfo {author} {\bibfnamefont {B.~T.}\ \bibnamefont
  {Jonker}}, \bibinfo {author} {\bibfnamefont {A.~D.}\ \bibnamefont {Kent}}, \
  and\ \bibinfo {author} {\bibfnamefont {Y.}~\bibnamefont {Suzuki}},\ }\href
  {\doibase 10.1063/1.5119726} {\bibfield  {journal} {\bibinfo  {journal}
  {Appl. Phys. Lett.}\ }\textbf {\bibinfo {volume} {115}},\ \bibinfo {pages}
  {122401} (\bibinfo {year} {2019})}\BibitemShut {NoStop}%
\bibitem [{\citenamefont {Liu}\ \emph {et~al.}(2011{\natexlab{b}})\citenamefont
  {Liu}, \citenamefont {Moriyama}, \citenamefont {Ralph},\ and\ \citenamefont
  {Buhrman}}]{Liu2011}%
  \BibitemOpen
  \bibfield  {author} {\bibinfo {author} {\bibfnamefont {L.}~\bibnamefont
  {Liu}}, \bibinfo {author} {\bibfnamefont {T.}~\bibnamefont {Moriyama}},
  \bibinfo {author} {\bibfnamefont {D.~C.}\ \bibnamefont {Ralph}}, \ and\
  \bibinfo {author} {\bibfnamefont {R.~A.}\ \bibnamefont {Buhrman}},\ }\href
  {\doibase 10.1103/PhysRevLett.106.036601} {\bibfield  {journal} {\bibinfo
  {journal} {Phys. Rev. Lett.}\ }\textbf {\bibinfo {volume} {106}},\ \bibinfo
  {pages} {036601} (\bibinfo {year} {2011}{\natexlab{b}})}\BibitemShut
  {NoStop}%
\bibitem [{\citenamefont {Kasai}\ \emph {et~al.}(2014)\citenamefont {Kasai},
  \citenamefont {Kondou}, \citenamefont {Sukegawa}, \citenamefont {Mitani},
  \citenamefont {Tsukagoshi},\ and\ \citenamefont {Otani}}]{Kasai2014}%
  \BibitemOpen
  \bibfield  {author} {\bibinfo {author} {\bibfnamefont {S.}~\bibnamefont
  {Kasai}}, \bibinfo {author} {\bibfnamefont {K.}~\bibnamefont {Kondou}},
  \bibinfo {author} {\bibfnamefont {H.}~\bibnamefont {Sukegawa}}, \bibinfo
  {author} {\bibfnamefont {S.}~\bibnamefont {Mitani}}, \bibinfo {author}
  {\bibfnamefont {K.}~\bibnamefont {Tsukagoshi}}, \ and\ \bibinfo {author}
  {\bibfnamefont {Y.}~\bibnamefont {Otani}},\ }\href {\doibase
  10.1063/1.4867649} {\bibfield  {journal} {\bibinfo  {journal} {Appl. Phys.
  Lett.}\ }\textbf {\bibinfo {volume} {104}},\ \bibinfo {pages} {092408}
  (\bibinfo {year} {2014})}\BibitemShut {NoStop}%
\bibitem [{\citenamefont {Nan}\ \emph {et~al.}(2015)\citenamefont {Nan},
  \citenamefont {Emori}, \citenamefont {Boone}, \citenamefont {Wang},
  \citenamefont {Oxholm}, \citenamefont {Jones}, \citenamefont {Howe},
  \citenamefont {Brown},\ and\ \citenamefont {Sun}}]{Nan2015a}%
  \BibitemOpen
  \bibfield  {author} {\bibinfo {author} {\bibfnamefont {T.}~\bibnamefont
  {Nan}}, \bibinfo {author} {\bibfnamefont {S.}~\bibnamefont {Emori}}, \bibinfo
  {author} {\bibfnamefont {C.~T.}\ \bibnamefont {Boone}}, \bibinfo {author}
  {\bibfnamefont {X.}~\bibnamefont {Wang}}, \bibinfo {author} {\bibfnamefont
  {T.~M.}\ \bibnamefont {Oxholm}}, \bibinfo {author} {\bibfnamefont {J.~G.}\
  \bibnamefont {Jones}}, \bibinfo {author} {\bibfnamefont {B.~M.}\ \bibnamefont
  {Howe}}, \bibinfo {author} {\bibfnamefont {G.~J.}\ \bibnamefont {Brown}}, \
  and\ \bibinfo {author} {\bibfnamefont {N.~X.}\ \bibnamefont {Sun}},\ }\href
  {\doibase 10.1103/PhysRevB.91.214416} {\bibfield  {journal} {\bibinfo
  {journal} {Phys. Rev. B}\ }\textbf {\bibinfo {volume} {91}},\ \bibinfo
  {pages} {214416} (\bibinfo {year} {2015})}\BibitemShut {NoStop}%
\bibitem [{\citenamefont {Tiwari}\ \emph {et~al.}(2017)\citenamefont {Tiwari},
  \citenamefont {Behera}, \citenamefont {Kumar}, \citenamefont
  {D{\"{u}}rrenfeld}, \citenamefont {Chaudhary}, \citenamefont {Pandya},
  \citenamefont {{\AA}kerman},\ and\ \citenamefont {Muduli}}]{Tiwari2017}%
  \BibitemOpen
  \bibfield  {author} {\bibinfo {author} {\bibfnamefont {D.}~\bibnamefont
  {Tiwari}}, \bibinfo {author} {\bibfnamefont {N.}~\bibnamefont {Behera}},
  \bibinfo {author} {\bibfnamefont {A.}~\bibnamefont {Kumar}}, \bibinfo
  {author} {\bibfnamefont {P.}~\bibnamefont {D{\"{u}}rrenfeld}}, \bibinfo
  {author} {\bibfnamefont {S.}~\bibnamefont {Chaudhary}}, \bibinfo {author}
  {\bibfnamefont {D.~K.}\ \bibnamefont {Pandya}}, \bibinfo {author}
  {\bibfnamefont {J.}~\bibnamefont {{\AA}kerman}}, \ and\ \bibinfo {author}
  {\bibfnamefont {P.~K.}\ \bibnamefont {Muduli}},\ }\href {\doibase
  10.1063/1.5007202} {\bibfield  {journal} {\bibinfo  {journal} {Appl. Phys.
  Lett.}\ }\textbf {\bibinfo {volume} {111}},\ \bibinfo {pages} {232407}
  (\bibinfo {year} {2017})}\BibitemShut {NoStop}%
\bibitem [{\citenamefont {Kim}\ \emph {et~al.}(2018)\citenamefont {Kim},
  \citenamefont {Kim}, \citenamefont {Chun}, \citenamefont {Moon},\ and\
  \citenamefont {Hwang}}]{Kim2018b}%
  \BibitemOpen
  \bibfield  {author} {\bibinfo {author} {\bibfnamefont {C.}~\bibnamefont
  {Kim}}, \bibinfo {author} {\bibfnamefont {D.}~\bibnamefont {Kim}}, \bibinfo
  {author} {\bibfnamefont {B.~S.}\ \bibnamefont {Chun}}, \bibinfo {author}
  {\bibfnamefont {K.-W.}\ \bibnamefont {Moon}}, \ and\ \bibinfo {author}
  {\bibfnamefont {C.}~\bibnamefont {Hwang}},\ }\href {\doibase
  10.1103/PhysRevApplied.9.054035} {\bibfield  {journal} {\bibinfo  {journal}
  {Phys. Rev. Appl.}\ }\textbf {\bibinfo {volume} {9}},\ \bibinfo {pages}
  {054035} (\bibinfo {year} {2018})}\BibitemShut {NoStop}%
\bibitem [{\citenamefont {Gambardella}\ and\ \citenamefont
  {Miron}(2011)}]{Gambardella2011a}%
  \BibitemOpen
  \bibfield  {author} {\bibinfo {author} {\bibfnamefont {P.}~\bibnamefont
  {Gambardella}}\ and\ \bibinfo {author} {\bibfnamefont {I.~M.}\ \bibnamefont
  {Miron}},\ }\href {\doibase 10.1098/rsta.2010.0336} {\bibfield  {journal}
  {\bibinfo  {journal} {Philos. Trans. A. Math. Phys. Eng. Sci.}\ }\textbf
  {\bibinfo {volume} {369}},\ \bibinfo {pages} {3175} (\bibinfo {year}
  {2011})}\BibitemShut {NoStop}%
\bibitem [{\citenamefont {Karnad}\ \emph {et~al.}(2018)\citenamefont {Karnad},
  \citenamefont {Gorini}, \citenamefont {Lee}, \citenamefont {Schulz},
  \citenamefont {{Lo Conte}}, \citenamefont {Wells}, \citenamefont {Han},
  \citenamefont {Shahbazi}, \citenamefont {Kim}, \citenamefont {Moore},
  \citenamefont {Swagten}, \citenamefont {Eckern}, \citenamefont {Raimondi},\
  and\ \citenamefont {Kl{\"{a}}ui}}]{Karnad2018a}%
  \BibitemOpen
  \bibfield  {author} {\bibinfo {author} {\bibfnamefont {G.~V.}\ \bibnamefont
  {Karnad}}, \bibinfo {author} {\bibfnamefont {C.}~\bibnamefont {Gorini}},
  \bibinfo {author} {\bibfnamefont {K.}~\bibnamefont {Lee}}, \bibinfo {author}
  {\bibfnamefont {T.}~\bibnamefont {Schulz}}, \bibinfo {author} {\bibfnamefont
  {R.}~\bibnamefont {{Lo Conte}}}, \bibinfo {author} {\bibfnamefont {A.~W.~J.}\
  \bibnamefont {Wells}}, \bibinfo {author} {\bibfnamefont {D.-S.}\ \bibnamefont
  {Han}}, \bibinfo {author} {\bibfnamefont {K.}~\bibnamefont {Shahbazi}},
  \bibinfo {author} {\bibfnamefont {J.-S.}\ \bibnamefont {Kim}}, \bibinfo
  {author} {\bibfnamefont {T.~A.}\ \bibnamefont {Moore}}, \bibinfo {author}
  {\bibfnamefont {H.~J.~M.}\ \bibnamefont {Swagten}}, \bibinfo {author}
  {\bibfnamefont {U.}~\bibnamefont {Eckern}}, \bibinfo {author} {\bibfnamefont
  {R.}~\bibnamefont {Raimondi}}, \ and\ \bibinfo {author} {\bibfnamefont
  {M.}~\bibnamefont {Kl{\"{a}}ui}},\ }\href {\doibase
  10.1103/PhysRevB.97.100405} {\bibfield  {journal} {\bibinfo  {journal} {Phys.
  Rev. B}\ }\textbf {\bibinfo {volume} {97}},\ \bibinfo {pages} {100405}
  (\bibinfo {year} {2018})}\BibitemShut {NoStop}%
\bibitem [{\citenamefont {Ryu}\ \emph {et~al.}(2016)\citenamefont {Ryu},
  \citenamefont {Kohda},\ and\ \citenamefont {Nitta}}]{Ryu2016}%
  \BibitemOpen
  \bibfield  {author} {\bibinfo {author} {\bibfnamefont {J.}~\bibnamefont
  {Ryu}}, \bibinfo {author} {\bibfnamefont {M.}~\bibnamefont {Kohda}}, \ and\
  \bibinfo {author} {\bibfnamefont {J.}~\bibnamefont {Nitta}},\ }\href
  {\doibase 10.1103/PhysRevLett.116.256802} {\bibfield  {journal} {\bibinfo
  {journal} {Phys. Rev. Lett.}\ }\textbf {\bibinfo {volume} {116}},\ \bibinfo
  {pages} {256802} (\bibinfo {year} {2016})}\BibitemShut {NoStop}%
\bibitem [{\citenamefont {Freeman}\ \emph {et~al.}(2018)\citenamefont
  {Freeman}, \citenamefont {Zholud}, \citenamefont {Dun}, \citenamefont
  {Zhou},\ and\ \citenamefont {Urazhdin}}]{Freeman2018}%
  \BibitemOpen
  \bibfield  {author} {\bibinfo {author} {\bibfnamefont {R.}~\bibnamefont
  {Freeman}}, \bibinfo {author} {\bibfnamefont {A.}~\bibnamefont {Zholud}},
  \bibinfo {author} {\bibfnamefont {Z.}~\bibnamefont {Dun}}, \bibinfo {author}
  {\bibfnamefont {H.}~\bibnamefont {Zhou}}, \ and\ \bibinfo {author}
  {\bibfnamefont {S.}~\bibnamefont {Urazhdin}},\ }\href {\doibase
  10.1103/PhysRevLett.120.067204} {\bibfield  {journal} {\bibinfo  {journal}
  {Phys. Rev. Lett.}\ }\textbf {\bibinfo {volume} {120}},\ \bibinfo {pages}
  {067204} (\bibinfo {year} {2018})}\BibitemShut {NoStop}%
\bibitem [{\citenamefont {Du}\ \emph {et~al.}(2020)\citenamefont {Du},
  \citenamefont {Gamou}, \citenamefont {Takahashi}, \citenamefont {Karube},
  \citenamefont {Kohda},\ and\ \citenamefont {Nitta}}]{Du2020}%
  \BibitemOpen
  \bibfield  {author} {\bibinfo {author} {\bibfnamefont {Y.}~\bibnamefont
  {Du}}, \bibinfo {author} {\bibfnamefont {H.}~\bibnamefont {Gamou}}, \bibinfo
  {author} {\bibfnamefont {S.}~\bibnamefont {Takahashi}}, \bibinfo {author}
  {\bibfnamefont {S.}~\bibnamefont {Karube}}, \bibinfo {author} {\bibfnamefont
  {M.}~\bibnamefont {Kohda}}, \ and\ \bibinfo {author} {\bibfnamefont
  {J.}~\bibnamefont {Nitta}},\ }\href {\doibase
  10.1103/PhysRevApplied.13.054014} {\bibfield  {journal} {\bibinfo  {journal}
  {Phys. Rev. Appl.}\ }\textbf {\bibinfo {volume} {13}},\ \bibinfo {pages}
  {054014} (\bibinfo {year} {2020})}\BibitemShut {NoStop}%
\bibitem [{\citenamefont {Wisser}\ \emph
  {et~al.}(2019{\natexlab{a}})\citenamefont {Wisser}, \citenamefont {Emori},
  \citenamefont {Riddiford}, \citenamefont {Altman}, \citenamefont {Li},
  \citenamefont {Mahalingam}, \citenamefont {Urwin}, \citenamefont {Howe},
  \citenamefont {Page}, \citenamefont {Grutter}, \citenamefont {Kirby},\ and\
  \citenamefont {Suzuki}}]{Wisser2019}%
  \BibitemOpen
  \bibfield  {author} {\bibinfo {author} {\bibfnamefont {J.~J.}\ \bibnamefont
  {Wisser}}, \bibinfo {author} {\bibfnamefont {S.}~\bibnamefont {Emori}},
  \bibinfo {author} {\bibfnamefont {L.}~\bibnamefont {Riddiford}}, \bibinfo
  {author} {\bibfnamefont {A.}~\bibnamefont {Altman}}, \bibinfo {author}
  {\bibfnamefont {P.}~\bibnamefont {Li}}, \bibinfo {author} {\bibfnamefont
  {K.}~\bibnamefont {Mahalingam}}, \bibinfo {author} {\bibfnamefont {B.~T.}\
  \bibnamefont {Urwin}}, \bibinfo {author} {\bibfnamefont {B.~M.}\ \bibnamefont
  {Howe}}, \bibinfo {author} {\bibfnamefont {M.~R.}\ \bibnamefont {Page}},
  \bibinfo {author} {\bibfnamefont {A.~J.}\ \bibnamefont {Grutter}}, \bibinfo
  {author} {\bibfnamefont {B.~J.}\ \bibnamefont {Kirby}}, \ and\ \bibinfo
  {author} {\bibfnamefont {Y.}~\bibnamefont {Suzuki}},\ }\href {\doibase
  10.1063/1.5111326} {\bibfield  {journal} {\bibinfo  {journal} {Appl. Phys.
  Lett.}\ }\textbf {\bibinfo {volume} {115}},\ \bibinfo {pages} {132404}
  (\bibinfo {year} {2019}{\natexlab{a}})}\BibitemShut {NoStop}%
\bibitem [{\citenamefont {Lustikova}\ \emph {et~al.}(2014)\citenamefont
  {Lustikova}, \citenamefont {Shiomi}, \citenamefont {Qiu}, \citenamefont
  {Kikkawa}, \citenamefont {Iguchi}, \citenamefont {Uchida},\ and\
  \citenamefont {Saitoh}}]{Lustikova2014}%
  \BibitemOpen
  \bibfield  {author} {\bibinfo {author} {\bibfnamefont {J.}~\bibnamefont
  {Lustikova}}, \bibinfo {author} {\bibfnamefont {Y.}~\bibnamefont {Shiomi}},
  \bibinfo {author} {\bibfnamefont {Z.}~\bibnamefont {Qiu}}, \bibinfo {author}
  {\bibfnamefont {T.}~\bibnamefont {Kikkawa}}, \bibinfo {author} {\bibfnamefont
  {R.}~\bibnamefont {Iguchi}}, \bibinfo {author} {\bibfnamefont
  {K.}~\bibnamefont {Uchida}}, \ and\ \bibinfo {author} {\bibfnamefont
  {E.}~\bibnamefont {Saitoh}},\ }\href {\doibase 10.1063/1.4898161} {\bibfield
  {journal} {\bibinfo  {journal} {J. Appl. Phys.}\ }\textbf {\bibinfo {volume}
  {116}},\ \bibinfo {pages} {153902} (\bibinfo {year} {2014})}\BibitemShut
  {NoStop}%
\bibitem [{\citenamefont {Chang}\ \emph {et~al.}(2017)\citenamefont {Chang},
  \citenamefont {Liu}, \citenamefont {{Reifsnyder Hickey}}, \citenamefont
  {Janantha}, \citenamefont {Mkhoyan},\ and\ \citenamefont {Wu}}]{Chang2017a}%
  \BibitemOpen
  \bibfield  {author} {\bibinfo {author} {\bibfnamefont {H.}~\bibnamefont
  {Chang}}, \bibinfo {author} {\bibfnamefont {T.}~\bibnamefont {Liu}}, \bibinfo
  {author} {\bibfnamefont {D.}~\bibnamefont {{Reifsnyder Hickey}}}, \bibinfo
  {author} {\bibfnamefont {P.~A.~P.}\ \bibnamefont {Janantha}}, \bibinfo
  {author} {\bibfnamefont {K.~A.}\ \bibnamefont {Mkhoyan}}, \ and\ \bibinfo
  {author} {\bibfnamefont {M.}~\bibnamefont {Wu}},\ }\href {\doibase
  10.1063/1.5013626} {\bibfield  {journal} {\bibinfo  {journal} {APL Mater.}\
  }\textbf {\bibinfo {volume} {5}},\ \bibinfo {pages} {126104} (\bibinfo {year}
  {2017})}\BibitemShut {NoStop}%
\bibitem [{\citenamefont {Schreier}\ \emph {et~al.}(2015)\citenamefont
  {Schreier}, \citenamefont {Chiba}, \citenamefont {Niedermayr}, \citenamefont
  {Lotze}, \citenamefont {Huebl}, \citenamefont {Gepr{\"{a}}gs}, \citenamefont
  {Takahashi}, \citenamefont {Bauer}, \citenamefont {Gross},\ and\
  \citenamefont {Goennenwein}}]{Schreier2015}%
  \BibitemOpen
  \bibfield  {author} {\bibinfo {author} {\bibfnamefont {M.}~\bibnamefont
  {Schreier}}, \bibinfo {author} {\bibfnamefont {T.}~\bibnamefont {Chiba}},
  \bibinfo {author} {\bibfnamefont {A.}~\bibnamefont {Niedermayr}}, \bibinfo
  {author} {\bibfnamefont {J.}~\bibnamefont {Lotze}}, \bibinfo {author}
  {\bibfnamefont {H.}~\bibnamefont {Huebl}}, \bibinfo {author} {\bibfnamefont
  {S.}~\bibnamefont {Gepr{\"{a}}gs}}, \bibinfo {author} {\bibfnamefont
  {S.}~\bibnamefont {Takahashi}}, \bibinfo {author} {\bibfnamefont {G.~E.~W.}\
  \bibnamefont {Bauer}}, \bibinfo {author} {\bibfnamefont {R.}~\bibnamefont
  {Gross}}, \ and\ \bibinfo {author} {\bibfnamefont {S.~T.~B.}\ \bibnamefont
  {Goennenwein}},\ }\href {\doibase 10.1103/PhysRevB.92.144411} {\bibfield
  {journal} {\bibinfo  {journal} {Phys. Rev. B}\ }\textbf {\bibinfo {volume}
  {92}},\ \bibinfo {pages} {144411} (\bibinfo {year} {2015})}\BibitemShut
  {NoStop}%
\bibitem [{\citenamefont {Sklenar}\ \emph {et~al.}(2015)\citenamefont
  {Sklenar}, \citenamefont {Zhang}, \citenamefont {Jungfleisch}, \citenamefont
  {Jiang}, \citenamefont {Chang}, \citenamefont {Pearson}, \citenamefont {Wu},
  \citenamefont {Ketterson},\ and\ \citenamefont {Hoffmann}}]{Sklenar2015a}%
  \BibitemOpen
  \bibfield  {author} {\bibinfo {author} {\bibfnamefont {J.}~\bibnamefont
  {Sklenar}}, \bibinfo {author} {\bibfnamefont {W.}~\bibnamefont {Zhang}},
  \bibinfo {author} {\bibfnamefont {M.~B.}\ \bibnamefont {Jungfleisch}},
  \bibinfo {author} {\bibfnamefont {W.}~\bibnamefont {Jiang}}, \bibinfo
  {author} {\bibfnamefont {H.}~\bibnamefont {Chang}}, \bibinfo {author}
  {\bibfnamefont {J.~E.}\ \bibnamefont {Pearson}}, \bibinfo {author}
  {\bibfnamefont {M.}~\bibnamefont {Wu}}, \bibinfo {author} {\bibfnamefont
  {J.~B.}\ \bibnamefont {Ketterson}}, \ and\ \bibinfo {author} {\bibfnamefont
  {A.}~\bibnamefont {Hoffmann}},\ }\href {\doibase 10.1103/PhysRevB.92.174406}
  {\bibfield  {journal} {\bibinfo  {journal} {Phys. Rev. B}\ }\textbf {\bibinfo
  {volume} {92}},\ \bibinfo {pages} {174406} (\bibinfo {year}
  {2015})}\BibitemShut {NoStop}%
\bibitem [{\citenamefont {Gon{\c{c}}alves}\ \emph {et~al.}(2013)\citenamefont
  {Gon{\c{c}}alves}, \citenamefont {Barsukov}, \citenamefont {Chen},
  \citenamefont {Yang}, \citenamefont {Katine},\ and\ \citenamefont
  {Krivorotov}}]{Goncalves2013b}%
  \BibitemOpen
  \bibfield  {author} {\bibinfo {author} {\bibfnamefont {A.~M.}\ \bibnamefont
  {Gon{\c{c}}alves}}, \bibinfo {author} {\bibfnamefont {I.}~\bibnamefont
  {Barsukov}}, \bibinfo {author} {\bibfnamefont {Y.-J.}\ \bibnamefont {Chen}},
  \bibinfo {author} {\bibfnamefont {L.}~\bibnamefont {Yang}}, \bibinfo {author}
  {\bibfnamefont {J.~A.}\ \bibnamefont {Katine}}, \ and\ \bibinfo {author}
  {\bibfnamefont {I.~N.}\ \bibnamefont {Krivorotov}},\ }\href {\doibase
  10.1063/1.4826927} {\bibfield  {journal} {\bibinfo  {journal} {Appl. Phys.
  Lett.}\ }\textbf {\bibinfo {volume} {103}},\ \bibinfo {pages} {172406}
  (\bibinfo {year} {2013})}\BibitemShut {NoStop}%
\bibitem [{\citenamefont {Okada}\ \emph {et~al.}(2019)\citenamefont {Okada},
  \citenamefont {Takeuchi}, \citenamefont {Furuya}, \citenamefont {Zhang},
  \citenamefont {Sato}, \citenamefont {Fukami},\ and\ \citenamefont
  {Ohno}}]{Okada2019}%
  \BibitemOpen
  \bibfield  {author} {\bibinfo {author} {\bibfnamefont {A.}~\bibnamefont
  {Okada}}, \bibinfo {author} {\bibfnamefont {Y.}~\bibnamefont {Takeuchi}},
  \bibinfo {author} {\bibfnamefont {K.}~\bibnamefont {Furuya}}, \bibinfo
  {author} {\bibfnamefont {C.}~\bibnamefont {Zhang}}, \bibinfo {author}
  {\bibfnamefont {H.}~\bibnamefont {Sato}}, \bibinfo {author} {\bibfnamefont
  {S.}~\bibnamefont {Fukami}}, \ and\ \bibinfo {author} {\bibfnamefont
  {H.}~\bibnamefont {Ohno}},\ }\href {\doibase
  10.1103/PhysRevApplied.12.014040} {\bibfield  {journal} {\bibinfo  {journal}
  {Phys. Rev. Appl.}\ }\textbf {\bibinfo {volume} {12}},\ \bibinfo {pages}
  {014040} (\bibinfo {year} {2019})}\BibitemShut {NoStop}%
\bibitem [{\citenamefont {Schultheiss}\ \emph {et~al.}(2012)\citenamefont
  {Schultheiss}, \citenamefont {Pearson}, \citenamefont {Bader},\ and\
  \citenamefont {Hoffmann}}]{Schultheiss2012}%
  \BibitemOpen
  \bibfield  {author} {\bibinfo {author} {\bibfnamefont {H.}~\bibnamefont
  {Schultheiss}}, \bibinfo {author} {\bibfnamefont {J.~E.}\ \bibnamefont
  {Pearson}}, \bibinfo {author} {\bibfnamefont {S.~D.}\ \bibnamefont {Bader}},
  \ and\ \bibinfo {author} {\bibfnamefont {A.}~\bibnamefont {Hoffmann}},\
  }\href {\doibase 10.1103/PhysRevLett.109.237204} {\bibfield  {journal}
  {\bibinfo  {journal} {Phys. Rev. Lett.}\ }\textbf {\bibinfo {volume} {109}},\
  \bibinfo {pages} {237204} (\bibinfo {year} {2012})}\BibitemShut {NoStop}%
\bibitem [{\citenamefont {Safranski}\ \emph {et~al.}(2017)\citenamefont
  {Safranski}, \citenamefont {Barsukov}, \citenamefont {Lee}, \citenamefont
  {Schneider}, \citenamefont {Jara}, \citenamefont {Smith}, \citenamefont
  {Chang}, \citenamefont {Lenz}, \citenamefont {Lindner}, \citenamefont
  {Tserkovnyak}, \citenamefont {Wu},\ and\ \citenamefont
  {Krivorotov}}]{Safranski2017}%
  \BibitemOpen
  \bibfield  {author} {\bibinfo {author} {\bibfnamefont {C.}~\bibnamefont
  {Safranski}}, \bibinfo {author} {\bibfnamefont {I.}~\bibnamefont {Barsukov}},
  \bibinfo {author} {\bibfnamefont {H.~K.}\ \bibnamefont {Lee}}, \bibinfo
  {author} {\bibfnamefont {T.}~\bibnamefont {Schneider}}, \bibinfo {author}
  {\bibfnamefont {A.~A.}\ \bibnamefont {Jara}}, \bibinfo {author}
  {\bibfnamefont {A.}~\bibnamefont {Smith}}, \bibinfo {author} {\bibfnamefont
  {H.}~\bibnamefont {Chang}}, \bibinfo {author} {\bibfnamefont
  {K.}~\bibnamefont {Lenz}}, \bibinfo {author} {\bibfnamefont {J.}~\bibnamefont
  {Lindner}}, \bibinfo {author} {\bibfnamefont {Y.}~\bibnamefont
  {Tserkovnyak}}, \bibinfo {author} {\bibfnamefont {M.}~\bibnamefont {Wu}}, \
  and\ \bibinfo {author} {\bibfnamefont {I.~N.}\ \bibnamefont {Krivorotov}},\
  }\href {\doibase 10.1038/s41467-017-00184-5} {\bibfield  {journal} {\bibinfo
  {journal} {Nat. Commun.}\ }\textbf {\bibinfo {volume} {8}},\ \bibinfo {pages}
  {117} (\bibinfo {year} {2017})}\BibitemShut {NoStop}%
\bibitem [{\citenamefont {Kalitsov}\ \emph {et~al.}(2017)\citenamefont
  {Kalitsov}, \citenamefont {Nikolaev}, \citenamefont {Velev}, \citenamefont
  {Chshiev},\ and\ \citenamefont {Mryasov}}]{Kalitsov2017}%
  \BibitemOpen
  \bibfield  {author} {\bibinfo {author} {\bibfnamefont {A.}~\bibnamefont
  {Kalitsov}}, \bibinfo {author} {\bibfnamefont {S.~A.}\ \bibnamefont
  {Nikolaev}}, \bibinfo {author} {\bibfnamefont {J.}~\bibnamefont {Velev}},
  \bibinfo {author} {\bibfnamefont {M.}~\bibnamefont {Chshiev}}, \ and\
  \bibinfo {author} {\bibfnamefont {O.}~\bibnamefont {Mryasov}},\ }\href
  {\doibase 10.1103/PhysRevB.96.214430} {\bibfield  {journal} {\bibinfo
  {journal} {Phys. Rev. B}\ }\textbf {\bibinfo {volume} {96}},\ \bibinfo
  {pages} {214430} (\bibinfo {year} {2017})}\BibitemShut {NoStop}%
\bibitem [{\citenamefont {Roy}(2020)}]{Roy2020}%
  \BibitemOpen
  \bibfield  {author} {\bibinfo {author} {\bibfnamefont {K.}~\bibnamefont
  {Roy}},\ }\href {\doibase 10.1063/5.0014270} {\bibfield  {journal} {\bibinfo
  {journal} {Appl. Phys. Lett.}\ }\textbf {\bibinfo {volume} {117}},\ \bibinfo
  {pages} {022404} (\bibinfo {year} {2020})},\ \Eprint
  {http://arxiv.org/abs/2011.11055} {arXiv:2011.11055} \BibitemShut {NoStop}%
\bibitem [{\citenamefont {Tserkovnyak}\ \emph {et~al.}(2002)\citenamefont
  {Tserkovnyak}, \citenamefont {Brataas},\ and\ \citenamefont
  {Bauer}}]{Tserkovnyak2002}%
  \BibitemOpen
  \bibfield  {author} {\bibinfo {author} {\bibfnamefont {Y.}~\bibnamefont
  {Tserkovnyak}}, \bibinfo {author} {\bibfnamefont {A.}~\bibnamefont
  {Brataas}}, \ and\ \bibinfo {author} {\bibfnamefont {G.}~\bibnamefont
  {Bauer}},\ }\href {\doibase 10.1103/PhysRevB.66.224403} {\bibfield  {journal}
  {\bibinfo  {journal} {Phys. Rev. B}\ }\textbf {\bibinfo {volume} {66}},\
  \bibinfo {pages} {224403} (\bibinfo {year} {2002})}\BibitemShut {NoStop}%
\bibitem [{\citenamefont {Brataas}\ \emph {et~al.}(2011)\citenamefont
  {Brataas}, \citenamefont {Tserkovnyak}, \citenamefont {Bauer},\ and\
  \citenamefont {Kelly}}]{Brataas2011}%
  \BibitemOpen
  \bibfield  {author} {\bibinfo {author} {\bibfnamefont {A.}~\bibnamefont
  {Brataas}}, \bibinfo {author} {\bibfnamefont {Y.}~\bibnamefont
  {Tserkovnyak}}, \bibinfo {author} {\bibfnamefont {G.~E.~W.}\ \bibnamefont
  {Bauer}}, \ and\ \bibinfo {author} {\bibfnamefont {P.~J.}\ \bibnamefont
  {Kelly}},\ }\href {http://arxiv.org/abs/1108.0385} {\  (\bibinfo {year}
  {2011})},\ \Eprint {http://arxiv.org/abs/1108.0385} {arXiv:1108.0385}
  \BibitemShut {NoStop}%
\bibitem [{\citenamefont {Wang}\ \emph {et~al.}(2016)\citenamefont {Wang},
  \citenamefont {Wesselink}, \citenamefont {Liu}, \citenamefont {Yuan},
  \citenamefont {Xia},\ and\ \citenamefont {Kelly}}]{Wang2016}%
  \BibitemOpen
  \bibfield  {author} {\bibinfo {author} {\bibfnamefont {L.}~\bibnamefont
  {Wang}}, \bibinfo {author} {\bibfnamefont {R.~J.~H.}\ \bibnamefont
  {Wesselink}}, \bibinfo {author} {\bibfnamefont {Y.}~\bibnamefont {Liu}},
  \bibinfo {author} {\bibfnamefont {Z.}~\bibnamefont {Yuan}}, \bibinfo {author}
  {\bibfnamefont {K.}~\bibnamefont {Xia}}, \ and\ \bibinfo {author}
  {\bibfnamefont {P.~J.}\ \bibnamefont {Kelly}},\ }\href {\doibase
  10.1103/PhysRevLett.116.196602} {\bibfield  {journal} {\bibinfo  {journal}
  {Phys. Rev. Lett.}\ }\textbf {\bibinfo {volume} {116}},\ \bibinfo {pages}
  {196602} (\bibinfo {year} {2016})}\BibitemShut {NoStop}%
\bibitem [{\citenamefont {Long}\ \emph {et~al.}(2016)\citenamefont {Long},
  \citenamefont {Mavropoulos}, \citenamefont {Bauer}, \citenamefont
  {Zimmermann}, \citenamefont {Mokrousov},\ and\ \citenamefont
  {Bl{\"{u}}gel}}]{Long2016}%
  \BibitemOpen
  \bibfield  {author} {\bibinfo {author} {\bibfnamefont {N.~H.}\ \bibnamefont
  {Long}}, \bibinfo {author} {\bibfnamefont {P.}~\bibnamefont {Mavropoulos}},
  \bibinfo {author} {\bibfnamefont {D.~S.~G.}\ \bibnamefont {Bauer}}, \bibinfo
  {author} {\bibfnamefont {B.}~\bibnamefont {Zimmermann}}, \bibinfo {author}
  {\bibfnamefont {Y.}~\bibnamefont {Mokrousov}}, \ and\ \bibinfo {author}
  {\bibfnamefont {S.}~\bibnamefont {Bl{\"{u}}gel}},\ }\href {\doibase
  10.1103/PhysRevB.94.180406} {\bibfield  {journal} {\bibinfo  {journal} {Phys.
  Rev. B}\ }\textbf {\bibinfo {volume} {94}},\ \bibinfo {pages} {180406}
  (\bibinfo {year} {2016})}\BibitemShut {NoStop}%
\bibitem [{\citenamefont {Gray}\ \emph
  {et~al.}(2018{\natexlab{b}})\citenamefont {Gray}, \citenamefont {Emori},
  \citenamefont {Gray}, \citenamefont {Jeon}, \citenamefont {{Van 'T Erve}},
  \citenamefont {Jonker}, \citenamefont {Kim}, \citenamefont {Suzuki},
  \citenamefont {Ono}, \citenamefont {Howe},\ and\ \citenamefont
  {Suzuki}}]{Gray2018b}%
  \BibitemOpen
  \bibfield  {author} {\bibinfo {author} {\bibfnamefont {M.}~\bibnamefont
  {Gray}}, \bibinfo {author} {\bibfnamefont {S.}~\bibnamefont {Emori}},
  \bibinfo {author} {\bibfnamefont {B.}~\bibnamefont {Gray}}, \bibinfo {author}
  {\bibfnamefont {H.}~\bibnamefont {Jeon}}, \bibinfo {author} {\bibfnamefont
  {O.}~\bibnamefont {{Van 'T Erve}}}, \bibinfo {author} {\bibfnamefont
  {B.}~\bibnamefont {Jonker}}, \bibinfo {author} {\bibfnamefont
  {S.}~\bibnamefont {Kim}}, \bibinfo {author} {\bibfnamefont {M.}~\bibnamefont
  {Suzuki}}, \bibinfo {author} {\bibfnamefont {T.}~\bibnamefont {Ono}},
  \bibinfo {author} {\bibfnamefont {B.}~\bibnamefont {Howe}}, \ and\ \bibinfo
  {author} {\bibfnamefont {Y.}~\bibnamefont {Suzuki}},\ }\href {\doibase
  10.1103/PhysRevApplied.9.064039} {\bibfield  {journal} {\bibinfo  {journal}
  {Phys. Rev. Appl.}\ }\textbf {\bibinfo {volume} {9}},\ \bibinfo {pages}
  {064039} (\bibinfo {year} {2018}{\natexlab{b}})}\BibitemShut {NoStop}%
\bibitem [{\citenamefont {Althammer}\ \emph {et~al.}(2013)\citenamefont
  {Althammer}, \citenamefont {Meyer}, \citenamefont {Nakayama}, \citenamefont
  {Schreier}, \citenamefont {Altmannshofer}, \citenamefont {Weiler},
  \citenamefont {Huebl}, \citenamefont {Gepr{\"{a}}gs}, \citenamefont {Opel},
  \citenamefont {Gross}, \citenamefont {Meier}, \citenamefont {Klewe},
  \citenamefont {Kuschel}, \citenamefont {Schmalhorst}, \citenamefont {Reiss},
  \citenamefont {Shen}, \citenamefont {Gupta}, \citenamefont {Chen},
  \citenamefont {Bauer}, \citenamefont {Saitoh},\ and\ \citenamefont
  {Goennenwein}}]{Althammer2013}%
  \BibitemOpen
  \bibfield  {author} {\bibinfo {author} {\bibfnamefont {M.}~\bibnamefont
  {Althammer}}, \bibinfo {author} {\bibfnamefont {S.}~\bibnamefont {Meyer}},
  \bibinfo {author} {\bibfnamefont {H.}~\bibnamefont {Nakayama}}, \bibinfo
  {author} {\bibfnamefont {M.}~\bibnamefont {Schreier}}, \bibinfo {author}
  {\bibfnamefont {S.}~\bibnamefont {Altmannshofer}}, \bibinfo {author}
  {\bibfnamefont {M.}~\bibnamefont {Weiler}}, \bibinfo {author} {\bibfnamefont
  {H.}~\bibnamefont {Huebl}}, \bibinfo {author} {\bibfnamefont
  {S.}~\bibnamefont {Gepr{\"{a}}gs}}, \bibinfo {author} {\bibfnamefont
  {M.}~\bibnamefont {Opel}}, \bibinfo {author} {\bibfnamefont {R.}~\bibnamefont
  {Gross}}, \bibinfo {author} {\bibfnamefont {D.}~\bibnamefont {Meier}},
  \bibinfo {author} {\bibfnamefont {C.}~\bibnamefont {Klewe}}, \bibinfo
  {author} {\bibfnamefont {T.}~\bibnamefont {Kuschel}}, \bibinfo {author}
  {\bibfnamefont {J.-M.}\ \bibnamefont {Schmalhorst}}, \bibinfo {author}
  {\bibfnamefont {G.}~\bibnamefont {Reiss}}, \bibinfo {author} {\bibfnamefont
  {L.}~\bibnamefont {Shen}}, \bibinfo {author} {\bibfnamefont {A.}~\bibnamefont
  {Gupta}}, \bibinfo {author} {\bibfnamefont {Y.-T.}\ \bibnamefont {Chen}},
  \bibinfo {author} {\bibfnamefont {G.~E.~W.}\ \bibnamefont {Bauer}}, \bibinfo
  {author} {\bibfnamefont {E.}~\bibnamefont {Saitoh}}, \ and\ \bibinfo {author}
  {\bibfnamefont {S.~T.~B.}\ \bibnamefont {Goennenwein}},\ }\href {\doibase
  10.1103/PhysRevB.87.224401} {\bibfield  {journal} {\bibinfo  {journal} {Phys.
  Rev. B}\ }\textbf {\bibinfo {volume} {87}},\ \bibinfo {pages} {224401}
  (\bibinfo {year} {2013})}\BibitemShut {NoStop}%
\bibitem [{\citenamefont {Chen}\ \emph {et~al.}(2013)\citenamefont {Chen},
  \citenamefont {Takahashi}, \citenamefont {Nakayama}, \citenamefont
  {Althammer}, \citenamefont {Goennenwein}, \citenamefont {Saitoh},\ and\
  \citenamefont {Bauer}}]{Chen2013}%
  \BibitemOpen
  \bibfield  {author} {\bibinfo {author} {\bibfnamefont {Y.-T.}\ \bibnamefont
  {Chen}}, \bibinfo {author} {\bibfnamefont {S.}~\bibnamefont {Takahashi}},
  \bibinfo {author} {\bibfnamefont {H.}~\bibnamefont {Nakayama}}, \bibinfo
  {author} {\bibfnamefont {M.}~\bibnamefont {Althammer}}, \bibinfo {author}
  {\bibfnamefont {S.}~\bibnamefont {Goennenwein}}, \bibinfo {author}
  {\bibfnamefont {E.}~\bibnamefont {Saitoh}}, \ and\ \bibinfo {author}
  {\bibfnamefont {G.}~\bibnamefont {Bauer}},\ }\href {\doibase
  10.1103/PhysRevB.87.144411} {\bibfield  {journal} {\bibinfo  {journal} {Phys.
  Rev. B}\ }\textbf {\bibinfo {volume} {87}},\ \bibinfo {pages} {144411}
  (\bibinfo {year} {2013})}\BibitemShut {NoStop}%
\bibitem [{\citenamefont {Sun}\ \emph {et~al.}(2013)\citenamefont {Sun},
  \citenamefont {Chang}, \citenamefont {Kabatek}, \citenamefont {Song},
  \citenamefont {Wang}, \citenamefont {Jantz}, \citenamefont {Schneider},
  \citenamefont {Wu}, \citenamefont {Montoya}, \citenamefont {Kardasz},
  \citenamefont {Heinrich}, \citenamefont {te~Velthuis}, \citenamefont
  {Schultheiss},\ and\ \citenamefont {Hoffmann}}]{Sun2013}%
  \BibitemOpen
  \bibfield  {author} {\bibinfo {author} {\bibfnamefont {Y.}~\bibnamefont
  {Sun}}, \bibinfo {author} {\bibfnamefont {H.}~\bibnamefont {Chang}}, \bibinfo
  {author} {\bibfnamefont {M.}~\bibnamefont {Kabatek}}, \bibinfo {author}
  {\bibfnamefont {Y.-Y.}\ \bibnamefont {Song}}, \bibinfo {author}
  {\bibfnamefont {Z.}~\bibnamefont {Wang}}, \bibinfo {author} {\bibfnamefont
  {M.}~\bibnamefont {Jantz}}, \bibinfo {author} {\bibfnamefont
  {W.}~\bibnamefont {Schneider}}, \bibinfo {author} {\bibfnamefont
  {M.}~\bibnamefont {Wu}}, \bibinfo {author} {\bibfnamefont {E.}~\bibnamefont
  {Montoya}}, \bibinfo {author} {\bibfnamefont {B.}~\bibnamefont {Kardasz}},
  \bibinfo {author} {\bibfnamefont {B.}~\bibnamefont {Heinrich}}, \bibinfo
  {author} {\bibfnamefont {S.~G.~E.}\ \bibnamefont {te~Velthuis}}, \bibinfo
  {author} {\bibfnamefont {H.}~\bibnamefont {Schultheiss}}, \ and\ \bibinfo
  {author} {\bibfnamefont {A.}~\bibnamefont {Hoffmann}},\ }\href {\doibase
  10.1103/PhysRevLett.111.106601} {\bibfield  {journal} {\bibinfo  {journal}
  {Phys. Rev. Lett.}\ }\textbf {\bibinfo {volume} {111}},\ \bibinfo {pages}
  {106601} (\bibinfo {year} {2013})}\BibitemShut {NoStop}%
\bibitem [{\citenamefont {Hahn}\ \emph {et~al.}(2013)\citenamefont {Hahn},
  \citenamefont {de~Loubens}, \citenamefont {Klein}, \citenamefont {Viret},
  \citenamefont {Naletov},\ and\ \citenamefont {{Ben Youssef}}}]{Hahn2013a}%
  \BibitemOpen
  \bibfield  {author} {\bibinfo {author} {\bibfnamefont {C.}~\bibnamefont
  {Hahn}}, \bibinfo {author} {\bibfnamefont {G.}~\bibnamefont {de~Loubens}},
  \bibinfo {author} {\bibfnamefont {O.}~\bibnamefont {Klein}}, \bibinfo
  {author} {\bibfnamefont {M.}~\bibnamefont {Viret}}, \bibinfo {author}
  {\bibfnamefont {V.~V.}\ \bibnamefont {Naletov}}, \ and\ \bibinfo {author}
  {\bibfnamefont {J.}~\bibnamefont {{Ben Youssef}}},\ }\href {\doibase
  10.1103/PhysRevB.87.174417} {\bibfield  {journal} {\bibinfo  {journal} {Phys.
  Rev. B}\ }\textbf {\bibinfo {volume} {87}},\ \bibinfo {pages} {174417}
  (\bibinfo {year} {2013})}\BibitemShut {NoStop}%
\bibitem [{\citenamefont {Weiler}\ \emph {et~al.}(2014)\citenamefont {Weiler},
  \citenamefont {Shaw}, \citenamefont {Nembach},\ and\ \citenamefont
  {Silva}}]{Weiler2014b}%
  \BibitemOpen
  \bibfield  {author} {\bibinfo {author} {\bibfnamefont {M.}~\bibnamefont
  {Weiler}}, \bibinfo {author} {\bibfnamefont {J.~M.}\ \bibnamefont {Shaw}},
  \bibinfo {author} {\bibfnamefont {H.~T.}\ \bibnamefont {Nembach}}, \ and\
  \bibinfo {author} {\bibfnamefont {T.~J.}\ \bibnamefont {Silva}},\ }\href
  {\doibase 10.1103/PhysRevLett.113.157204} {\bibfield  {journal} {\bibinfo
  {journal} {Phys. Rev. Lett.}\ }\textbf {\bibinfo {volume} {113}},\ \bibinfo
  {pages} {157204} (\bibinfo {year} {2014})}\BibitemShut {NoStop}%
\bibitem [{\citenamefont {Keller}\ \emph {et~al.}(2019)\citenamefont {Keller},
  \citenamefont {Gerace}, \citenamefont {Arora}, \citenamefont
  {Delczeg-Czirjak}, \citenamefont {Shaw},\ and\ \citenamefont
  {Silva}}]{Keller2019}%
  \BibitemOpen
  \bibfield  {author} {\bibinfo {author} {\bibfnamefont {M.~W.}\ \bibnamefont
  {Keller}}, \bibinfo {author} {\bibfnamefont {K.~S.}\ \bibnamefont {Gerace}},
  \bibinfo {author} {\bibfnamefont {M.}~\bibnamefont {Arora}}, \bibinfo
  {author} {\bibfnamefont {E.~K.}\ \bibnamefont {Delczeg-Czirjak}}, \bibinfo
  {author} {\bibfnamefont {J.~M.}\ \bibnamefont {Shaw}}, \ and\ \bibinfo
  {author} {\bibfnamefont {T.~J.}\ \bibnamefont {Silva}},\ }\href {\doibase
  10.1103/PhysRevB.99.214411} {\bibfield  {journal} {\bibinfo  {journal} {Phys.
  Rev. B}\ }\textbf {\bibinfo {volume} {99}},\ \bibinfo {pages} {214411}
  (\bibinfo {year} {2019})}\BibitemShut {NoStop}%
\bibitem [{\citenamefont {Huo}\ \emph {et~al.}(2017)\citenamefont {Huo},
  \citenamefont {Zeng}, \citenamefont {Zhou},\ and\ \citenamefont
  {Wu}}]{Huo2017}%
  \BibitemOpen
  \bibfield  {author} {\bibinfo {author} {\bibfnamefont {Y.}~\bibnamefont
  {Huo}}, \bibinfo {author} {\bibfnamefont {F.~L.}\ \bibnamefont {Zeng}},
  \bibinfo {author} {\bibfnamefont {C.}~\bibnamefont {Zhou}}, \ and\ \bibinfo
  {author} {\bibfnamefont {Y.~Z.}\ \bibnamefont {Wu}},\ }\href {\doibase
  10.1063/1.4976957} {\bibfield  {journal} {\bibinfo  {journal} {AIP Adv.}\
  }\textbf {\bibinfo {volume} {7}},\ \bibinfo {pages} {056024} (\bibinfo {year}
  {2017})}\BibitemShut {NoStop}%
\bibitem [{\citenamefont {Keller}\ \emph {et~al.}(2018)\citenamefont {Keller},
  \citenamefont {Mihalceanu}, \citenamefont {Schweizer}, \citenamefont {Lang},
  \citenamefont {Heinz}, \citenamefont {Geilen}, \citenamefont {Br{\"{a}}cher},
  \citenamefont {Pirro}, \citenamefont {Meyer}, \citenamefont {Conca},
  \citenamefont {Karfaridis}, \citenamefont {Vourlias}, \citenamefont
  {Kehagias}, \citenamefont {Hillebrands},\ and\ \citenamefont
  {Papaioannou}}]{Keller2018}%
  \BibitemOpen
  \bibfield  {author} {\bibinfo {author} {\bibfnamefont {S.}~\bibnamefont
  {Keller}}, \bibinfo {author} {\bibfnamefont {L.}~\bibnamefont {Mihalceanu}},
  \bibinfo {author} {\bibfnamefont {M.~R.}\ \bibnamefont {Schweizer}}, \bibinfo
  {author} {\bibfnamefont {P.}~\bibnamefont {Lang}}, \bibinfo {author}
  {\bibfnamefont {B.}~\bibnamefont {Heinz}}, \bibinfo {author} {\bibfnamefont
  {M.}~\bibnamefont {Geilen}}, \bibinfo {author} {\bibfnamefont
  {T.}~\bibnamefont {Br{\"{a}}cher}}, \bibinfo {author} {\bibfnamefont
  {P.}~\bibnamefont {Pirro}}, \bibinfo {author} {\bibfnamefont
  {T.}~\bibnamefont {Meyer}}, \bibinfo {author} {\bibfnamefont
  {A.}~\bibnamefont {Conca}}, \bibinfo {author} {\bibfnamefont
  {D.}~\bibnamefont {Karfaridis}}, \bibinfo {author} {\bibfnamefont
  {G.}~\bibnamefont {Vourlias}}, \bibinfo {author} {\bibfnamefont
  {T.}~\bibnamefont {Kehagias}}, \bibinfo {author} {\bibfnamefont
  {B.}~\bibnamefont {Hillebrands}}, \ and\ \bibinfo {author} {\bibfnamefont
  {E.~T.}\ \bibnamefont {Papaioannou}},\ }\href {\doibase
  10.1088/1367-2630/aabc46} {\bibfield  {journal} {\bibinfo  {journal} {New J.
  Phys.}\ }\textbf {\bibinfo {volume} {20}},\ \bibinfo {pages} {053002}
  (\bibinfo {year} {2018})}\BibitemShut {NoStop}%
\bibitem [{\citenamefont {Guillemard}\ \emph {et~al.}(2018)\citenamefont
  {Guillemard}, \citenamefont {Petit-Watelot}, \citenamefont {Andrieu},\ and\
  \citenamefont {Rojas-S{\'{a}}nchez}}]{Guillemard2018}%
  \BibitemOpen
  \bibfield  {author} {\bibinfo {author} {\bibfnamefont {C.}~\bibnamefont
  {Guillemard}}, \bibinfo {author} {\bibfnamefont {S.}~\bibnamefont
  {Petit-Watelot}}, \bibinfo {author} {\bibfnamefont {S.}~\bibnamefont
  {Andrieu}}, \ and\ \bibinfo {author} {\bibfnamefont {J.-C.}\ \bibnamefont
  {Rojas-S{\'{a}}nchez}},\ }\href {\doibase 10.1063/1.5079236} {\bibfield
  {journal} {\bibinfo  {journal} {Appl. Phys. Lett.}\ }\textbf {\bibinfo
  {volume} {113}},\ \bibinfo {pages} {262404} (\bibinfo {year}
  {2018})}\BibitemShut {NoStop}%
\bibitem [{\citenamefont {Ryu}\ \emph {et~al.}(2019)\citenamefont {Ryu},
  \citenamefont {Avci}, \citenamefont {Karube}, \citenamefont {Kohda},
  \citenamefont {Beach},\ and\ \citenamefont {Nitta}}]{Ryu2019}%
  \BibitemOpen
  \bibfield  {author} {\bibinfo {author} {\bibfnamefont {J.}~\bibnamefont
  {Ryu}}, \bibinfo {author} {\bibfnamefont {C.~O.}\ \bibnamefont {Avci}},
  \bibinfo {author} {\bibfnamefont {S.}~\bibnamefont {Karube}}, \bibinfo
  {author} {\bibfnamefont {M.}~\bibnamefont {Kohda}}, \bibinfo {author}
  {\bibfnamefont {G.~S.~D.}\ \bibnamefont {Beach}}, \ and\ \bibinfo {author}
  {\bibfnamefont {J.}~\bibnamefont {Nitta}},\ }\href {\doibase
  10.1063/1.5090610} {\bibfield  {journal} {\bibinfo  {journal} {Appl. Phys.
  Lett.}\ }\textbf {\bibinfo {volume} {114}},\ \bibinfo {pages} {142402}
  (\bibinfo {year} {2019})}\BibitemShut {NoStop}%
\bibitem [{\citenamefont {Wisser}\ \emph
  {et~al.}(2019{\natexlab{b}})\citenamefont {Wisser}, \citenamefont {Grutter},
  \citenamefont {Gilbert}, \citenamefont {N'Diaye}, \citenamefont {Klewe},
  \citenamefont {Shafer}, \citenamefont {Arenholz}, \citenamefont {Suzuki},\
  and\ \citenamefont {Emori}}]{Wisser2019b}%
  \BibitemOpen
  \bibfield  {author} {\bibinfo {author} {\bibfnamefont {J.~J.}\ \bibnamefont
  {Wisser}}, \bibinfo {author} {\bibfnamefont {A.~J.}\ \bibnamefont {Grutter}},
  \bibinfo {author} {\bibfnamefont {D.~A.}\ \bibnamefont {Gilbert}}, \bibinfo
  {author} {\bibfnamefont {A.~T.}\ \bibnamefont {N'Diaye}}, \bibinfo {author}
  {\bibfnamefont {C.}~\bibnamefont {Klewe}}, \bibinfo {author} {\bibfnamefont
  {P.}~\bibnamefont {Shafer}}, \bibinfo {author} {\bibfnamefont
  {E.}~\bibnamefont {Arenholz}}, \bibinfo {author} {\bibfnamefont
  {Y.}~\bibnamefont {Suzuki}}, \ and\ \bibinfo {author} {\bibfnamefont
  {S.}~\bibnamefont {Emori}},\ }\href {\doibase
  10.1103/PhysRevApplied.12.054044} {\bibfield  {journal} {\bibinfo  {journal}
  {Phys. Rev. Appl.}\ }\textbf {\bibinfo {volume} {12}},\ \bibinfo {pages}
  {054044} (\bibinfo {year} {2019}{\natexlab{b}})}\BibitemShut {NoStop}%
\bibitem [{\citenamefont {Zhu}\ \emph {et~al.}(2019{\natexlab{c}})\citenamefont
  {Zhu}, \citenamefont {Ralph},\ and\ \citenamefont {Buhrman}}]{Zhu2019b}%
  \BibitemOpen
  \bibfield  {author} {\bibinfo {author} {\bibfnamefont {L.}~\bibnamefont
  {Zhu}}, \bibinfo {author} {\bibfnamefont {D.~C.}\ \bibnamefont {Ralph}}, \
  and\ \bibinfo {author} {\bibfnamefont {R.~A.}\ \bibnamefont {Buhrman}},\
  }\href {\doibase 10.1103/PhysRevLett.123.057203} {\bibfield  {journal}
  {\bibinfo  {journal} {Phys. Rev. Lett.}\ }\textbf {\bibinfo {volume} {123}},\
  \bibinfo {pages} {057203} (\bibinfo {year} {2019}{\natexlab{c}})}\BibitemShut
  {NoStop}%
\bibitem [{\citenamefont {Liu}\ \emph {et~al.}(2014)\citenamefont {Liu},
  \citenamefont {Yuan}, \citenamefont {Wesselink}, \citenamefont {Starikov},\
  and\ \citenamefont {Kelly}}]{Liu2014c}%
  \BibitemOpen
  \bibfield  {author} {\bibinfo {author} {\bibfnamefont {Y.}~\bibnamefont
  {Liu}}, \bibinfo {author} {\bibfnamefont {Z.}~\bibnamefont {Yuan}}, \bibinfo
  {author} {\bibfnamefont {R.~J.~H.}\ \bibnamefont {Wesselink}}, \bibinfo
  {author} {\bibfnamefont {A.~A.}\ \bibnamefont {Starikov}}, \ and\ \bibinfo
  {author} {\bibfnamefont {P.~J.}\ \bibnamefont {Kelly}},\ }\href {\doibase
  10.1103/PhysRevLett.113.207202} {\bibfield  {journal} {\bibinfo  {journal}
  {Phys. Rev. Lett.}\ }\textbf {\bibinfo {volume} {113}},\ \bibinfo {pages}
  {207202} (\bibinfo {year} {2014})}\BibitemShut {NoStop}%
\bibitem [{\citenamefont {Mahfouzi}\ \emph {et~al.}(2017)\citenamefont
  {Mahfouzi}, \citenamefont {Kim},\ and\ \citenamefont
  {Kioussis}}]{Mahfouzi2017}%
  \BibitemOpen
  \bibfield  {author} {\bibinfo {author} {\bibfnamefont {F.}~\bibnamefont
  {Mahfouzi}}, \bibinfo {author} {\bibfnamefont {J.}~\bibnamefont {Kim}}, \
  and\ \bibinfo {author} {\bibfnamefont {N.}~\bibnamefont {Kioussis}},\ }\href
  {\doibase 10.1103/PhysRevB.96.214421} {\bibfield  {journal} {\bibinfo
  {journal} {Phys. Rev. B}\ }\textbf {\bibinfo {volume} {96}},\ \bibinfo
  {pages} {214421} (\bibinfo {year} {2017})}\BibitemShut {NoStop}%
\bibitem [{\citenamefont {Zhao}\ \emph {et~al.}(2018)\citenamefont {Zhao},
  \citenamefont {Liu}, \citenamefont {Tang}, \citenamefont {Jiang},
  \citenamefont {Yuan},\ and\ \citenamefont {Xia}}]{Zhao2018}%
  \BibitemOpen
  \bibfield  {author} {\bibinfo {author} {\bibfnamefont {Y.}~\bibnamefont
  {Zhao}}, \bibinfo {author} {\bibfnamefont {Y.}~\bibnamefont {Liu}}, \bibinfo
  {author} {\bibfnamefont {H.}~\bibnamefont {Tang}}, \bibinfo {author}
  {\bibfnamefont {H.}~\bibnamefont {Jiang}}, \bibinfo {author} {\bibfnamefont
  {Z.}~\bibnamefont {Yuan}}, \ and\ \bibinfo {author} {\bibfnamefont
  {K.}~\bibnamefont {Xia}},\ }\href {\doibase 10.1103/PhysRevB.98.174412}
  {\bibfield  {journal} {\bibinfo  {journal} {Phys. Rev. B}\ }\textbf {\bibinfo
  {volume} {98}},\ \bibinfo {pages} {174412} (\bibinfo {year}
  {2018})}\BibitemShut {NoStop}%
\bibitem [{\citenamefont {Zhu}\ \emph {et~al.}(2021)\citenamefont {Zhu},
  \citenamefont {Zhu},\ and\ \citenamefont {Buhrman}}]{Zhu2021}%
  \BibitemOpen
  \bibfield  {author} {\bibinfo {author} {\bibfnamefont {L.}~\bibnamefont
  {Zhu}}, \bibinfo {author} {\bibfnamefont {L.}~\bibnamefont {Zhu}}, \ and\
  \bibinfo {author} {\bibfnamefont {R.~A.}\ \bibnamefont {Buhrman}},\ }\href
  {\doibase 10.1103/physrevlett.126.107204} {\bibfield  {journal} {\bibinfo
  {journal} {Phys. Rev. Lett.}\ }\textbf {\bibinfo {volume} {126}},\ \bibinfo
  {pages} {107204} (\bibinfo {year} {2021})},\ \Eprint
  {http://arxiv.org/abs/2102.08487} {arXiv:2102.08487} \BibitemShut {NoStop}%
\bibitem [{\citenamefont {Brataas}\ \emph {et~al.}(2012)\citenamefont
  {Brataas}, \citenamefont {Tserkovnyak}, \citenamefont {Bauer},\ and\
  \citenamefont {Kelly}}]{Brataas2012a}%
  \BibitemOpen
  \bibfield  {author} {\bibinfo {author} {\bibfnamefont {A.}~\bibnamefont
  {Brataas}}, \bibinfo {author} {\bibfnamefont {Y.}~\bibnamefont
  {Tserkovnyak}}, \bibinfo {author} {\bibfnamefont {G.~E.~W.}\ \bibnamefont
  {Bauer}}, \ and\ \bibinfo {author} {\bibfnamefont {P.~J.}\ \bibnamefont
  {Kelly}},\ }in\ \href@noop {} {\emph {\bibinfo {booktitle} {Spin Current}}}\
  (\bibinfo {year} {2012})\ Chap.~\bibinfo {chapter} {8}, pp.\ \bibinfo {pages}
  {87--135}\BibitemShut {NoStop}%
\bibitem [{\citenamefont {Tanaka}\ \emph {et~al.}(2008)\citenamefont {Tanaka},
  \citenamefont {Kontani}, \citenamefont {Naito}, \citenamefont {Naito},
  \citenamefont {Hirashima}, \citenamefont {Yamada},\ and\ \citenamefont
  {Inoue}}]{Tanaka2008}%
  \BibitemOpen
  \bibfield  {author} {\bibinfo {author} {\bibfnamefont {T.}~\bibnamefont
  {Tanaka}}, \bibinfo {author} {\bibfnamefont {H.}~\bibnamefont {Kontani}},
  \bibinfo {author} {\bibfnamefont {M.}~\bibnamefont {Naito}}, \bibinfo
  {author} {\bibfnamefont {T.}~\bibnamefont {Naito}}, \bibinfo {author}
  {\bibfnamefont {D.}~\bibnamefont {Hirashima}}, \bibinfo {author}
  {\bibfnamefont {K.}~\bibnamefont {Yamada}}, \ and\ \bibinfo {author}
  {\bibfnamefont {J.}~\bibnamefont {Inoue}},\ }\href {\doibase
  10.1103/PhysRevB.77.165117} {\bibfield  {journal} {\bibinfo  {journal} {Phys.
  Rev. B}\ }\textbf {\bibinfo {volume} {77}},\ \bibinfo {pages} {165117}
  (\bibinfo {year} {2008})}\BibitemShut {NoStop}%
\bibitem [{\citenamefont {Guo}\ \emph {et~al.}(2008)\citenamefont {Guo},
  \citenamefont {Murakami}, \citenamefont {Chen},\ and\ \citenamefont
  {Nagaosa}}]{Guo2008}%
  \BibitemOpen
  \bibfield  {author} {\bibinfo {author} {\bibfnamefont {G.~Y.}\ \bibnamefont
  {Guo}}, \bibinfo {author} {\bibfnamefont {S.}~\bibnamefont {Murakami}},
  \bibinfo {author} {\bibfnamefont {T.-W.}\ \bibnamefont {Chen}}, \ and\
  \bibinfo {author} {\bibfnamefont {N.}~\bibnamefont {Nagaosa}},\ }\href
  {\doibase 10.1103/PhysRevLett.100.096401} {\bibfield  {journal} {\bibinfo
  {journal} {Phys. Rev. Lett.}\ }\textbf {\bibinfo {volume} {100}},\ \bibinfo
  {pages} {096401} (\bibinfo {year} {2008})}\BibitemShut {NoStop}%
\bibitem [{\citenamefont {Obstbaum}\ \emph {et~al.}(2016)\citenamefont
  {Obstbaum}, \citenamefont {Decker}, \citenamefont {Greitner}, \citenamefont
  {Haertinger}, \citenamefont {Meier}, \citenamefont {Kronseder}, \citenamefont
  {Chadova}, \citenamefont {Wimmer}, \citenamefont {K{\"{o}}dderitzsch},
  \citenamefont {Ebert},\ and\ \citenamefont {Back}}]{Obstbaum2016}%
  \BibitemOpen
  \bibfield  {author} {\bibinfo {author} {\bibfnamefont {M.}~\bibnamefont
  {Obstbaum}}, \bibinfo {author} {\bibfnamefont {M.}~\bibnamefont {Decker}},
  \bibinfo {author} {\bibfnamefont {A.~K.}\ \bibnamefont {Greitner}}, \bibinfo
  {author} {\bibfnamefont {M.}~\bibnamefont {Haertinger}}, \bibinfo {author}
  {\bibfnamefont {T.~N.~G.}\ \bibnamefont {Meier}}, \bibinfo {author}
  {\bibfnamefont {M.}~\bibnamefont {Kronseder}}, \bibinfo {author}
  {\bibfnamefont {K.}~\bibnamefont {Chadova}}, \bibinfo {author} {\bibfnamefont
  {S.}~\bibnamefont {Wimmer}}, \bibinfo {author} {\bibfnamefont
  {D.}~\bibnamefont {K{\"{o}}dderitzsch}}, \bibinfo {author} {\bibfnamefont
  {H.}~\bibnamefont {Ebert}}, \ and\ \bibinfo {author} {\bibfnamefont {C.~H.}\
  \bibnamefont {Back}},\ }\href {\doibase 10.1103/PhysRevLett.117.167204}
  {\bibfield  {journal} {\bibinfo  {journal} {Phys. Rev. Lett.}\ }\textbf
  {\bibinfo {volume} {117}},\ \bibinfo {pages} {167204} (\bibinfo {year}
  {2016})}\BibitemShut {NoStop}%
\bibitem [{\citenamefont {Go}\ \emph {et~al.}(2018)\citenamefont {Go},
  \citenamefont {Jo}, \citenamefont {Kim},\ and\ \citenamefont {Lee}}]{Go2018}%
  \BibitemOpen
  \bibfield  {author} {\bibinfo {author} {\bibfnamefont {D.}~\bibnamefont
  {Go}}, \bibinfo {author} {\bibfnamefont {D.}~\bibnamefont {Jo}}, \bibinfo
  {author} {\bibfnamefont {C.}~\bibnamefont {Kim}}, \ and\ \bibinfo {author}
  {\bibfnamefont {H.-W.}\ \bibnamefont {Lee}},\ }\href {\doibase
  10.1103/PhysRevLett.121.086602} {\bibfield  {journal} {\bibinfo  {journal}
  {Phys. Rev. Lett.}\ }\textbf {\bibinfo {volume} {121}},\ \bibinfo {pages}
  {086602} (\bibinfo {year} {2018})}\BibitemShut {NoStop}%
\bibitem [{\citenamefont {Derunova}\ \emph {et~al.}(2019)\citenamefont
  {Derunova}, \citenamefont {Sun}, \citenamefont {Felser}, \citenamefont
  {Parkin}, \citenamefont {Yan},\ and\ \citenamefont {Ali}}]{Derunova2019}%
  \BibitemOpen
  \bibfield  {author} {\bibinfo {author} {\bibfnamefont {E.}~\bibnamefont
  {Derunova}}, \bibinfo {author} {\bibfnamefont {Y.}~\bibnamefont {Sun}},
  \bibinfo {author} {\bibfnamefont {C.}~\bibnamefont {Felser}}, \bibinfo
  {author} {\bibfnamefont {S.~S.~P.}\ \bibnamefont {Parkin}}, \bibinfo {author}
  {\bibfnamefont {B.}~\bibnamefont {Yan}}, \ and\ \bibinfo {author}
  {\bibfnamefont {M.~N.}\ \bibnamefont {Ali}},\ }\href {\doibase
  10.1126/sciadv.aav8575} {\bibfield  {journal} {\bibinfo  {journal} {Sci.
  Adv.}\ }\textbf {\bibinfo {volume} {5}},\ \bibinfo {pages} {eaav8575}
  (\bibinfo {year} {2019})}\BibitemShut {NoStop}%
\bibitem [{Note1()}]{Note1}%
  \BibitemOpen
  \bibinfo {note} {We used the MATLAB function of nlinmultifit, which is a
  wrapper for nlinfit that allows for simultaneous fitting of multiple data
  sets with shared parameters, available at: \protect \url
  {https://www.mathworks.com/matlabcentral/fileexchange/40613-multiple-curve-fitting-with-common-parameters-using-nlinfit}}\BibitemShut
  {NoStop}%
\end{thebibliography}

%

\end{document}